\documentclass[preprint,12pt,authoryear]{elsarticle}
\usepackage[margin=2.54cm]{geometry}
\usepackage{microtype}
\usepackage{amssymb}
\usepackage{amsmath}
\usepackage{gensymb}
\usepackage[colorlinks=true]{hyperref}
\usepackage[nameinlink,noabbrev]{cleveref}
\usepackage{subcaption}
\usepackage{caption}
\usepackage{lineno}
\usepackage{empheq}
\usepackage{changes}
\definechangesauthor[name={Mathias}, color=orange]{km}
\definechangesauthor[name={Damiano}, color=red]{kd}
\definechangesauthor[name={Xuwen}, color=purple]{kx}
\usepackage[table,dvipsnames]{xcolor}
\definecolor{blue}{RGB}{0, 0, 130}

\usepackage{siunitx}

\usepackage{enumitem}
 \setlist[description]{nosep}

\usetikzlibrary{shapes.geometric}
\usepackage{pgfplots}
\pgfplotsset{compat=newest}

\DeclareRobustCommand{\legendline}[1]{(\tikz[baseline=-\the\dimexpr\fontdimen22\textfont2\relax,inner sep=0pt]\draw[#1,line width=1.25pt](0,0)--(5mm,0);)
}

\DeclareRobustCommand{\legendlinestar}[1]{%
    (\begin{tikzpicture}[baseline=-\the\dimexpr\fontdimen22\textfont2\relax,inner sep=0pt]
    \draw[#1,line width=1.25pt](0,0)--(5mm,0);
    \node[star, star points=5, star point ratio=2.25, minimum size=8pt, 
        inner sep=0pt, fill=#1, draw=#1] at (0.25,0) {};
    \end{tikzpicture})%
}

\DeclareRobustCommand{\legendlinetriangle}[1]{
(\begin{tikzpicture}
  \draw[#1,line width=1.25pt] (0,0) -- (5mm,0);
  \filldraw[#1] (0.25,0.1) -- (0.15,-0.1) -- (0.35,-0.1) -- cycle;
\end{tikzpicture})
}

\DeclareRobustCommand{\legendlinediamond}[1]{
(\begin{tikzpicture}
  \draw[#1,line width=1.25pt] (0,0) -- (5mm,0);
  \filldraw[#1,line width=1.25pt] (0.25,0.1) -- (0.15,0) -- (0.25,-0.1) -- (0.35,0) -- cycle;
\end{tikzpicture})
}

\DeclareRobustCommand{\legendmarker}[1]{%
    (~\begin{tikzpicture}[baseline=-\the\dimexpr\fontdimen22\textfont2\relax,inner sep=0pt]
    \filldraw[#1] (0,0) circle (3pt);
    \end{tikzpicture}~)%
}

\DeclareRobustCommand{\StmSurface}[1]{%
 (%
 \tikz[baseline=-\the\dimexpr\fontdimen22\textfont2\relax,inner sep=0pt, outer xsep=0pt]{%
  \filldraw[#1,draw=black](0,-1mm) rectangle (4mm,1mm);%
 }%
 )%
}

\DeclareRobustCommand{\StmRectangle}[1]{%
 (%
 \tikz[baseline=-\the\dimexpr\fontdimen22\textfont2\relax,inner sep=0pt, outer xsep=0pt]{%
  \filldraw[#1,draw=none](0,-1mm) rectangle (4mm,1mm);%
 }%
 )%
}

\DeclareRobustCommand\StmCourbe[2]{%
 (%
 \tikz[baseline=-\the\dimexpr\fontdimen22\textfont2\relax,inner sep=0pt] {%
  \draw[line width=1pt,text height = \textheight,#1] plot coordinates {(0,0)} -- plot[#2] coordinates {(2.5mm,0)} -- plot coordinates {(5mm,0)};%
 }%
 )%
}
\definecolor{asymptotic}{RGB}{255,0,0}
\definecolor{energy}{RGB}{200, 112, 214}
\definecolor{detailedFE}{RGB}{31, 119, 180}
\definecolor{BH}{RGB}{168, 159, 46}
\definecolor{rigid}{RGB}{0, 158, 115}

\usepackage{tocloft}
\journal{}

\makeatletter
\renewcommand\appendix{\par
  \setcounter{section}{0}%
  \setcounter{figure}{0}%
  \setcounter{table}{0}%
  \setcounter{equation}{0}%
  \gdef\thesection{\@Alph\c@section}
  \renewcommand{\theHsection}{appendixsection.\Alph{section}}%
  \numberwithin{equation}{section}%
  \crefalias{section}{appendix}		%
  \crefalias{subsection}{appendix}	%
  \renewcommand\thefigure{\thesection.\arabic{figure}}%
  \renewcommand\thetable{\thesection.\arabic{table}}%
}
\makeatother

\newcommand*{\diffdchar}{\mathrm{d}}
\newcommand*{\dd}{\mathop{\diffdchar\!}}

\begin{document}
\begin{frontmatter}
\title{Homogenization framework for rigid and non-rigid foldable origami metamaterials}
\author{Xuwen Li, Amin Jamalimehr, Mathias Legrand, Damiano Pasini}
\affiliation{organization={Department of Mechanical Engineering, McGill University},
            addressline={845 Sherbrooke Street West}, 
            city={Montreal},
            postcode={H3A0G4}, 
            state={Quebec},
            country={Canada}}

\begin{abstract}
Origami metamaterials typically consist of folded sheets with periodic patterns, conferring them with remarkable mechanical properties. In the context of Continuum Mechanics, the majority of existing predictive methods are mechanism analogs which favor rigid folding and panel bending. While effective in predicting primary deformation modes, existing methods fall short in capturing the full spectrum of deformation of non-rigid foldable origami, such as the emergence of curvature along straight creases, local strain at vertices and warpage in panels. To fully capture the entire deformation spectrum and enhance the accuracy of existing methods, this paper introduces a homogenization framework for origami metamaterials where the faces are modeled as plate elements. Both asymptotic and energy-based homogenization methods are formulated and implemented. As a representative crease pattern, we examine the Miura origami sheet homogenized as an equivalent Kirchhoff-Love plate. The results reveal that certain effective elastic properties are nonlinearly related to both the initial fold angle and the crease stiffness. When benchmarked with results from fully resolved simulations, our framework yields errors up to \qty{12.9}{\%}, while existing models, including the bar-and-hinge model and the rigid-panel model, show up to \qty{161}{\%} error. The differences in errors are associated with the complex modes of crease and panel deformation in non-rigid origami, unexplored by the existing models. This work demonstrates a precise and efficient continuum framework for origami metamaterials as an effective strategy for predicting their elastic properties, understanding their mechanics, and designing their functionalities.

\end{abstract}

\begin{highlights}
\item Homogenization platform for property prediction of origami metamaterials.
\item Asymptotic and energy-based homogenization with numerical implementation.
\item Equivalent continuum of origami metamaterials with high panel flexibility.
\end{highlights}

\begin{keyword}
Homogenization framework \sep Effective elastic properties \sep Origami metamaterials \sep Kirchhoff–Love plate \sep Finite-Element modeling \sep Rigid foldable origami \sep Non-rigid foldable origami
\end{keyword}

\end{frontmatter}


\section{Introduction}
\label{intro}

\begin{figure}[!ht]%
    \centering%
    \includegraphics[width=0.5\linewidth]{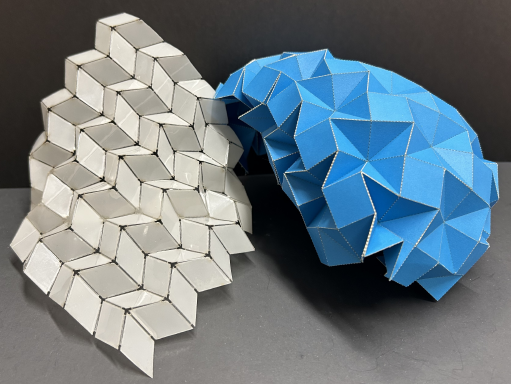}%
    \caption{Origami metamaterials. Left is a Miura origami sheet made of tape and Mylar. Right is a waterbomb origami sheet made of cardboard.}
    \label{fig:origamiFigures}%
\end{figure}

Origami, the art of paper folding, enables the folding of a flat, thin sheet into a three-dimensional structure with remarkable kinematics and mechanical properties. Initially considered as an art, origami has so far inspired the development of multiple technologies for applications, such as soft robotic actuators~\citep{chi2022bistable,li2019vacuum,yang2024complex,xue2024rigid}, deployable structures~\citep{Dang2025,jamalimehr2022rigidly,Overvelde2017,melancon2021multistable,Fang2024,li2023general}, energy absorbers~\citep{Jamalimehr2026,JIANG2025113433,Almessabi}, and foldable electronics~\citep{zhang2022kirigami,li2019origamiMetawall}. Crease patterns with period tessellation enable the design of mechanical metamaterials, namely origami metamaterials, whose physical properties are primarily governed by the architecture of the pattern rather than the base material. \Cref{fig:origamiFigures} shows some examples of origami metamaterials. A wide range of properties have been investigated thus far, such as energy absorption capacity~\citep{gattas2015behaviour,harris2021impact}, wave propagation~\citep{oudghiri2022effective,zhang2024propagation,yasuda2019origami}, and tunable or re-programmable geometry~\citep{Mirzajanzadeh2025,Nassar2022,castle2016additive,gao2023pneumatic}.

Despite the wealth of research in origami metamaterials, modeling their mechanical behavior still poses significant challenges. One of the earliest approaches, the rigid-panel model~\citep{schenk2013geometry,wei2013geometric,hu2020folding}, treats an origami metamaterial as an assembly of rigid panels connected with revolute joints forming a mechanism. The kinematics can thus be studied through the theory of mechanisms, and the structural mechanics through the force-displacement relation. Later, the bar-and-hinge model~\citep{Schenk2011,Filipov2017,liu2017nonlinear,Overvelde2017} was proposed to include the role of panel deformation. The origami metamaterial was thus treated as a combination of a truss lattice and rotational springs~\citep{lahiri2023folding}. The former allows both extension in truss elements and revolution around pin joints, capturing the primary deformation modes at the global level. The bar-and-hinge model has been used to capture the overall shape change under large deformation~\citep{xu2024derivation,xu2025modeling}. However, local stress and strain are often neglected or simplified by this method. For non-rigid foldable origami patterns, another popular option is to model their panels with plate Finite Elements (FE)~\citep{Hu2021,Filipov2015,feng2022simplified}. This method gives an accurate description of the stress and strain distribution but is computationally demanding. The computational requirement becomes overwhelming when there is a large origami tessellation with a complex fold pattern. For this reason, the FE analysis of origami specimens is limited to a few unit cells of relatively simple patterns.

To bypass the fully resolved, detailed simulation of a periodic metamaterial, homogenization is often used to approximate the global response using an equivalent continuum. Multiple theories exist for homogenization, and are capable of simplifying the mechanical analysis of large, periodic tessellations, such as cellular origami metamaterials~\citep{Cheung2014,LYU2021106791}, origami sheets~\citep{Heimbs2013} and origami tubes~\citep{turco2024long}, corresponding to 3D, 2D and 1D tessellations, respectively, for both the in-plane and out-of-plane mechanical properties. \cite{xu2024derivation,xu2025modeling} proposed an effective surface theory and its numerical implementation on nonlinear homogenization to characterize origami soft modes using the bar-and-hinge model. \cite{Cheung2014} applied periodic boundary conditions to an interleaved tube cellular structure to examine its elastic properties with classical dimensional scaling analysis. \cite{vasudevan2024homogenization} presented an energy-based homogenization framework with bar-and-hinge to determine the effective elastic constants of a couple-stress continuum representing origami sheets. \cite{Nassar2017} leveraged asymptotic homogenization to fit smooth surfaces with bar-and-hinge origami tessellations. Regardless of the method, the idea of homogenization proves to be a powerful tool in revealing the unique kinematics and mechanics of origami metamaterials.

Existing methods to homogenize the mechanical properties of origami metamaterials mainly capture rigid-foldable modes and rely on specific assumptions for folding. One is that creases and panels have contrasting properties, enabling the metamaterial to fold as a mechanism. Another is to treat the metamaterial as a non-reconfigurable structure that deforms in the panels rather than folding, similarly to a foldcore sandwich plate designed for load bearing and energy absorption~\citep{zheng2022vam,sturm2015multiscale,zhang2021study}. Most works adopt either of these assumptions as they are easily implemented with the modeling methods reviewed above. However, origami systems may encompass a wide range of panel and crease characteristics. An example is soft origami actuators made of elastomer~\citep{martinez2012elastomeric,Jin2022}. The soft and stretchable base material allows panels to bulge and wrinkle. Describing such mechanical responses is challenging for the rigid panel model and the bar-and-hinge model. At the same time, elastomeric creases may not show contrasting mechanical properties compared to the surrounding panels. Although this does not prevent the origami structure from folding into a wide range of motion, the assumption of rigid folding no longer applies. This is one among several other examples that call for the development of a property prediction method which is compatible with a broad variety of attributes in panels and creases so as to capture the mechanics of non-rigid foldable origami metamaterials. To address this gap, we investigate the previously uncharacterized deformation modes of origami metamaterials under varying crease stiffness.

This paper presents a homogenization framework for both non-rigid foldable and rigid foldable origami metamaterials that bridges the dichotomy between a mechanism and a structure. The framework predicts with accuracy when there is both folding kinematics and strain in the constituent material. As a demonstrative case, we examine a single-layered planar origami tessellation under stretching, shear, bending and twisting, which is of interest to a wide range of structural applications. Our method aims to homogenize origami metamaterials under the assumption that they can be treated as Kirchhoff-Love plates. We explore their linear elastic behaviors by computing the effective stiffness coefficients. To illustrate the versatility of the proposed framework, we apply two independent homogenization methods, namely asymptotic homogenization and energy-based homogenization. By comparing the results with those obtained with a detailed FE model of the corresponding origami tessellation, we determine the accuracy of effective properties for a wide range of the initial fold states. Next, we compare the accuracy of our method with the state-of-the-art, namely the bar-and-hinge model~\citep{vasudevan2024homogenization} and the rigid-panel model~\citep{wei2013geometric}. The homogenization methods can be straightforwardly extended to 3D for cellular origami metamaterials, such as the stacked Miura-origami and the Tachi-Miura polyhedron, but we present the case of a planar origami for brevity. Although the framework is general to a variety of origami patterns, we examine the Miura pattern as a representative pattern to demonstrate the procedure. 

The remainder of the article is as follows: \Cref{homogSection} presents the fundamentals for the homogenization methods, namely the asymptotic and the energy-based homogenization, which are suitable for both rigid and non-rigid origami patterns. The benchmark result of a Miura sheet is validated in \Cref{validation}.  \Cref{result} investigates the roles of initial fold angle and crease stiffness and compares results with two well-known origami models in the literature. \Cref{discussion} highlights the differences between models in terms of crease and panel deformation under specific loads and boundary conditions. \Cref{conclusion} concludes the study with final remarks.

\section{Homogenization formulations for origami metamaterials}
\label{homogSection}
With a focus on a single origami sheet, we present the homogenization formulations for determining the effective elastic constants of an equivalent Kirchhoff-Love plate. A common approach of all homogenization methods is to take a representative region, which in our paper is a unit cell, and assume that its properties are equivalent to those of the entire metamaterial~\citep{hollister1992comparison}. This section presents two formulations, i.e., asymptotic and energy-based methods. They tackle the homogenization problem from dissimilar perspectives. The former approximates the field variables in the governing equation, such as the displacement, by an asymptotic expansion. The latter, on the other hand, approximates the \emph{in-situ} stress or strain of the metamaterial with applied boundary conditions of a unit cell. Both methods are applicable to origami metamaterials under linear elastic deformation. The pattern should exhibit translational periodicity, with the unit cell size much smaller than the full tessellation. Finally, the unit cell height, length, and width should be on the same order of magnitude. By comparing both methods, we can appreciate their common and distinct traits, obtain the effective properties, and identify the strain energy density of the metamaterial under prescribed deformation modes.

\subsection{Homogenized Kirchhoff-Love plate model for origami sheets}
\label{KLplate}
\begin{figure}[!ht]
    \centering
    \includegraphics[]{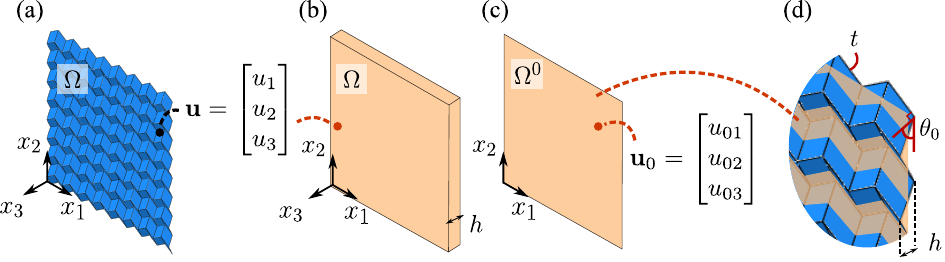}
    \begin{subfigure}{0.1\textwidth}
        \phantomsubcaption
        \label{fig:frameworka}
        \phantomsubcaption
        \label{fig:frameworkb}
        \phantomsubcaption
        \label{fig:frameworkc}
        \phantomsubcaption
        \label{fig:frameworkd}
    \end{subfigure}
    \caption{Homogenization scheme of a single-layered, periodic origami sheet. (a)~Detailed Miura sheet $\Omega$. (b)~Homogeneous Kirchhoff-Love plate. (c)~Mid-surface of homogenized plate $\Omega^0$. (d)~Zoom-in of origami sheet showing the location of the neutral surface (superimposed orange plane) with respect to the detailed origami sheet.}
    \label{fig:framework}
\end{figure}

\Cref{fig:frameworka} shows in blue the physical domain $\Omega$ of an origami sheet. \Cref{fig:frameworkb} shows its homogenized counterpart with equal length, width and height. The homogenization framework aims to find the effective properties of the homogeneous thin plate $\Omega$. \Cref{fig:frameworkc} introduces the 2D domain $\Omega^0$, which is both the neutral surface of the true origami sheet and the mid-surface of the homogenized one. 
We select the neutral surface $\Omega^0$ such that it does not stretch or compress under pure bending~\citep{vasudevan2024homogenization}. \Cref{fig:frameworkd} presents the location of $\Omega^0$. The origami panels have a thickness of $t$. After folding, the sheet assumes an apparent global height of $h$, which increases with the folding angle $\theta_0$, shown in \Cref{fig:frameworkd}. A Cartesian coordinate system is introduced with axes $x_1$, $x_2$, and $x_3$ where $x_1$ and $x_2$ lie on the neutral surface of the origami sheet and the mid-surface of the homogenized plate. The displacements of the homogenized plate are described by the displacements of its mid-surface as
\begin{subequations}
\begin{align}
    &u_1(x_1,x_2,x_3) = u_{01}(x_1,x_2)-x_3\partial_{x_1} u_{03}(x_1,x_2)\\
    &u_2(x_1,x_2,x_3) = u_{02}(x_1,x_2)-x_3\partial_{x_2} u_{03}(x_1,x_2),\\
    & u_3(x_1,x_2,x_3) = u_{03}(x_1,x_2),
\end{align}
\end{subequations}
where $u_{01}$ and $u_{02}$ are the in-plane displacements of a point on the mid-surface, and $u_{03}$ is the out-of-plane deflection. Assuming small displacement gradients, we can express the strain components $\tilde{e}_{ij}$ in terms of the displacements $u_i$ as $2\tilde{e}_{ij}=\partial_iu_j+\partial_ju_i$. The following expressions adopt index notations with Greek letters $\alpha,\beta,\gamma,\delta=1,2$. The plate theory assumes that the strains $\tilde{e}_{13}$, $\tilde{e}_{23}$, and $\tilde{e}_{33}$ are zeros. The strains $\tilde{e}_{11}$, $\tilde{e}_{22}$, and $\tilde{e}_{12}$ can be reduced to functions of the mid-surface strains $e_{\alpha\beta}$ and curvatures $\kappa_{\alpha\beta}$ following 
$\tilde{e}_{\alpha\beta}=e_{\alpha\beta}+x_3\kappa_{\alpha\beta}$, where the mid-surface strains and curvatures read
\begin{subequations}
\begin{align}
    &e_{11}=\partial_{x_1} u_{01},&& e_{22}=\partial_{x_2} u_{02},&&
    e_{12}=\tfrac{1}{2}(\partial_{x_2} u_{01}+\partial_{x_1} u_{02}),\\
    &\kappa_{11}=-\partial_{x_1}^2 u_{03},&& \kappa_{22}=-\partial_{x_2}^2 u_{03},&& \kappa_{12}=-\partial_{x_1}\partial_{x_2} u_{03}.
\end{align}
\end{subequations}
Adopting the Kirchhoff-Love plate model, the plate constitutive relations are~\citep{reddy1996mechanics}
\begin{equation}
\label{plateConst}
\begin{bmatrix}
\mathbf{N} \\ 
\mathbf{M}
\end{bmatrix}
=
\begin{bmatrix}
\mathbf{A}^H & \mathbf{B}^H \\ 
\mathbf{B}^H & \mathbf{D}^H
\end{bmatrix}
\begin{bmatrix}
\mathbf{e} \\ 
\boldsymbol{\kappa}
\end{bmatrix},
\end{equation}
where $\mathbf{N}$ is the resultant force vector, $\mathbf{M}$ is the resultant moment vector, \smash{$\mathbf{A}^H$} is the extensional stiffness coefficient matrix with components \smash{$A^H_{\alpha\beta\gamma\delta}$}, \smash{$\mathbf{B}^H$} is the bending-extensional coupling stiffness coefficient matrix with components \smash{$B^H_{\alpha\beta\gamma\delta}$}, and \smash{$\mathbf{D}^H$} is the bending stiffness coefficient matrix with components \smash{$D^H_{\alpha\beta\gamma\delta}$}. The matrix of the stiffness coefficients, known as the ABD matrix, is symmetric with 21 independent components.

For homogeneous or laminated plates, there exist relationships between the stiffness coefficients \smash{$A^H_{\alpha\beta\gamma\delta}$}, \smash{$B^H_{\alpha\beta\gamma\delta}$}, \smash{$D^H_{\alpha\beta\gamma\delta}$} and the base material properties obtained from the Cauchy continuum theory~\citep{reddy2006theory}. However, this simplification does not apply to a nonhomogeneous medium such as an origami metamaterial with a general fold pattern. Rather, \smash{$A^H_{\alpha\beta\gamma\delta}$}, \smash{$B^H_{\alpha\beta\gamma\delta}$}, and \smash{$D^H_{\alpha\beta\gamma\delta}$} should all be computed independently~\citep{vasudevan2024homogenization}. The strain energy density of a homogeneous thin plate is
\begin{equation}
\label{homogU2}
    \bar{U}^H = \tfrac{1}{2}\begin{bmatrix}
\mathbf{e} \\ 
\boldsymbol{\kappa}
\end{bmatrix}^T\begin{bmatrix}
\mathbf{A}^H & \mathbf{B}^H \\ 
\mathbf{B}^H & \mathbf{D}^H
\end{bmatrix}
\begin{bmatrix}
\mathbf{e} \\ 
\boldsymbol{\kappa}
\end{bmatrix}.
\end{equation}

The following introduces the formulations for asymptotic and energy-based homogenization, enabling us to determine the effective constants of a Kirchhoff-Love plate equivalent to the origami sheet under investigation. 
\subsection{Asymptotic homogenization}
\label{AHsubsection}
\subsubsection{Theory}
\label{AH theory}
To apply the asymptotic homogenization theory to origami metamaterials, we assume that the apparent plate height $h$ is small and comparable to the length and width of a unit cell~\citep{kalamkarov1997analysis}. We start with a review of a modified asymptotic homogenization method for periodic thin plates~\citep{Caillerie1984,kohn1984new}.

\begin{figure}[!ht]
    \centering
    \includegraphics[]{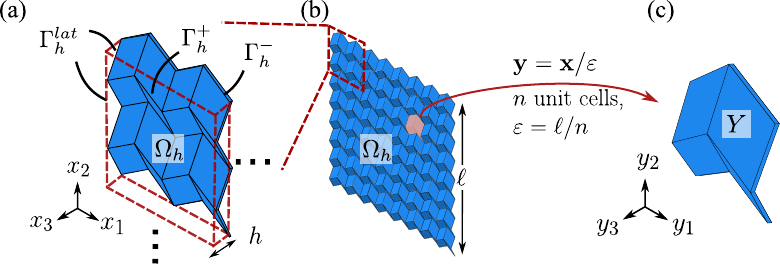}
    \begin{subfigure}{0.1\textwidth}
        \phantomsubcaption
        \label{fig:AHProba}
        \phantomsubcaption
        \label{fig:AHProbb}
        \phantomsubcaption
        \label{fig:AHProbc}
    \end{subfigure}
\caption{Coordinate systems used in asymptotic homogenization. (a)~A zoomed-in section of the origami sheet showing the macroscopic coordinates $\mathbf{x}$ and domain boundaries $\Gamma_h^+$, $\Gamma_h^-$, $\Gamma_h^{\textup{lat}}$. (b)~Domain $\Omega_h$ of the origami sheet. (c)~A unit cell $Y$ scaled up by $1/\varepsilon$ and the microscopic coordinates~$\mathbf{y}$.}\label{fig:AHProb}
\end{figure}
Consider the representative origami sheet shown in \Cref{fig:AHProba,fig:AHProbb}. $\Omega_h$ denotes the domain of the sheet. A small parameter $h$ denotes the domain height of the equivalent homogenized plate. The upper, lower, and lateral surfaces are respectively $\Gamma_h^+$, $\Gamma_h^-$, $\Gamma_h^{\textup{lat}}$. $\Gamma_h^+$ and $\Gamma_h^-$ consist of all points $(x_1,x_2,x_3)$ such that $(x_1,x_2)$ lies in the mid-surface $\Omega^0$. On $\Gamma_h^+$ $x_3=h$, while on $\Gamma_h^-$ $x_3=0$. $\Gamma_h^{\textup{lat}}$ has $(x_1,x_2)$ lying on the boundary of $\Omega^0$ and $x_3$ varying from 0 to $h$. There are $n$ unit cells along one given side of the sheet, with a total length of $\ell$. The origami pattern is periodic with a unit cell size of $\varepsilon=\ell/n$. Since the origami sheet is single-layered, its height $h$ and unit cell size $\varepsilon$ have the same order of magnitude. They are small parameters, meaning $h\to 0$ and $\varepsilon\to 0$ simultaneously as the origami tessellation becomes infinitely large. 

To analyze the origami pattern, we rescale one of its unit cells $Y$ with a scaling factor $\varepsilon$. In addition to the coordinate system $\mathbf{x}=(x_1,x_2,x_3)$ of the original problem, a new coordinate system $\mathbf{y}=(y_1,y_2,y_3)$ is introduced, whose origin lies on the bottom left corner of the unit cell mid-surface. The macroscopic coordinates $\mathbf{x}$ for the origami sheet and the microscopic coordinates $\mathbf{y}$ for a unit cell are related through $y_1=x_1/\varepsilon$, $y_2=x_2/\varepsilon$, and $y_3=x_3/h$. The elastic stiffness of the material in the unit cell $C_{ijk\ell}$ is a function of $y_1$, $y_2$, and $y_3$. Before homogenization, the elastic stiffness of the full origami metamaterial $C_{ijk\ell}^{h\varepsilon}(x_1,x_2,x_3)$ is periodic in the macroscopic coordinates $x_1$ and $x_2$ with its period identical to the size of the unit cell. It is assumed that
\begin{equation}
    C_{ijk\ell}^{h\varepsilon}(x_1,x_2,x_3)=\frac{1}{h^3}C_{ijk\ell}\Big(\frac{x_1}{\varepsilon},\frac{x_2}{\varepsilon},\frac{x_3}{h}\Big),
\end{equation}
where the superscripts $h,\varepsilon$ signify that the elastic stiffness is a function of the height and the periodicity. The indices $i,j, k, \ell$ range from 1 to 3. The following change of variables is now adopted to simplify the analysis: all functions of the origami sheet are made independent of $h$, i.e., $z_1=x_1$, $z_2=x_2$, and $z_3=x_3/h$.
This results in a domain $\Omega$ for the origami sheet, an upper surface $\Gamma^+$, a lower surface $\Gamma^-$, and a lateral surface $\Gamma^{\textup{lat}}$. The elastic stiffness $C_{ijk\ell}^{h\varepsilon}(x_1,x_2,x_3)$ becomes $C_{ijk\ell}^{\varepsilon}(z_1,z_2,z_3)$, which is a function of the periodicity $\varepsilon$ but not a function of the height $h$. The full problem of elasticity of the origami sheet is
\begin{subequations}\label{staticEqn}
  \begin{empheq}[left=\empheqlbrace]{align}
&\partial_{z_1} \sigma_{i1}^\varepsilon + \partial_{z_2} \sigma_{i2}^\varepsilon + \tfrac{1}{h} \partial_{z_3} \sigma_{i3}^\varepsilon  = 0 && \text{in } \Omega\label{staticEqna}\\
&\sigma_{ij}^\varepsilon = C_{ijk1}^\varepsilon\partial_{z_1} u_k^{\varepsilon} + C_{ijk2}^\varepsilon\partial_{z_2} u_k^{\varepsilon} + \tfrac{1}{h} C_{ijk3}^\varepsilon \partial_{z_3} u_k^{\varepsilon} && \text{in } \Omega\label{staticEqnb}\\
&u_i^\varepsilon = 0 && \text{on } \Gamma^{\text{lat}}\label{staticEqnc}\\
&\sigma_{1 j}^\varepsilon n_j = t_1/h, \quad \sigma_{2 j}^\varepsilon n_j = t_2/h, \quad \sigma_{3j}^\varepsilon n_j = t_3 && \text{on } \Gamma^{\pm}\label{staticEqnd}
  \end{empheq}
\end{subequations}
where $\sigma_{ij}^\varepsilon$ is the stress tensor, $u_i^\varepsilon$ is the displacement, $n_j$ is the unit normal to the surface, and $t_i$ is the surface force. In \Cref{staticEqna,staticEqnb}, summation is only on the indices $i$, $j$, and $k$ but not on the fourth index. This is due to the $1/h$ term introduced by the change in variables from $x_3$ to $z_3$. We search for the solution to \Cref{staticEqn} under the limit $\varepsilon\to 0$. The displacement is represented by the asymptotic expansion
\begin{equation}\label{asymExpanU}
    u_i^\varepsilon(\mathbf{z}) = u_i^{(0)}(\mathbf{x}, \mathbf{y}) + \varepsilon u_i^{(1)}(\mathbf{x}, \mathbf{y}) + \varepsilon^2 u_i^{(2)}(\mathbf{x}, \mathbf{y}) + \varepsilon^3 u_i^{(3)}(\mathbf{x}, \mathbf{y}) + \varepsilon^4 u_i^{(4)}(\mathbf{x}, \mathbf{y}) + \dots
\end{equation}
where each term $u_i^{(0)}$, $u_i^{(1)}$... are functions of both coordinates $\mathbf{x}$ and $\mathbf{y}$. These terms are smooth in $\mathbf{x}$ and periodic in $y_1$ and $y_2$. On the other hand, the stress tensor expands to
\begin{equation}
    \sigma_{ij}^\varepsilon(\mathbf{z}) = \varepsilon^{-4}\sigma_{ij}^{(-4)}(\mathbf{x},\mathbf{y})+\varepsilon^{-3}\sigma_{ij}^{(-3)}(\mathbf{x},\mathbf{y})+\varepsilon^{-2}\sigma_{ij}^{(-2)}(\mathbf{x},\mathbf{y})+\varepsilon^{-1}\sigma_{ij}^{(-1)}(\mathbf{x},\mathbf{y})+\dots
\end{equation}
where the terms $\sigma_{ij}^{(-4)}$, $\sigma_{ij}^{(-3)}$... have the same properties on $\mathbf{x}$ and $\mathbf{y}$ as the displacement terms. Applying the above expansions to \Cref{staticEqn} gives an expression with various powers of $\varepsilon$. Grouping the terms with the same power of $\varepsilon$ yields a set of equations that simplifies to two microscopic problems. 

The first one is
\begin{equation}
\label{microProbIP}
\partial_{y_j}\bigl[ C_{ijpq}\bigl(\delta_{pk}\delta_{q\delta}+\tfrac{1}{2}( \partial_{y_q} w^{k\delta}_p + \partial_{y_p} w^{k\delta}_q)\bigr) \bigr] = 0 \text{ on }Y, \quad 
\mathbf{w}^{k\delta} \text{ is } (y_1, y_2)\text{-periodic,}
\end{equation}
where $\mathbf{w}^{k\delta}$ is a displacement field. The first term $\partial_{y_j}(C_{ijpq}\delta_{pk}\delta_{q\delta})$ represents the reaction force in the unit cell when it is extended or sheared to a strain equal to 1. $\mathbf{w}^{k\delta}$ is thus the solution of the microscopic problem corresponding to in-plane deformation modes. 

The second microscopic problem is
\begin{equation}
\label{microProbOOP}
\partial_{y_j}\bigl[ C_{ijpq}\bigl( y_3\delta_{p\gamma}\delta_{q\delta} + \tfrac{1}{2}( \partial_{y_q} p^{\gamma\delta}_p + \partial_{y_p} p^{\gamma\delta}_q) \bigr) \bigr] = 0 \text{ on }Y, \quad
\mathbf{p}^{\gamma\delta} \text{ is } (y_1,y_2)\text{-periodic,}
\end{equation}
where $\mathbf{p}^{\gamma\delta}$ is a displacement field. The first term $\partial_{y_j}(C_{ijpq}y_3\delta_{p\gamma}\delta_{q\delta})$ represents the reaction force in the unit cell when it is bent or twisted to a curvature equal to 1. $\mathbf{p}^{\gamma\delta}$ is the solution of the microscopic problem corresponding to out-of-plane deformation modes. 

Both microscopic problems are then rewritten in their weak forms before being solved through FEA, as detailed in the following section. The solutions $\mathbf{w}^{k\delta}$ and $\mathbf{p}^{\gamma\delta}$ are also known as the characteristic displacement fields~\citep{Hassani1998}.

\begin{figure}[!ht]
    \centering
    \includegraphics[]{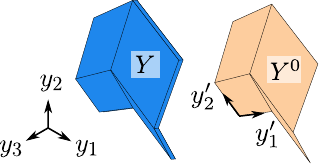}
    \caption{Origami unit cell $Y$ consisting of thin plates and their mid-surface $Y^0$.}
    \label{fig:UCmidsurface}
\end{figure}
After homogenization, the effective extensional stiffness coefficients are
\begin{equation}\label{AHAH}
A_{\alpha \beta \gamma \delta}^H = \frac{1}{L_1L_2} \int_Y C_{ijk\ell} \bigl( \delta_{k\gamma} \delta_{\ell\delta} + \tfrac{1}{2}( \partial_{y_\ell} w^{\gamma \delta}_k + \partial_{y_k} w^{\gamma \delta}_\ell)\bigr)\bigl( \delta_{i\alpha} \delta_{j\beta} + \tfrac{1}{2}( \partial_{y_j} w^{\alpha \beta}_i + \partial_{y_i} w^{\alpha \beta}_j  \bigr) \dd\mathbf{y},
\end{equation}
where $L_1$ and $L_2$ are the lengths of the unit cell along $y_1$ and $y_2$, respectively. \Cref{AHAH} applies to unit cells with a general shape. Since an origami unit cell consists of thin panels, we simplify the expression by assuming that the panels are Kirchhoff-Love plates. Their collective mid-surface $Y^0$ is shown in \Cref{fig:UCmidsurface}. $Y^0$ is referred to the coordinates $(y_1',y_2')$. Note that the plates in a unit cell are \emph{not} the homogenized plate introduced in \Cref{KLplate}. With a displacement field $\mathbf{w}$, the local strain field in the unit cell panels is $\mathbf{e}= \boldsymbol{\mu} + y_3 \boldsymbol{\kappa}$ with
\begin{equation}
\mathbf{e}=\begin{bmatrix}
e_{11} & e_{12} \\
e_{12} & e_{22}
\end{bmatrix},\quad 2\boldsymbol{\mu}=
\begin{bmatrix}
2\partial_{y_1} w_1 & \partial_{y_2} w_1 + \partial_{y_1} w_2 \\
\partial_{y_2} w_1 + \partial_{y_1} w_2 & 2\partial_{y_2} w_2
\end{bmatrix}, \quad \boldsymbol{\kappa}= -
\begin{bmatrix}
\partial_{y_1}^2 w_3 & \partial_{y_1}\partial_{y_2} w_3\\
\partial_{y_1}\partial_{y_2} w_3 & \partial_{y_2}^2 w_3
\end{bmatrix}
\end{equation}
where $\boldsymbol{\mu}$ is the in-plane strain and $\boldsymbol{\kappa}$, the curvature. The base material properties reduce to $C_{\zeta\eta\theta\lambda}$, with $\zeta,\eta,\theta,\lambda=1,2$, on the mid-surface $Y^0$~\citep{lewinski1988asymptotic}. Applying the above to \Cref{AHAH} gives
\begin{equation}
\begin{aligned}
    A_{\alpha \beta \gamma \delta}^H = \frac{1}{L_1L_2} \!\int_{Y^0}\!\!C_{\zeta\eta\theta\lambda} \Bigl( \delta_{\theta\gamma} \delta_{\lambda\delta} + \tfrac{1}{2}( \partial_{y_\lambda} w^{\gamma \delta}_\theta + \partial_{y_\theta} w^{\gamma \delta}_\lambda ) +y_3\partial _{y_\theta}\partial_{y_\lambda} w_3^{\gamma\delta} \Bigr)\\
    \Bigl( \delta_{\zeta\alpha} \delta_{\eta\beta} + \tfrac{1}{2}( \partial_{y_\eta} w^{\alpha \beta}_\zeta + \partial_{y_\zeta} w^{\alpha \beta}_\eta ) +y_3\partial_{y_\zeta}\partial _{y_\eta} w_3^{\alpha \beta} \Bigr)\dd y_1'\dd y_2'.
\end{aligned}
\end{equation}
Its shorthand expression is
\begin{equation}\label{AHAHs}
    A_{\alpha \beta \gamma \delta}^H = \frac{1}{L_1L_2}\! \int_{Y^0}\!\! \mathbf{C}\bigl(\mathbf{e}^{0e,(\gamma\delta)}+(\boldsymbol{\mu}^{*e,(\gamma\delta)}+y_3\boldsymbol{\kappa}^{*e,(\gamma\delta)})\bigr):\bigl(\mathbf{e}^{0e,(\alpha\beta)}+(\boldsymbol{\mu}^{*e,(\alpha\beta)}+y_3\boldsymbol{\kappa}^{*e,(\alpha\beta)})\bigr)\dd\mathbf{y}',
\end{equation}
where $\mathbf{e}^{0e,(\alpha\beta)}$ denotes unit extension or shear strain in direction $\alpha\beta$, $\boldsymbol{\mu}^{*e,(\alpha\beta)}$ is the strain of the characteristic displacement $\mathbf{w}^{\alpha\beta}$ from the first microscopic problem in \Cref{microProbIP}, and $\boldsymbol{\kappa}^{*e,(\alpha\beta)}$ is the curvature of the characteristic displacement $\mathbf{w}^{\alpha\beta}$. The integral part of the expression is twice the strain energy of the unit cell. The relation between strain energy and effective properties applies to various types of structures, as seen later in our origami assembly of thin plates and torsional springs. 

The strain energy is computed differently for diagonal terms $A^H_{1111},A^H_{2222},A^H_{1212}$ and off-diagonal terms $A^H_{1122},A^H_{1112},A^H_{2212}$. To find the diagonal terms $A^H_{\alpha\beta\alpha\beta}$, we sum the unit strain $\mathbf{e}^{0e,(\alpha\beta)}$ and the characteristic strain \smash{$\boldsymbol{\mu}^{*e,(\alpha\beta)}+y_3\boldsymbol{\kappa}^{*e,(\alpha\beta)}$} everywhere in the unit cell. The sum of strains has a corresponding strain energy $U$, which is expressed as a function of the effective stiffness coefficient, namely~\citep{Cheng2013}
\begin{equation}
\label{AHdiagonal}
    A^H_{\alpha\beta\alpha\beta}=\frac{2}{L_1L_2}U.
\end{equation} 
The rest of $A_{\alpha \beta \gamma \delta}^H$ is associated to two unit strains $\mathbf{e}^{0e,(\alpha\beta)}$ and $\mathbf{e}^{0e,(\gamma\delta)}$. Since our problem satisfies the principle of superposition, a linear combination of the two unit strains has a corresponding strain energy $U$, which gives
\begin{equation}
A^H_{\alpha\beta\alpha\beta}+A^H_{\alpha\beta\gamma\delta}+A^H_{\gamma\delta\alpha\beta}+A^H_{\gamma\delta\gamma\delta}=\frac{2}{L_1L_2}U,
\end{equation}
where $A^H_{\alpha\beta\alpha\beta},A^H_{\gamma\delta\gamma\delta}$ are diagonal terms obtained from \Cref{AHdiagonal}, and $A^H_{\alpha\beta\gamma\delta}=A^H_{\gamma\delta\alpha\beta}$ from the symmetry of the stiffness matrix. The off-diagonal terms are expressed as
\begin{equation}
\label{AHoffDiagonal}
    A^H_{\alpha\beta\gamma\delta}=\frac{1}{L_1L_2}U-\frac{A^H_{\alpha\beta\alpha\beta}}{2}-\frac{A^H_{\gamma\delta\gamma\delta}}{2}.
\end{equation}
Similarly,
\begin{equation}
    D_{\alpha \beta \gamma \delta}^H = \frac{1}{L_1L_2} \int_Y C_{ijk\ell}\bigl( y_3 \delta_{k\gamma} \delta_{\ell\delta} + \tfrac{1}{2}( \partial_{y_\ell} p^{\gamma \delta}_k + \partial_{y_k} p^{\gamma \delta}_\ell )\bigr)
\bigl( y_3 \delta_{i\alpha} \delta_{j\beta} + \tfrac{1}{2}( \partial_{y_j} p^{\alpha \beta}_i + \partial_{y_i} p^{\alpha \beta}_j )\bigr) \dd\mathbf{y}
\end{equation}
are the homogenized bending stiffness coefficients. Assuming the unit cell consists of thin plates, the expression simplifies to
\begin{equation}\label{AHDHs}
    D_{\alpha \beta \gamma \delta}^H = \frac{1}{L_1L_2} \!\int_{Y^0}\!\! \mathbf{C}\bigl(\mathbf e^{0\kappa,(\gamma\delta)}+(\boldsymbol \mu^{*\kappa,(\gamma\delta)}+y_3\boldsymbol \kappa^{*\kappa,(\gamma\delta)})\bigr):\bigl(\mathbf e^{0\kappa,(\alpha\beta)}+(\boldsymbol \mu^{*\kappa,(\alpha\beta)}+y_3\boldsymbol \kappa^{*\kappa,(\alpha\beta)})\bigr)\dd\mathbf{y},
\end{equation}
where $\mathbf e^{0\kappa,(\alpha\beta)}$ denotes unit curvature or twist in direction $\alpha\beta$, $\boldsymbol\mu^{*\kappa,(\alpha\beta)}$ is the strain of characteristic displacement $\mathbf{p}^{\alpha\beta}$ from the second microscopic problem in \Cref{microProbOOP}, and $\boldsymbol\kappa^{*\kappa,(\alpha\beta)}$ is the curvature of the characteristic displacement $\mathbf{p}^{\alpha\beta}$.  The unit cell strain energy $U$ is calculated from the sum of unit curvature $\mathbf e^{0\kappa,(\alpha\beta)}$ and the strain of corresponding characteristic displacement $\boldsymbol \mu^{*\kappa,(\alpha\beta)}+y_3\boldsymbol \kappa^{*\kappa,(\alpha\beta)}$. This strain energy relates to the diagonal terms as
\begin{equation}
\label{DHdiagonal}
     D^H_{\alpha\beta\alpha\beta}=\frac{2}{L_1L_2}U.
\end{equation}
Note that \Cref{DHdiagonal} differs from \Cref{AHdiagonal} in the values of $U$. Similar to the in-plane case, there exists 3 linear combinations of unit curvature, each corresponding to a strain energy $U$. The off-diagonal terms are
\begin{equation}
\label{DHoffDiagonal}
    D^H_{\alpha\beta\gamma\delta}=\frac{1}{L_1L_2}U-\frac{D^H_{\alpha\beta\alpha\beta}}{2}-\frac{D^H_{\gamma\delta\gamma\delta}}{2}.
\end{equation}

Another set of effective stiffness coefficients
\begin{equation}
    B_{\alpha \beta \gamma \delta}^H = \frac{1}{L_1L_2} \int_Y C_{ijk\ell}\bigl( y_3 \delta_{k\gamma} \delta_{\ell\delta} + \tfrac{1}{2}( \partial_{y_\ell} p^{\gamma \delta}_k + \partial_{y_k} p^{\gamma \delta}_\ell )\bigr) 
    \bigl( \delta_{i\alpha} \delta_{j\beta} + \tfrac{1}{2}( \partial_{y_j} w^{\alpha \beta}_i + \partial_{ y_i} w^{\alpha \beta}_j ) \bigr) \dd\mathbf{y},
\end{equation}
describes the coupling effect between in-plane and out-of-plane responses. It also simplifies to
\begin{equation}\label{AHBHs}
    B_{\alpha \beta \gamma \delta}^H = \frac{1}{L_1L_2} \!\int_{Y^0}\!\! \mathbf C\bigl(\mathbf e^{0\kappa,(\gamma\delta)}+(\boldsymbol \mu^{*\kappa,(\gamma\delta)}+y_3\boldsymbol \kappa^{*\kappa,(\gamma\delta)})\bigr):\bigl(\mathbf e^{0e,(\alpha\beta)}+(\boldsymbol \mu^{*e,(\alpha\beta)}+y_3\boldsymbol \kappa^{*e,(\alpha\beta)})\bigr)\dd\mathbf{y}.
\end{equation}
The integral part is twice the strain energy of a unit cell. All of $ B_{\alpha \beta \gamma \delta}^H$ are off-diagonal and associated to the linear combinations of both in-plane strain and curvature. The sum of unit strains $\mathbf e^{0e,(\alpha\beta)}$, $\mathbf e^{0\kappa,(\gamma\delta)}$ and characteristic strains $\boldsymbol \mu^{*e,(\alpha\beta)}+y_3\boldsymbol \kappa^{*e,(\alpha\beta)}$, $\boldsymbol \mu^{*\kappa,(\gamma\delta)}+y_3\boldsymbol \kappa^{*\kappa,(\gamma\delta)}$ relates to the strain energy $U$. The coupling coefficients are
\begin{equation}
\label{BHoffDiagonal}
    B^H_{\alpha\beta\gamma\delta}=\frac{1}{L_1L_2}U-\frac{A^H_{\alpha\beta\alpha\beta}}{2}-\frac{D^H_{\gamma\delta\gamma\delta}}{2}.
\end{equation}

With the above effective stiffness coefficients, we can now approximate the global response of an origami sheet as a thin plate. The governing equation \Cref{staticEqn} homogenizes to
\begin{equation}\label{homogMacro}
\left\{\begin{aligned}
&\partial_{x_\beta} N_{\alpha\beta} + T_\alpha = 0 && \text{in } \Omega^0, \\
&\partial_{x_\alpha}\partial_{x_\beta} M_{\alpha\beta} + \partial_{x_\alpha} Q_\alpha - T_3 = 0 && \text{in } \Omega^0, \\
&N_{\alpha\beta} = A^H_{\alpha\beta\gamma\delta}\tfrac{1}{2}(\partial_{x_\delta} \bar{u}_{\gamma}+\partial_{x_\gamma} \bar{u}_{\delta}) + B^H_{\alpha\beta\gamma\delta}(-\partial^2_{x_\gamma x_\delta} u^{(0)}_3) && \text{in } \Omega^0, \\
&M_{\alpha\beta} = B^H_{\gamma\delta\alpha\beta}\tfrac{1}{2}(\partial_{x_\delta} \bar{u}_{\gamma}+\partial_{x_\gamma} \bar{u}_{\delta})
+ D^H_{\alpha\beta\gamma\delta}(-\partial^2_{x_\gamma x_\delta} u^{(0)}_3) && \text{in } \Omega^0, \\
&u^{(0)}_3 = 0, \quad \bar{u}_\alpha = 0 && \text{in } \Gamma^0,
\end{aligned}
\right.
\end{equation}
where $\Omega^0$ is the mid-surface of the homogenized plate, $\Gamma^0$ is the boundary of the mid-surface, $N_{\alpha\beta}$ are the force resultants, $M_{\alpha\beta}$ are the moment resultants, $T_1$, $T_2$, and $T_3$ are averaged tractions defined as \smash{$T_i = \frac{h}{|Y|} \int_{\Gamma^\pm} t_i^\pm \dd s$}, $Q_\alpha$ are averaged bending moments defined as \smash{$Q_\alpha = \frac{h}{|Y|} \int_{\Gamma^\pm} y_3t_\alpha^\pm \dd s$}, \smash{$\bar{u}_\alpha$} are the homogenized displacements, and \smash{$u^{(0)}_3$} is the homogenized deflection.

\subsubsection{Numerical implementation}\label{AHimplementation}
Given the geometry of origami patterns, which can be complex, it is often not feasible to use closed-form expressions to perform asymptotic homogenization. Rather, a numerical approach becomes preferable. Our method here relies on FE analysis to find numerical solutions to the microscopic problems (\Cref{microProbIP,microProbOOP}) and the effective stiffness coefficients (\Cref{AHAHs,AHDHs,AHBHs}). The following procedure is fully implemented through ABAQUS python scripting, with python codes and input files available~\citep{github_wendy}. However, it is possible to generalize the procedure to any FE software package with minor modifications. Our implementation differs from the literature \citep{cai2014novel,eskandari2024unravelling} in the strain energy calculation. Rather than calculating the strain energy from the nodal quantities during postprocessing, this step is performed fully using the selected FE software. This modification adds flexibility to the choice of element types. It is relevant to models of origami tessellations that require various types of plate elements, shell elements and torsional springs.

\begin{figure}[!ht]
    \centering
    \includegraphics[]{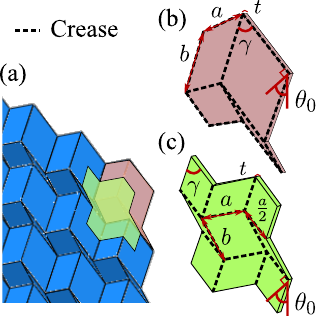}
    \begin{subfigure}{0.1\textwidth}
        \phantomsubcaption
        \label{fig:UCmodelinga}
        \phantomsubcaption
        \label{fig:UCmodelingb}
        \phantomsubcaption
        \label{fig:UCmodelingc}
    \end{subfigure}
    \caption{Miura unit cells and geometric parameters. (a)~A Miura tessellation. (b)~A common choice of Miura unit cell. (c)~Equivalent unit cell adopted in this work.}
    \label{fig:UCmodeling}
\end{figure}

Taking the Miura sheet as an example, the procedure begins with identifying the geometry and material of the unit cell, specified in \Cref{MiuraProperties} along with \Cref{fig:UCmodeling}. We integrate this information into an FE model of the unit cell. We construct the unit cell~\citep{Filipov2015} with panels meshed as plate elements and creases as torsional springs. Since the choice of unit cell origin is arbitrary, in this example the bottom left corner of the Miura unit cell is selected as the center of a panel, shown in \Cref{fig:UCmodelingc}. This choice simplifies the specification of boundary conditions in the microscopic problem.
\begin{table}[ht]
    \centering\small
    \begin{tabular}{lccc} \hline 
         Panel side length& $a$ &mm  & 20\\ 
         Panel side length& $b$ &mm  & 20\\ 
         Sector angle& $\gamma$ &$^\circ$ & 60\\ 
         Initial fold angle& $\theta_0$ &$^\circ$ & 30\\
         Panel thickness& $t$ &mm & 0.13\\ 
         Unit length crease stiffness& $K_{\textup{cr}}/a$ &N & 0.1\\ 
         Base material Young's modulus& $E$ &MPa & 4000\\
         Base material Poisson's ratio& $\nu$& & 0.38\\ \hline 
    \end{tabular}
    \caption{Geometric and material parameters of Miura origami metamaterial.}
    \label{MiuraProperties}
\end{table}

\begin{figure}[!ht]
    \centering
    \includegraphics[]{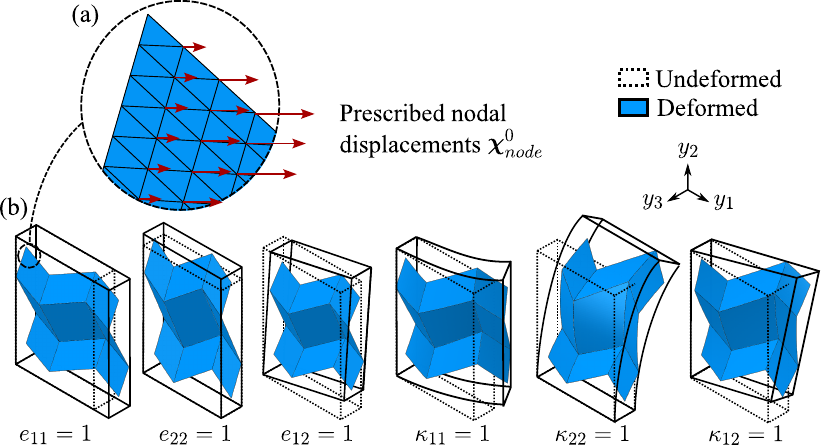}
    \caption{Step 1 of asymptotic homogenization showing six of the deformation modes with unit strain or unit curvature. (a)~Displacements are prescribed nodally to ensure uniform strain or curvature distribution. (b)~Displacement fields $\boldsymbol\chi^0$ corresponding to unit strain ($e_{11}=1$ or $e_{22}=1$ or $e_{12}=1$) or unit curvature ($\kappa_{11}=1$ or $\kappa_{22}=1$ or $\kappa_{12}=1$). Displacements are down-scaled for plotting purposes.}
    \label{AHunitStrain}
\end{figure}

Discretizing the microscopic problems exposed in \Cref{microProbIP,microProbOOP} leads to solving two systems of linear equations $\mathbf{K}\mathbf{w}^{\gamma\delta}=\mathbf{f}^{0}$ and $\mathbf{K}\mathbf{q}^{\gamma\delta}=\mathbf{f}^{0}$, 
where $\mathbf{K}$ is the stiffness matrix of the unit cell, $\mathbf{w}^{\gamma\delta}$ and $\mathbf{q}^{\gamma\delta}$ are the discretized characteristic displacement fields, and $\mathbf{f}^{0}$ is the initial strain loading. The latter has a physical meaning, which is nodal reactions induced by unit strains $e_{\gamma\delta}$ or unit curvatures $\kappa_{\gamma\delta}$. At the continuous level, the corresponding strain or curvature equals $1$ while all others are $0$ everywhere in the unit cell. $\mathbf{f}^0$ is expressed as $\mathbf{f}^0=\int_Y\mathbf{B}^T\mathbf{E}\boldsymbol{\chi}^0\dd\mathbf{y}$~\citep{cai2014novel}, where $\mathbf{B}$ is the strain-displacement matrix, $\mathbf{E}$ is the constitutive matrix, and $\boldsymbol{\chi}^0$ is the displacement corresponding to $e_{\gamma\delta}$ or $\kappa_{\gamma\delta}$. $\boldsymbol{\chi}^0$ is obtained by constraining the displacement of each node, shown on the left of \Cref{AHunitStrain}. The constraints are expressed as
\begin{equation}
\label{unitStrainDisp}
\begin{aligned}
   &e_{11}=1,\; \boldsymbol{\chi}^{0}_{\textup{node}}=\begin{bmatrix}
        y_1\\
        0\\
        0
    \end{bmatrix};&&e_{22}=1,\; \boldsymbol{\chi}^{0}_{\textup{node}}=\begin{bmatrix}
        0\\
        y_2\\
        0
    \end{bmatrix};&&e_{12}=1,\; \boldsymbol{\chi}^{0}_{\textup{node}}=\tfrac{1}{2}\begin{bmatrix}
        y_2\\
        y_1\\
        0
    \end{bmatrix}\\
    &\kappa_{11}=1,\; \boldsymbol{\chi}^{0}_{\textup{node}}=\tfrac{1}{2}\begin{bmatrix}
        2y_1y_3\\
        0\\
        -y_1^2
    \end{bmatrix};&
    &\kappa_{22}=1,\; \boldsymbol{\chi}^{0}_{\textup{node}}=\tfrac{1}{2}\begin{bmatrix}
        0\\
        2y_2y_3\\
        -y_2^2
    \end{bmatrix};&
    &\kappa_{12}=1,\; \boldsymbol{\chi}^{0}_{\textup{node}}=\tfrac{1}{2}\begin{bmatrix}
        y_2y_3\\
        y_1y_3\\
        -y_1y_2
    \end{bmatrix}
    \end{aligned}
\end{equation}
where $\boldsymbol{\chi}^{0}_{\textup{node}}$ is the displacement vector of a node in the unit cell, and $(y_1,y_2,y_3)$ are the coordinates of the node. Such constraints ensure that each displacement field $\boldsymbol\chi^0$ satisfies unit strain or curvature. The resulting unit cell with unit strain or unit curvature is presented in \Cref{AHunitStrain}.

In addition, the six deformation modes in \Cref{unitStrainDisp} have 15 linear combinations, as for instance
\begin{equation}
\label{unitStrainDisp2}
    e_{11}=1,\quad e_{22}=1,\quad \boldsymbol{\chi}^{0}_{\textup{node}}=\begin{bmatrix}
        y_1\\
        0\\
        0
    \end{bmatrix}+\begin{bmatrix}
        0\\
        y_2\\
        0
    \end{bmatrix}=\begin{bmatrix}
        y_1\\
        y_2\\
        0
    \end{bmatrix}.
\end{equation}
The initial strain loading $\mathbf{f}^{0}$ is then found from
\begin{equation}
\label{initial strain loading}
    \mathbf{K}\boldsymbol{\chi}^{0}=\mathbf{f}^{0}.
\end{equation}
\Cref{unitStrainDisp} and their linear combinations can be implemented by specifying a displacement boundary condition at each node. A subsequent static analysis in any FE software solves \Cref{initial strain loading}. The initial strain loading $\mathbf{f}^{0}$ is extracted from the results as nodal reactions. The procedure is repeated 21 times ($6+15=21$)  for the deformation modes in \Cref{unitStrainDisp} and their linear combinations. 

The next step is to solve the microscopic problems
\begin{equation}\label{microProbFE}
    \mathbf{K}\boldsymbol{\chi}^{*}=\mathbf{f}^{0},
\end{equation}
where $\mathbf{f}^{0}$ is the initial strain loading obtained from the previous step, and $\boldsymbol{\chi}^{*}$ is the unknown characteristic displacement field. Different from the unit strains and curvatures, the characteristic displacements are assumed to be periodic in $y_1$ and $y_2$, as specified in \Cref{microProbIP,microProbOOP}. This means that opposite unit cell boundaries have identical displacement. Because the Miura pattern tessellates in $(y_1,y_2)$, each unit cell has four boundaries, with the right boundary connected to the left boundary of an adjacent unit cell, and similar for the top and bottom boundaries. Accordingly the periodic boundary condition constrains the displacements of each pair of boundary nodes as
\begin{subequations} \label{AHPBC} \begin{align}
    &\boldsymbol{\chi}^{*(\textup{l})}-\boldsymbol{\chi}^{*(\textup{r})}=\mathbf 0,\\
    &\boldsymbol{\chi}^{*(\textup{t})}-\boldsymbol{\chi}^{*(\textup{b})}=\mathbf 0,
\end{align} \end{subequations}
where $\boldsymbol{\chi}^{*(\textup{l})}$, $\boldsymbol{\chi}^{*(\textup{r})}$, $\boldsymbol{\chi}^{*(\textup{t})}$, and $\boldsymbol{\chi}^{*(\textup{b})}$ are displacements on the left, right, top, and bottom boundaries, respectively. 

\begin{figure}[!ht]
    \centering
    \includegraphics[]{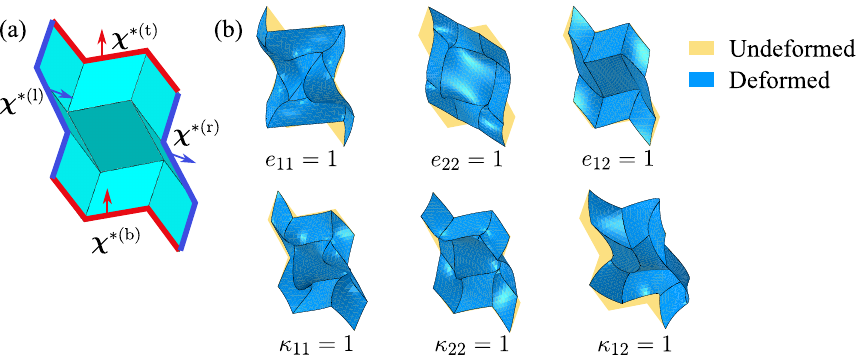}
    \caption{Step 2 of asymptotic homogenization solving the microscopic problem. (a)~Periodic boundary conditions are applied on the boundary displacements according to \Cref{AHPBC}. (b)~Numerical solutions of the characteristic displacement field $\boldsymbol{\chi}^{*}$. Displacements are down-scaled for plotting purposes.}
    \label{AHcharacteristic}
\end{figure}

The initial strain loadings $\mathbf{f}^0$ is applied to each node of the unit cell as a nodal force. Periodic boundary conditions are applied to the four sides of the unit cell in \Cref{AHcharacteristic} as nodal displacement constraints. Fixed boundary conditions to selected nodes are also enforced to prevent rigid body translation and rotation. Performing a static analysis in the FE software solves \Cref{microProbFE}. The procedure is repeated for all 21 initial strain loadings from the previous step to obtain the characteristic displacement fields $\boldsymbol{\chi}^{*}$. \Cref{AHcharacteristic} shows  $\boldsymbol{\chi}^{*}$ corresponding to the deformation mode $e_{11}=1$.

\begin{figure}[!ht]
    \centering
    \includegraphics[]{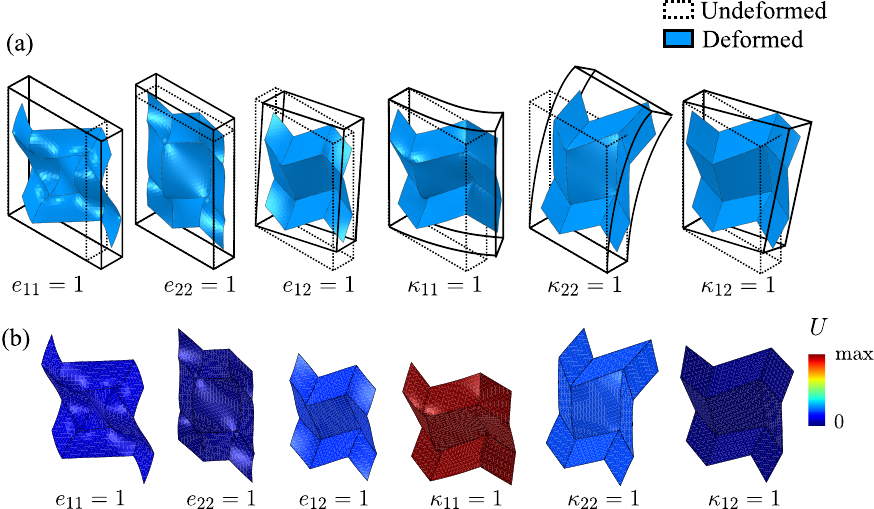}
    \caption{Step 3 of asymptotic homogenization computing unit cell strain energy $U$. (a)~Difference between the unit strain displacement field $\boldsymbol{\chi}^{0}$ and the characteristic displacement field $\boldsymbol{\chi}^{*}$. Only 6 of the deformation modes are shown with lighting effects in 3D rendering. Displacements are down-scaled for plotting purposes. (b)~Unit cell strain energy $U$. Colors represent the strain energy magnitudes of the corresponding deformation modes.}
    \label{fig:AHdiff}
\end{figure}

The third step begins with finding the difference between the unit strain displacement field and the characteristic displacement field, 
\begin{equation}
\tilde{\boldsymbol{\chi}}=\boldsymbol{\chi}^{0}-\boldsymbol{\chi}^{*}.
\end{equation}
$\tilde{\boldsymbol{\chi}}$ is applied to the unit cell to find its associated strain energy $U$. \Cref{fig:AHdiff} shows the displacement fields of six deformation modes as well as their strain energy. We specify $\tilde{\boldsymbol{\chi}}$ at each node of the unit cell through displacement boundary conditions. A static analysis is performed to find the total strain energy $U$. The analysis is repeated for the 6 deformation modes and their 15 linear combinations.

The fourth step is to find the effective stiffness coefficients $A_{\alpha \beta \gamma \delta}^H$, $B_{\alpha \beta \gamma \delta}^H$, $D_{\alpha \beta \gamma \delta}^H$ from \Cref{AHdiagonal,AHoffDiagonal,DHdiagonal,DHoffDiagonal,BHoffDiagonal}.

The final step is the homogenized problem. The origami sheet is simplified into a thin plate. The homogenized plate $\Omega$, shown in \Cref{fig:frameworkc}, is modeled with triangular or quadrilateral thin plate elements with a plate thickness of $h$. The effective elastic properties are specified in the FE model as elastic and anisotropic material properties. After applying appropriate loads and boundary conditions on the homogenized model, one last static analysis is performed, which solves the homogenized governing equation, see \Cref{homogMacro}.

A flowchart of the numerical implementation above is summarized in \Cref{AHFlowchart}.

\subsection{Energy-based homogenization}
\label{EHsubsection}
\subsubsection{Theory}
To apply the energy-based homogenization method, we assume that the origami metamaterial is represented by a unit cell, whose strain equals the average strain of the entire metamaterial. We apply periodic boundary conditions on the displacement to generate the desired strains, before solving the boundary value problem. Finally, we derive the effective stiffness coefficients from the unit cell strain energy.

A microscopic coordinate system $\mathbf{y}$ is introduced with its origin at the bottom left corner of the unit cell and $(y_1,y_2)$ axes on the mid-surface of the unit cell. Periodic boundary conditions are applied on the four sides of the unit cell. Specifically, the relative displacements are specified between opposite boundaries, generating specified strain or curvature in a given direction. The expressions for periodic boundary conditions, which generate in-plane strains, are~\citep{Wang2009}
\begin{subequations} \begin{align}
    &u_{\alpha}^{(\textup{r})}=u_{\alpha}^{(\textup{l})}+e_{\alpha\beta}L_{\beta},\\
    &u_{3}^{(\textup{r})}=u_{3}^{(\textup{l})},\\
    &u_{\alpha}^{(\textup{t})}=u_{\alpha}^{(\textup{b})}+e_{\alpha\beta}L_{\beta},\\    
    &u_{3}^{(\textup{t})}=u_{3}^{(\textup{b})},
\end{align} \end{subequations}
where $u_{\alpha}^{(\textup{r})}$, $u_{\alpha}^{(\textup{l})}$, $u_{\alpha}^{(\textup{t})}$, and $u_{\alpha}^{(\textup{b})}$ are displacements respectively on the right, left, top and bottom boundaries of the unit cell, $e_{\alpha\beta}$ are applied strains, and $L_\beta$ are the length of the unit cell along $y_\beta$. The periodic boundary conditions above differ from those of asymptotic homogenization (\Cref{AHPBC}) by a term $e_{\alpha\beta}L_{\beta}$, which generates an average strain of $e_{\alpha\beta}$.

The periodic boundary conditions corresponding to an average curvature $\kappa_{\alpha\beta}$ are~\citep{vasudevan2024homogenization}
\begin{subequations} \begin{align}
\begin{bmatrix} u_1^{(\textup{r})}-u_1^{(\textup{l})}\\u_2^{(\textup{r})}-u_2^{(\textup{l})}\\u_3^{(\textup{r})}-u_3^{(\textup{l})} \end{bmatrix} = \frac{1}{2}
\begin{bmatrix}
(x_1^{(\textup{r})}-x_1^{(\textup{l})}) (h - 2x_3^{(\textup{r})}) & 0 & 0 \\
0 & 0 & (x_1^{(\textup{r})}-x_1^{(\textup{l})}) (h - 2x_3^{(\textup{r})})/2 \\
 (x_1^{(\textup{r})}-x_1^{(\textup{l})}) (x_1^{(\textup{r})}+x_1^{(\textup{l})}) & 0 & x_1^{(\textup{r})}x_2^{(\textup{r})}- x_1^{(\textup{l})}x_2^{(\textup{l})}
\end{bmatrix}
\begin{bmatrix}
\kappa_{11} \\
\kappa_{22} \\
2\kappa_{12}
\end{bmatrix},\\
\begin{bmatrix} u_1^{(\textup{t})}-u_1^{(\textup{b})}\\u_2^{(\textup{t})}-u_2^{(\textup{b})}\\u_3^{(\textup{t})}-u_3^{(\textup{b})} \end{bmatrix} = \frac{1}{2}
\begin{bmatrix}
0 & 0 & (x_2^{(\textup{t})}-x_2^{(\textup{b})}) (h - 2x_3^{(\textup{t})})/2 \\
0 & (x_2^{(\textup{t})}-x_2^{(\textup{b})}) (h - 2x_3^{(\textup{t})}) & 0 \\
 0 & (x_2^{(\textup{t})}-x_2^{(\textup{b})}) (x_2^{(\textup{t})}+x_2^{(\textup{b})}) & x_1^{(\textup{t})}x_2^{(\textup{t})}- x_1^{(\textup{b})}x_2^{(\textup{b})}
\end{bmatrix}
\begin{bmatrix}
\kappa_{11} \\
\kappa_{22} \\
2\kappa_{12}
\end{bmatrix}. 
\end{align} \end{subequations}
where $x_i^{(\textup{r})}$, $x_i^{(\textup{l})}$, $x_i^{(\textup{t})}$, $x_i^{(\textup{b})}$ are positions respectively on the right, left, top and bottom boundaries of the unit cell.

The resulting strain energy density of a unit cell is 
\begin{equation}
    \bar{U} = \frac{1}{2L_{1}L_{2}}\int_{Y}\hat\sigma_{ij}\hat e_{ij}\dd y_1\dd y_2
\end{equation}
where $Y$ is the mid-surface area, $\hat\sigma_{ij}$ is the local stress tensor, $\hat e_{ij}$ is the local strain tensor, and $y_1,y_2$ are the local coordinates at the unit cell level. The energy density is evaluated per area of the unit cell mid-surface. The energy-based homogenization method requires that
\begin{equation}
\label{enerEquiv}
    \bar{U}^H=\bar{U},
\end{equation}
with $\bar{U}^H$ expressed in \Cref{homogU2}. As per the symmetry of the stiffness coefficient matrix, 21 equations are required to compute the effective stiffness coefficients $A_{\alpha \beta \gamma \delta}^H$, $B_{\alpha \beta \gamma \delta}^H$, and $D_{\alpha \beta \gamma \delta}^H$. These equations are obtained from 21 unit strains, unit curvatures and their linear combinations applied to \Cref{enerEquiv}. In practice, appropriate values are set for the strains $e_{\alpha \beta}$ and curvatures $\kappa_{\alpha \beta}$ to identify the corresponding deformation mode of the unit cell.

\subsubsection{Numerical implementation}
\label{EHimplementation}
The numerical implementation of energy-based homogenization begins with constructing a model of the unit cell and identifying its boundaries. To obtain strains $e_{\alpha \beta}$, we apply constraints on each pair of nodes on opposite boundaries of the unit cell. For instance, to deform the unit cell into an average strain of $e_{11}=1$ on the boundaries of the unit cell, each node on the left boundary and its counterpart on the right boundary follow the constraints
\begin{subequations} 
\label{PBCe11lr}
\begin{align}
    u_{1}^{(\textup{r})} - u_{1}^{(\textup{l})}-(y_{1}^{(\textup{r})}-y_{1}^{(\textup{l})})=0,\\
    u_{2}^{(\textup{r})}- u_{2}^{(\textup{l})}=0,\\
    u_{3}^{(\textup{r})} - u_{3}^{(\textup{l})}=0,
\end{align} \end{subequations}
where $u_{i}^{(\textup{r})}$, $u_{i}^{(\textup{l})}$ are the right and left nodal displacements in $y_i$, and $y_{1}^{(\textup{r})}$, $y_{1}^{(\textup{l})}$ are the undeformed $y_1$ nodal coordinates. In parallel, each node pair on the top and bottom boundaries keeps its distance after deformation such that
\begin{equation}
\label{PBCe11tb}
    u_{i}^{(\textup{t})} - u_{i}^{(\textup{b})}=0.
\end{equation}
We apply the constraints in \Cref{PBCe11lr,PBCe11tb} to the 3 translational DoFs of the corresponding node pairs. The constraints for in-plane deformation modes summarize as
\begin{align}
\label{EHipe11}
   & e_{11}=1,&&\mathbf{u}^{(\textup{r})}-\mathbf{u}^{(\textup{l})}=\begin{bmatrix}
        y_{1}^{(\textup{r})}-y_{1}^{(\textup{l})}\\0\\0
    \end{bmatrix},&&\mathbf{u}^{(\textup{t})}-\mathbf{u}^{(\textup{b})}=\mathbf{0},\\
    \label{EHipe22}
   & e_{22}=1,&&\mathbf{u}^{(\textup{r})}-\mathbf{u}^{(\textup{l})}=\mathbf{0},&&\mathbf{u}^{(\textup{t})}-\mathbf{u}^{(\textup{b})}=\begin{bmatrix}
        0\\y_{2}^{(\textup{t})}-y_{2}^{(\textup{b})}\\0
    \end{bmatrix}\\
    \label{EHipe12}
   & e_{12}=2,&&\mathbf{u}^{(\textup{r})}-\mathbf{u}^{(\textup{l})}=\begin{bmatrix}
        0\\y_{1}^{(\textup{t})}-y_{1}^{(\textup{b})}\\0
    \end{bmatrix},&&\mathbf{u}^{(\textup{t})}-\mathbf{u}^{(\textup{b})}=\begin{bmatrix}
        y_{2,}^{(\textup{r})}-y_{2}^{(\textup{l})}\\0\\0
    \end{bmatrix}.
\end{align}
For out-of-plane deformation modes,
\begin{align}
\label{EHoopk11}
    &\kappa_{11}\!=\!1,&&\mathbf{u}^{(\textup{r})}-\mathbf{u}^{(\textup{l})}\!=\!\!\begin{bmatrix}
        (\frac{h}{2}-y_{3}^{(\textup{r})})(y_{1}^{(\textup{r})}-y_{1}^{(\textup{l})})\\0\\\tfrac{1}{2}(y_{1}^{(\textup{r})\,2}-y_{1}^{(\textup{l})2})
    \end{bmatrix}\!\!,&&\mathbf{u}^{(\textup{t})}-\mathbf{u}^{(\textup{b})}\!=\!\mathbf{0},\\
    \label{EHoopk22}
    &\kappa_{22}\!=\!1,&&\mathbf{u}^{(\textup{r})}-\mathbf{u}^{(\textup{l})}\!=\!\mathbf{0},&&\mathbf{u}^{(\textup{t})}-\mathbf{u}^{(\textup{b})}\!=\!\!\begin{bmatrix}
        0\\(\frac{h}{2}-y_{3}^{(\textup{t})})(y_{2}^{(\textup{t})}-y_{2}^{(\textup{b})})\\\tfrac{1}{2}(y_{2}^{(\textup{t})2}-y_{2}^{(\textup{b})2})
    \end{bmatrix}\!\!,\\
    \label{EHoopk12}
    &\kappa_{12}\!=\!2,&&\mathbf{u}^{(\textup{r})}-\mathbf{u}^{(\textup{l})}\!=\!\!\begin{bmatrix}
        0\\\tfrac{1}{2}(\frac{h}{2}-y_{3}^{(\textup{r})})(y_{1}^{(\textup{r})}-y_{1}^{(\textup{l})})\\\tfrac{1}{2}(y_{1}^{(\textup{r})}y_{2}^{(\textup{r})}-y_{1}^{(\textup{l})}y_{2}^{(\textup{l})})
    \end{bmatrix}\!\!,&&\mathbf{u}^{(\textup{t})}-\mathbf{u}^{(\textup{b})}\!=\!\!\begin{bmatrix}
        \tfrac{1}{2}(\frac{h}{2}-y_{3}^{(\textup{t})})(y_{2}^{(\textup{t})}-y_{2}^{(\textup{b})})\\0\\\tfrac{1}{2}(y_{1}^{(\textup{t})}y_{2}^{(\textup{t})}-y_{1}^{(\textup{b})}y_{2}^{(\textup{b})})
    \end{bmatrix}
\end{align}
where $y_{3}^{(\textup{r})}=y_{3}^{(\textup{l})}$ and $y_{3}^{(\textup{t})}=y_{3}^{(\textup{b})}$ due to geometric periodicity. In addition, we add appropriate boundary conditions to prevent rigid body motion. \Cref{AHFlowchart} compares the boundary conditions and loads of energy-based and asymptotic homogenization.

\begin{figure}[!ht]
\centering
\includegraphics[]{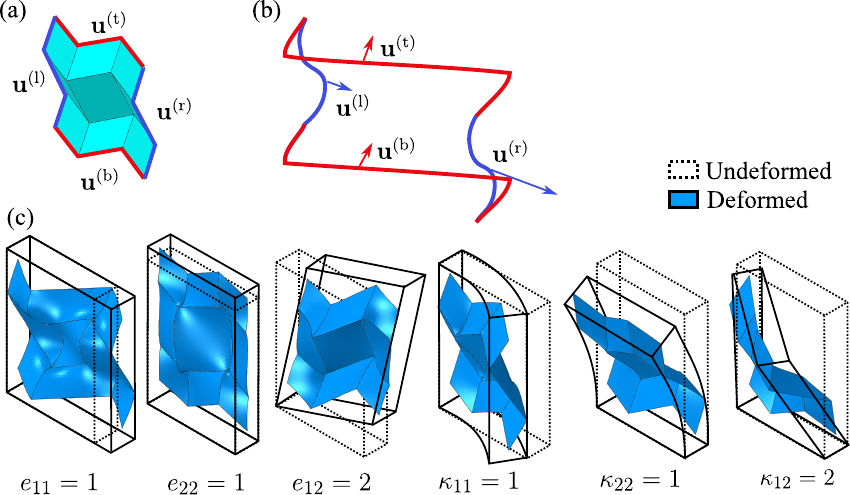}
\caption{Step 1 of energy-based homogenization identifying deformation modes. (a)~Four boundaries of the unit cell. (b)~Deformed boundaries after applying periodic boundary conditions. The overall strain is $e_{11}=1$. (c)~Six of the deformation modes under periodic boundary conditions with lighting effects in 3D rendering. Displacements are down-scaled for plotting purposes.}
\label{EHdefModes}
\end{figure}

\begin{figure}[!ht]
\centering
\includegraphics[]{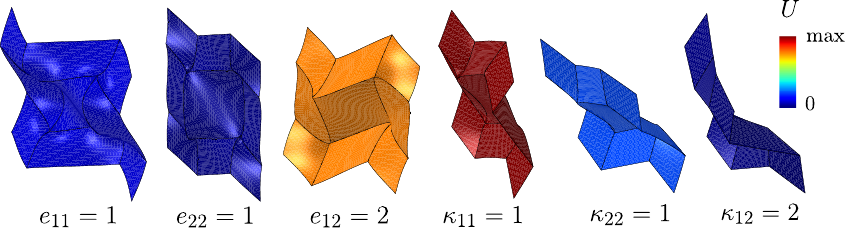}
\caption{Step 2 of energy-based homogenization computes the total unit cell strain energy~$U$ of each deformation mode in step 1. Colors represent the strain energy magnitudes of the corresponding deformation modes.}
\label{EHU}
\end{figure}

We numerically solve six boundary value problems with the above periodic boundary conditions enforced as nodal displacement constraints. \Cref{EHdefModes} shows the deformed unit cells. From the solutions of nodal displacements of each boundary value problem, we obtain the unit cell strain energy $U$. \Cref{EHU} compares the total strain energy of each deformation mode.

In addition, we write the 15 linear combinations of the periodic boundary conditions in \Cref{EHipe11,EHipe22,EHipe12,EHoopk11,EHoopk22,EHoopk12}. For example, the linear combination of $e_{11}$ and $e_{22}$ is known as biaxial extension. To find the corresponding boundary condition, we sum \Cref{EHipe11,EHipe22} to obtain
\begin{equation}\label{EHe11e22}
    e_{11}=e_{22}=1,\quad\mathbf{u}^{(\textup{r})}-\mathbf{u}^{(\textup{l})}=\begin{bmatrix}
        y_{1}^{(\textup{r})}-y_{1}^{(\textup{l})}\\0\\0
    \end{bmatrix},\quad \mathbf{u}^{(\textup{t})}-\mathbf{u}^{(\textup{b})}=\begin{bmatrix}
        0\\y_{2}^{(\textup{t})}-y_{2}^{(\textup{b})}\\0
    \end{bmatrix}.
\end{equation}
Similar to \Cref{EHipe11,EHipe22,EHipe12,EHoopk11,EHoopk22,EHoopk12}, we apply \Cref{EHe11e22} as nodal displacement constraints. Solving the boundary value problem numerically gives the displacement field as well as the unit cell strain energy $U$ of the biaxial extension test. We then obtain the strain energy of the rest of the deformation modes.

The next step is to compute the effective stiffness coefficients $A_{ijk\ell}^H$, $B_{ijk\ell}^H$, and $D_{ijk\ell}^H$ from \Cref{enerEquiv}~\citep{Wang2009}. Each strain energy $U$ computed from \Cref{EHipe11,EHipe22,EHipe12,EHoopk11,EHoopk22,EHoopk12} gives a diagonal term in \Cref{homogU2}, while the linear combinations of the periodic boundary conditions give the off-diagonal terms.

\paragraph{Uniaxial extension, shear, bending, and twisting}
Each set of periodic boundary conditions in \Cref{EHipe11,EHipe22,EHipe12,EHoopk11,EHoopk22,EHoopk12}, corresponds to a total strain energy in the unit cell $U$. From the energy equivalence, we have 
\begin{equation}
A^H_{\alpha\beta\gamma\delta}=\frac{2U}{L_{1}L_{2}e_{\alpha\beta}e_{\gamma\delta}}\quad\text{and}\quad D^H_{\alpha\beta\gamma\delta}=\frac{2U}{L_{1}L_{2}\kappa_{\alpha\beta}\kappa_{\gamma\delta}},
\end{equation}
where $L_{1}$ and $L_{2}$ are the lengths of the unit cell along $y_1$ and $y_2$, respectively.

\paragraph{Biaxial extension and other coupling terms}
Each off-diagonal term of the effective stiffness matrix in \Cref{homogU2} corresponds to a strain or curvature in its row and another strain or curvature in its column. For example, $B^H_{1111}$ describes the coupling effect between uniaxial extension in $e_{11}$ and bending $\kappa_{11}$. Energy equivalence indicates that $B^H_{1111}$ is twice the strain energy density of the unit cell whose strains and curvatures are $e_{11}=\kappa_{11}=1$ and $e_{22}=e_{12}=\kappa_{22}=\kappa_{12}=0$. The strains and curvatures are imposed using a set of periodic boundary conditions, the linear combination of \Cref{EHipe11,EHoopk11}. Then the strain energy $U$ is numerically computed. The effective properties are
\begin{equation}
    \begin{aligned}
        &A^H_{\alpha\beta\gamma\delta}= \frac{1}{e_{\alpha\beta}e_{\gamma\delta}}\Bigl(\frac{U}{L_{1}L_{2}}-\frac{1}{2e_{\alpha\beta}e_{\alpha\beta}}A^H_{\alpha\beta\alpha\beta}-\frac{1}{2e_{\gamma\delta}e_{\gamma\delta}}A^H_{\gamma\delta\gamma\delta}\Bigr),\\
        &B^H_{\alpha\beta\gamma\delta}= \frac{1}{e_{\alpha\beta}\kappa_{\gamma\delta}}\Bigl(\frac{U}{L_{1}L_{2}}-\frac{1}{2e_{\alpha\beta}e_{\alpha\beta}}A^H_{\alpha\beta\alpha\beta}-\frac{1}{2\kappa_{\gamma\delta}\kappa_{\gamma\delta}}D^H_{\gamma\delta\gamma\delta}\Bigr),\\
        &D^H_{\alpha\beta\gamma\delta}= \frac{1}{\kappa_{\alpha\beta}\kappa_{\gamma\delta}}\Bigl(\frac{U}{L_{1}L_{2}}-\frac{1}{2\kappa_{\alpha\beta}\kappa_{\alpha\beta}}D^H_{\alpha\beta\alpha\beta}-\frac{1}{2\kappa_{\gamma\delta}\kappa_{\gamma\delta}}D^H_{\gamma\delta\gamma\delta}\Bigr).
    \end{aligned}
\end{equation}
Note that the results from uniaxial extension, shear, bending, and twisting are required for $A^H_{\alpha\beta\alpha\beta}$ and $D^H_{\alpha\beta\alpha\beta}$ terms on the right-hand side of the equation above.

Finally, the homogenized stiffness matrix is applied to the homogenized model, as in the last step of \Cref{AHimplementation}.

A flowchart of the numerical implementation above is summarized in \Cref{EHFlowchart}.

\section{Validation}
\label{validation}
While an arbitrary origami sheet is described with 21 effective stiffness coefficients as in \Cref{plateConst}, it is possible to reduce the number of constants by enforcing the symmetry of the origami pattern. For example, a Miura sheet exhibits centrosymmetry, meaning that a Miura sheet rotated by $180^\circ$ is identical to the original one. For centrosymmetric origami patterns, the following constitutive relation can be derived from the couple-stress theory~\citep{vasudevan2024homogenization}:
\begin{equation}
\label{complMtx}
    \begin{bmatrix}
e_{11} \\ 
e_{22} \\ 
2e_{12} \\ 
\kappa_{11} \\ 
\kappa_{22} \\ 
2\kappa_{12}
\end{bmatrix}
=
\begin{bmatrix}
\frac{1}{E_{1}} & -\frac{\nu^s_{21}}{E_{2}} & 0 & 0 & 0 & 0 \\ 
-\frac{\nu^s_{12}}{E_{1}} & \frac{1}{E_{2}} & 0 & 0 & 0 & 0 \\ 
0 & 0 & \frac{1}{G_{12}} & 0 & 0 & 0 \\ 
0 & 0 & 0 & \frac{1}{M_{1}} & -\frac{\nu^b_{21}}{M_{2}} & 0 \\ 
0 & 0 & 0 & -\frac{\nu^b_{12}}{M_{1}} & \frac{1}{M_{2}} & 0 \\ 
0 & 0 & 0 & 0 & 0 & \frac{1}{T_{12}}
\end{bmatrix}
\begin{bmatrix}
\sigma_{11} \\ 
\sigma_{22} \\ 
\sigma_{12} \\ 
\mu_{11} \\ 
\mu_{22} \\ 
\mu_{12}
\end{bmatrix},
\end{equation}
where $\sigma_{\alpha\beta}$ are stresses, and $\mu_{\alpha\beta}$ are couple-stresses. The relationship involves only nine independent effective elastic constants, which are in-plane Poisson's ratio $\nu_{12}$, Young's moduli $E_1$ and $E_2$, shear modulus $G_{12}$, out-of-plane Poisson's ratio $\nu^b_{12}$, bending moduli $M_1$ and $M_2$, and twisting modulus $T_{12}$. The out-of-plane Poisson's ratio $\nu^b_{12}$ and $\nu^b_{21}$ are defined from the transverse curvatures, $\nu^b_{12}=-\kappa_{22}/\kappa_{11}$, and $\nu^b_{21}=-\kappa_{11}/\kappa_{22}$. 

The effective elastic constants can be determined from the effective stiffness coefficients as
\begin{equation}
    \mathbf{S}=h[\mathbf{ABD}]^{-1}
\end{equation}
where $\mathbf{S}$ is the compliance matrix in \Cref{complMtx}, $h$ is the height of the origami sheet, and $[\mathbf{ABD}]$ is the homogenized matrix in \Cref{plateConst}.

\begin{figure}[!ht]
    \centering
    \includegraphics[]{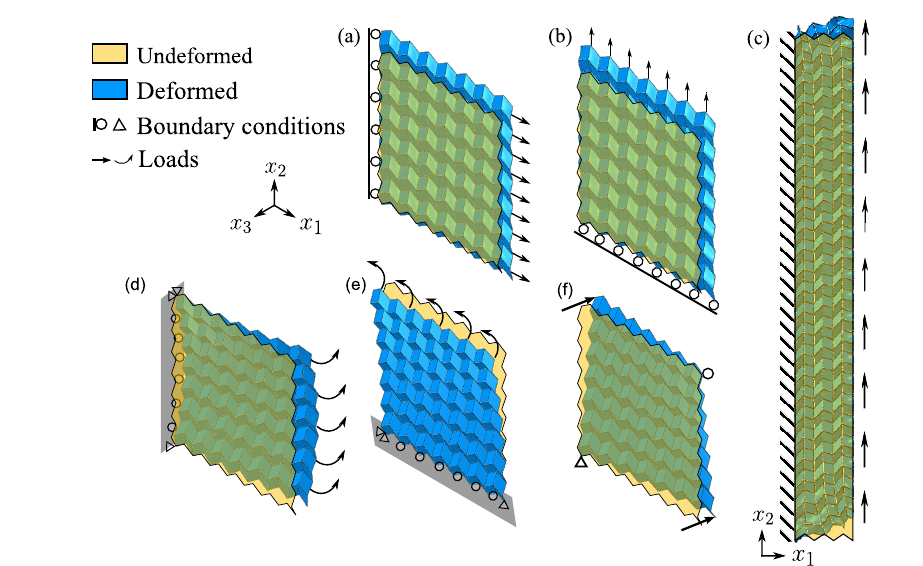}
    \begin{subfigure}{0.1\textwidth}
        \phantomsubcaption
        \label{fig:modelsErrora}
        \phantomsubcaption
        \label{fig:modelsErrorb}
        \phantomsubcaption
        \label{fig:modelsErrorc}
        \phantomsubcaption
        \label{fig:modelsErrord}
        \phantomsubcaption
        \label{fig:modelsErrore}
        \phantomsubcaption
        \label{fig:modelsErrorf}
    \end{subfigure}
\caption{Detailed plate FE model under mechanical tests: (a,b) uniaxial extension, (c) in-plane shear, (d,e) bending, and (f) twisting.}\label{fig:modelsError}
\end{figure}

To set a baseline for comparison, we solved a detailed FE model of the origami sheet with identical parameters of the homogenized models, as listed in \Cref{MiuraProperties}. Standard practice in homogenization applied to both periodic porous composites~\citep{hollister1992comparison} and origami metamaterials~\citep{vasudevan2024homogenization} shows that a detailed FE model is sufficient to validate the result from homogenization. To determine the elastic constants, we use a specimen with seven unit cells on each side of the sheet, adding up to 49 unit cells. On the other hand, for $G_{12}$ we examine a rectangular domain of 3 by 21 unit cells to avoid in-plane bending~\citep{vasudevan2024homogenization}. Although homogenization assumes an infinite tessellation of unit cells, a convergence test on the detailed FE model shows that 7 by 7 unit cells (or 3 by 21 for shear modulus) is a sufficient tessellation size. Simulations of six mechanical tests, shown in \Cref{fig:modelsError}, are required to determine all the elastic constants. To reduce the boundary effect, we measure the strain or curvature of the unit cell at the center of the tessellation and compute the reference elastic constants.

\Cref{fig:modelsErrora} shows a uniaxial extension test with a roller boundary condition on the left boundary and a small displacement on the right boundary to generate an in-plane strain $e_1$. The resulting stress $\sigma_1$ and transverse strain $e_2$ are measured to obtain the Young's modulus $E_1 = \sigma_1/e_1$ and the Poisson's ratio $\nu_{12}=-e_2/e_1$. \Cref{fig:modelsErrorb} shows the uniaxial extension along $x_2$, with a roller boundary condition on the bottom boundary and a displacement on the top. Measuring the stress $\sigma_2$ gives the Young's modulus $E_2 = \sigma_2/e_2$. \Cref{fig:modelsErrorc} is a shear test with the left boundary fully fixed and a displacement in $x_2$ applied on the right boundary. The shear modulus is $G_{12}=\sigma_{12}/(2e_{12})$. An additional roller boundary condition is required for the above tests to prevent global bending and twisting.

\Cref{fig:modelsErrord} is a bending test with rollers on the left boundary and an applied bending moment $m_2$ evenly distributed on the right. The curvatures $\kappa_1$ and $\kappa_2$ are measured at the center of the specimen using the finite difference method. The bending modulus is $M_1=m_2/(hL_2\kappa_1)$, where $H$ is the height, and $L_2$ is the width of the specimen in $x_2$-direction. The bending Poisson's ratio is $\nu^b_{12}=-\kappa_2/\kappa_1$. \Cref{fig:modelsErrore} is another bending test with rollers on the bottom boundary and a distributed moment $m_1$ on the top. We measure the curvature $\kappa_2$ at the center to find the bending modulus $M_2=m_1/(hL_1\kappa_2)$. \Cref{fig:modelsErrorf} is a twisting test. Two corners on the diagonal are fixed in $x_3$, while the other two corners undergo a small displacement in $x_3$. The reaction force $F$ is measured at each corner, as well as the twist $\kappa_{12}$ at the specimen center. The twisting modulus is calculated as $T_{12}=F/(4h\kappa_{12})$. Additionally, the bending and twisting tests require boundary conditions to prevent rigid body motion.

\Cref{errors} reports and discusses the errors with respect to the baseline results of the detailed FE models. Comparing the deviations of the plate FE asymptotic and energy-based homogenization, the bar-and-hinge model, and the rigid-panel model, we find that on average the errors are lowest for $\nu_{12}$, with the asymptotic plate FE \qty{3.47}{\%}, the energy-based plate FE \qty{3.20}{\%}, the bar-and-hinge \qty{3.83}{\%}, and the rigid-panel \qty{32.1}{\%}. The results of $G_{12}$ and $nu^b_{12}$ are similarly accurate. For $E_1$, $E_2$, $M_1$, and $M_2$, the averaged errors of the homogenized plate FE models are one third to one fifth those of other methods. In addition, the results for $T_{12}$ show that our framework is far more accurate (error of asymptotic plate FE \qty{11.1}{\%}, energy-based plate FE \qty{12.9}{\%}) than the bar-and-hinge model (error \qty{161}{\%}).

\begin{figure}[!ht]
    \centering
    \includegraphics[]{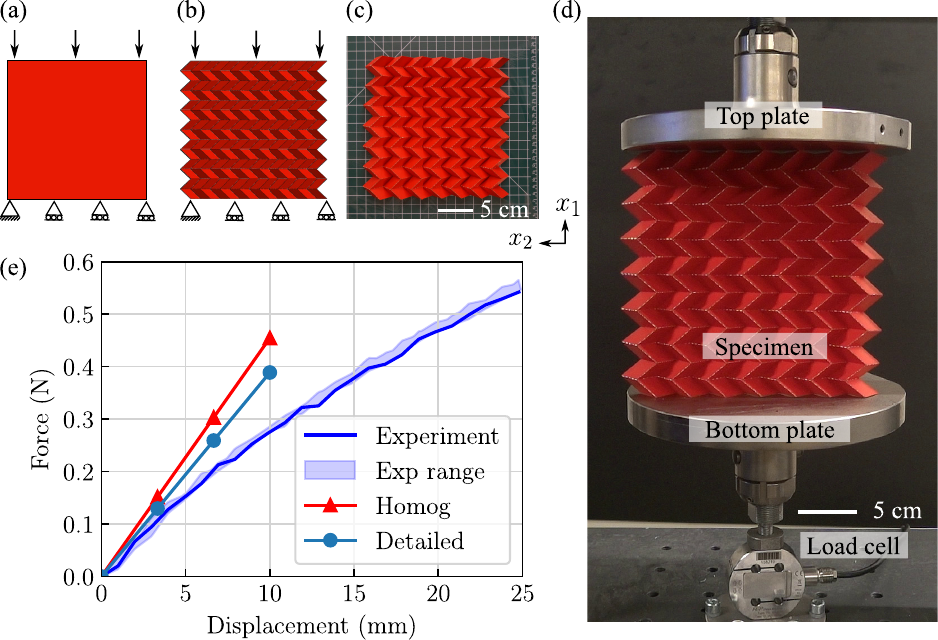}
    \begin{subfigure}{0.1\textwidth}
        \phantomsubcaption
        \label{fig:E11expa}
        \phantomsubcaption
        \label{fig:E11expb}
        \phantomsubcaption
        \label{fig:E11expc}
        \phantomsubcaption
        \label{fig:E11expd}
        \phantomsubcaption
        \label{fig:E11expe}
    \end{subfigure}
    \caption{Experimental, homogenized and detailed simulation of force-displacement curve of Miura origami. (a)~Homogenized model under roller boundary condition and compression load. (b)~Detailed model of Miura origami. (c)~Specimen used for experiment. (d)~Compression test setup. (e)~Force-displacement curve comparing experiment, homogenized model and detailed model. The experimental curve consists of a shade indicating the range of 6 tests, as well as a solid curve showing a representative test.}
    \label{fig:E11exp}
\end{figure}

To further validate our framework, we conducted a compression test on a Miura origami specimen to experimentally evaluate its effective Young's modulus $E_1$. The specimen in \Cref{fig:E11expc} is a 7 by 7 Miura tessellation made of cardboard. The fabrication procedure and specimen properties are detailed in \Cref{experiment}. The specimen is compressed with a universal testing machine between two flat plates with minimal friction. \Cref{fig:E11expd} shows the experimental setup. We simulate the experiment with a homogenized model using energy-based method in \Cref{fig:E11expa} and a detailed model in \Cref{fig:E11expb}. Both models have the same parameters as the specimen. The models share the same boundary conditions, with rollers on the bottom boundary and a uniform displacement of \qty{10}{mm} downwards on the top boundary. Comparing the results in \Cref{fig:E11expe}, we find the homogenized model in good agreement with both the detailed model and the experiment. In the linear elastic range, we compute the slope of the experimental curve at \qty{1.98}{mm} displacement, corresponding to a \qty{1.04}{\%} strain. The error of the homogenized model is \qty{34.7}{\%} compared with the experiment and \qty{16.9}{\%} compared with the detailed model. The discrepancy between the homogenized model and the experiments can be attributed to manufacturing defects: the specimen is folded manually and its fold angle varies slightly from cell to cell.  Other sources of error include a change of the crease properties due to repeated folding, the material anisotropy of the cardboard, and the friction during testing between the specimen and the holder.

\section{Results}
\label{result}
We can now apply both homogenization methods to determine the effective in-plane and out-of-plane properties of Miura origami sheets. Properties in the effective compliance matrix resemble those of a homogeneous thin plate made of an orthotropic material, with minor coupling between in-plane and out-of-plane responses. This observation confirms that the Miura pattern is centrosymmetric~\citep{vasudevan2024homogenization}.

\begin{figure}[!ht]
\centering
\includegraphics[]{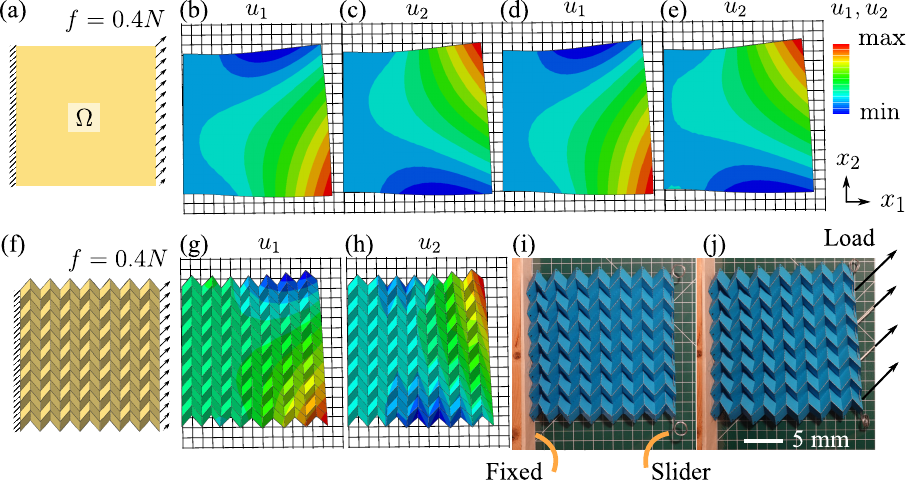}
\begin{subfigure}{0.1\textwidth}
        \phantomsubcaption
        \label{fig:macroa}
        \phantomsubcaption
        \label{fig:macrob}
        \phantomsubcaption
        \label{fig:macroc}
        \phantomsubcaption
        \label{fig:macrod}
        \phantomsubcaption
        \label{fig:macroe}
        \phantomsubcaption
        \label{fig:macrof}
        \phantomsubcaption
        \label{fig:macrog}
        \phantomsubcaption
        \label{fig:macroh}
        \phantomsubcaption
        \label{fig:macroi}
        \phantomsubcaption
        \label{fig:macroj}
    \end{subfigure}
\caption{Comparison of homogenized displacement fields against detailed simulation and experiment. (a)~Boundary condition and load chosen for demonstration on homogenized model. (b)~$u_1$ displacements of asymptotic homogenization model.(c)~$u_2$ displacements of asymptotic homogenization model. (d)~$u_1$ displacements of energy-based homogenization model. (e)~$u_2$ displacements of energy-based homogenization model. The homogenized models in (b)-(e) have 575 elements. (f)~Equivalent load and boundary condition on detailed model with 317,715 elements. (g)~$u_1$ displacements of detailed model. (h)~$u_2$ displacements of detailed model. (i)~Experimental setup. A cardboard Miura origami with left edge glued to a fixture and right edge attached to a slider. (j)~Deformed specimen under applied load on the slider. All simulation results are superimposed on a grid with a cell size equal to that used in the experiment.}
\label{fig:macro}
\end{figure}

To demonstrate the usage of the effective compliance matrices, we solve the global deformation of a Miura tessellation with seven unit cells on each side. \Cref{fig:macroa} shows a homogenized plate $\Omega$ with equal dimensions as the tessellation. The above compliance matrices are applied to the homogenized models as detailed in the last steps of \Cref{AHimplementation,EHimplementation}. The boundary condition on the left boundary is fixed in $u_1$ and $u_2$. There is a load of \qty{0.4}{N} evenly distributed on the right boundary, pointing \ang[]{45} to the upper right. \Cref{fig:macrob,fig:macroc,fig:macrod,fig:macroe} present the resulting displacement fields in the homogenized models. The detailed plate FE model, as the baseline, has a boundary condition of fixed $u_1$, $u_2$, and $u_3$ on the left boundary. There is a evenly distributed load of \qty{0.4}{N} on the right boundary, pointing \ang[]{45} to the upper right. \Cref{fig:macrog,fig:macroh} show the displacement fields. Both homogenized models agree well with the baseline in terms of the distribution of global displacement. There is discrepancy on the right boundary where the load is applied. Detailed model shows localized stretching of the panels that is not observed in the homogenized models.

The homogenized displacement fields are further validated against those from experiments. The cardboard Miura prototype and the simulations share the same geometric and material parameters, presented in \Cref{MiuraCardboard}. The experimental setup ensures a fixed boundary condition on the left and a distributed load on the right. \Cref{fig:macroj} shows the deformed specimen. Its upper right corner undergoes a displacement magnitude of \qty{10}{mm}, which is identical to the detailed simulation at the same location. \Cref{experiment} details the experimental procedure.

To demonstrate the applicability range of the framework across fold patterns, we report in \Cref{curveCreaseEff} the effective properties of a curve crease origami pattern adapted from the literature~\citep{liu2024design}.

\subsection{Role of initial fold angle}
\label{parametric}
\begin{figure}[!ht]
\centering
\includegraphics[]{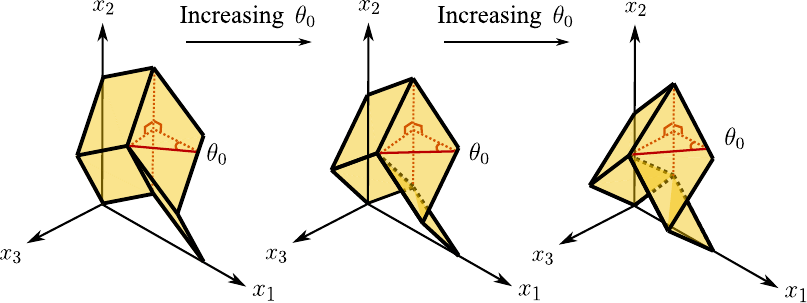}
\caption{Three configurations of Miura unit cell with increasing fold angle $\theta_0$.}
\label{fig:FoldAngle}
\end{figure}

Since the architecture of a metamaterial governs its properties, we now examine how geometry tuning impacts the effective elastic constants of a Miura sheet. In this section, the selected geometric parameter is the initial fold angle $\theta_0$, as shown in \Cref{fig:FoldAngle}. It is measured as the dihedral angle between the panel and the $x_1$-$x_2$ plane. All four panels have the identical angle, a result stemming from the kinematics of Miura origami. $\theta_0$ represents the state of the pre-folded origami sheet, defined as the fold angle when the origami sheet is stress-free. We investigate the linear properties since the analysis does not consider large displacement or large strain. Among the geometric parameters listed in \Cref{MiuraProperties}, we select $\theta_0$ as the varying parameter due to its strong influence on the folding kinematics. It also allows direct comparison with the bar-and-hinge model~\citep{vasudevan2024homogenization} and the rigid-panel model~\citep{wei2013geometric}, both of which have computed the effective properties of the Miura pattern under various fold angles. In this parametric study, $\theta_0$ varies from $1^\circ$ to $80^\circ$. Results from asymptotic and energy-based homogenization models are compared with those of the fully detailed FE model in \Cref{validation} to determine result accuracy. After applying boundary conditions as shown in \Cref{fig:modelsError}, we characterize the detailed FE model in terms of its in-plane Poisson's ratio $\nu_{12},$ Young's moduli $E_1$, $E_2,$ shear modulus $G_{12},$ out-of-plane Poisson's ratio $\nu^b_{12},$ bending moduli $M_1$, $M_2,$ and twisting modulus $T_{12}$. To reduce the boundary effect, we measure the above elastic constants using the unit cell at the center of the origami metamaterial.

\begin{figure}[!ht]
\centering
\includegraphics[width=0.9\linewidth]{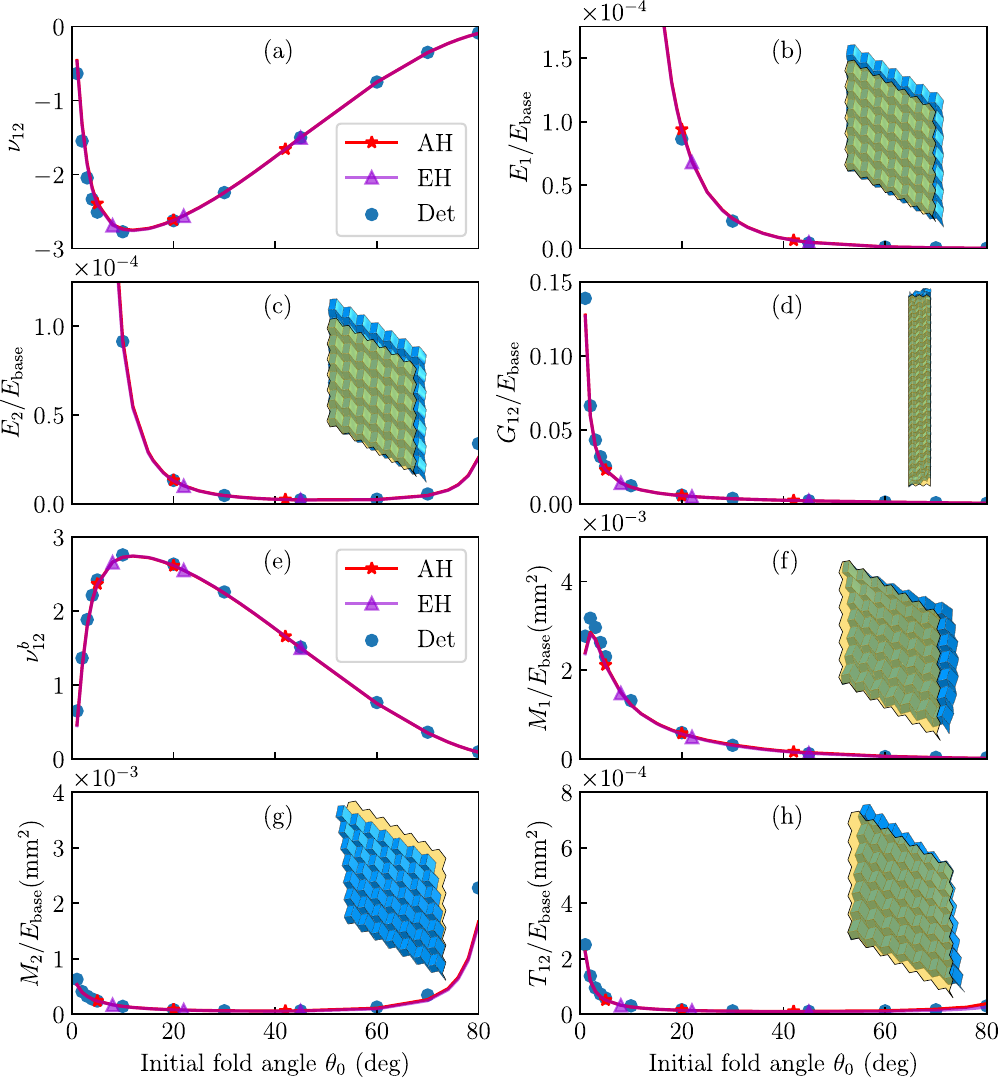}
    \begin{subfigure}{0.1\textwidth}
        \phantomsubcaption
        \label{fig:EHallConstsa}
        \phantomsubcaption
        \label{fig:EHallConstsb}
        \phantomsubcaption
        \label{fig:EHallConstsc}
        \phantomsubcaption
        \label{fig:EHallConstsd}
        \phantomsubcaption
        \label{fig:EHallConstse}
        \phantomsubcaption
        \label{fig:EHallConstsf}
        \phantomsubcaption
        \label{fig:EHallConstsg}
        \phantomsubcaption
        \label{fig:EHallConstsh}
   \end{subfigure}
\caption{In-plane and out-of-plane effective elastic constants of Miura origami metamaterial, including (a)~in-plane Poisson's ratio $\nu_{12}$, effective Young's moduli (b)~$E_1$, (c)~$E_2$, (d)~shear modulus $G_{12}$, (e)~out-of-plane Poisson's ratio $\nu^b_{12}$, bending moduli (f)~$M_1$, (g)~$M_2$, and (h)~twisting modulus $T_{12}$. $E_1$, $E_2$, $G_{12}$, $M_1$, $M_2$, and $T_{12}$ are normalized against the base material Young's modulus $E_{\text{base}}$. Insets show corresponding deformation modes of a detailed FE model. A subset of data points is shown for visual clarity. \legendlinestar{asymptotic} Asymptotic plate FE; \legendlinetriangle{energy}  Energy-based plate FE; \legendmarker{detailedFE} Detailed plate FE.}
\label{fig:EHallConsts}
\end{figure}

\Cref{fig:EHallConstsa,fig:EHallConstse} compare the in-plane and out-of-plane kinematics of Miura origami through its Poisson's ratio. The in-plane Poisson's ratio is negative, a result that parallels the well-known observation of the auxetic response of the Miura sheet. A minimum occurs at $\theta_0=12^\circ$ with $\nu_{12}=-2.76$, while $\nu_{12}$ approaches zero when fully folded. The bending Poisson's ratio is almost equal and opposite to the in-plane one, echoing the prediction of a pin-jointed truss framework~\citep{Schenk2011}. These synchronized properties confirm that the Miura sheet behaves similarly to a single-DOF mechanism.

The in-plane moduli $E_1$, $E_2$, and $G_{12}$ share similar features at low to medium fold angles. For $\theta_0 \approx 0^\circ$, the Miura sheet is almost fully flat, with properties reaching towards those of its base material, a thin sheet of Mylar. As the origami sheet folds, all effective moduli continuously decrease, because the relatively more flexible creases begin to play an increasingly influential role in the overall response. On the other hand, the kinematics of Miura pattern indicates that the total stiffness of the creases decreases for an increase in $\theta_0$~\citep{wei2013geometric}, a result that explains the trend observed on $E_1,E_2$ in \Cref{fig:EHallConstsb,fig:EHallConstsc}. \Cref{fig:EHallConstsd} shows that the shear modulus $G_{12}$ is notably higher than the other two moduli. Previous work~\citep{vasudevan2024homogenization} demonstrated that the shearing deformation mode involves mostly panel stretching, requiring a significant amount of strain energy. Finally, for $\theta_0>60^\circ$ there is a slight increase in $E_2$. This is explained by the alignment of panels along the $x_2$-direction at the highly folded state. In this case, loading along $x_2$ causes stretching in the panels and therefore an increase of $E_2$.

The out-of-plane moduli $M_1$, $M_2$, and $T_{12}$ follow a similar decreasing trend for $5^\circ<\theta_0<60^\circ$. However, at a small $\theta_0$ the bending modulus $M_1$ in \Cref{fig:EHallConstsf} shows a slight increase. This effect is similar to that of a corrugated board: the flutes makes it difficult to bend in one direction but easy in the other. The Miura pattern makes z-shaped flutes along $x_1$ that causes an increase in $M_1$, but as $\theta_0$ increases, this effect diminishes quickly since the folding kinematics has a strong influence on $M_1$. The 'flutes' in the Miura sheet also makes $M_1$ higher than $M_2$ for most $\theta_0$. $M_2$ and $T_{12}$ show a decreasing to increasing trend as $\theta_0$ increases, as seen in \Cref{fig:EHallConstsg,fig:EHallConstsh}, which is qualitatively in agreement with the response of a truss-based lattice~\citep{vasudevan2024homogenization}.

\subsection{Comparison with the literature}
\begin{figure}[!ht]\centering
\includegraphics[]{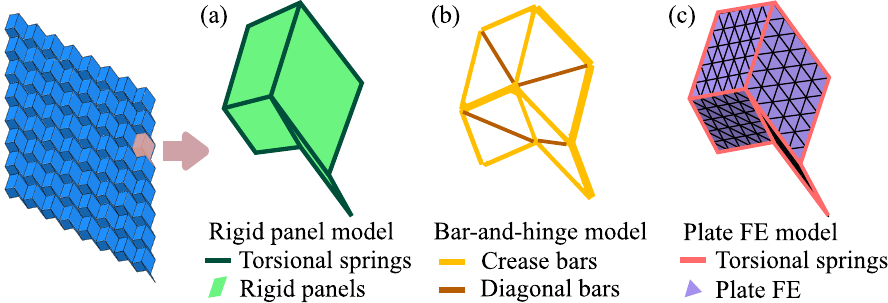}
\caption{Comparison of modeling methods applied to a Miura unit cell for demonstrative purposes. (a)~Rigid panel model has ideally rigid panels and creases represented with torsional springs. (b)~Bar-and-hinge model has crease bars to model crease folding and diagonal bars to model panel bending. (c)~Plate FE model has torsional springs as creases and panels meshed with plate elements.}
\label{fig:modelingMethods}
\end{figure}

To illustrate the applicable range of the proposed framework, we compare results with other origami modeling methods in the literature, including the bar-and-hinge model~\citep{vasudevan2024homogenization} and the rigid-panel model~\citep{wei2013geometric}. \Cref{fig:modelingMethods} summarizes all modeling methods under comparison.
As with the plate FE results presented in \Cref{parametric}, the bar-and-hinge model has been homogenized with the energy-based method. The homogenized properties have been validated against a detailed bar-and-hinge model of 20 by 20 unit cells (with the exception of shear modulus, which requires a 4 by 50 tessellation). Both the bar-and-hinge models share the same parameters as the plate FE model, as listed in \Cref{MiuraProperties}. The material properties are converted to the bar and spring properties as specified in previous work \citep{Filipov2017}. Another method in the literature is the rigid-panel model. It applies to the in-plane Poisson's ratio $\nu_{12}$ and Young's moduli $E_1$ and $E_2$. The shear modulus $G_{12}$ and all out-of-plane effective constants $\nu^b_{12}$, $M_1$, $M_2$, and $T_{12}$ are unbounded due to the assumption that the panels are perfectly rigid. The geometric parameters and crease stiffness are given in \Cref{MiuraProperties}.

\begin{figure}[htbp]\centering
\includegraphics[width=0.9\linewidth]{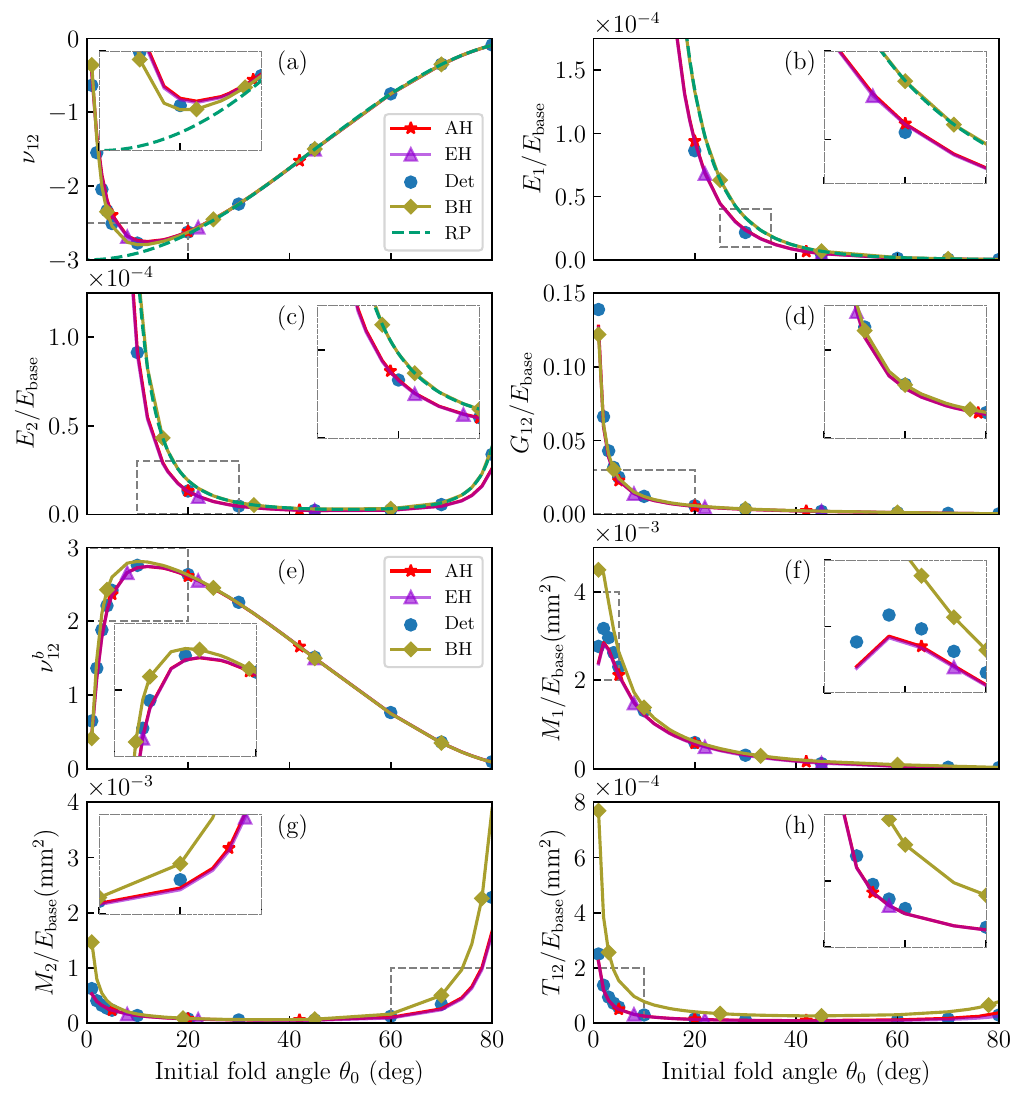}
\begin{subfigure}{0.1\textwidth}
        \phantomsubcaption
        \label{fig:compareMethodsa}
        \phantomsubcaption
        \label{fig:compareMethodsb}
        \phantomsubcaption
        \label{fig:compareMethodsc}
        \phantomsubcaption
        \label{fig:compareMethodsd}
        \phantomsubcaption
        \label{fig:compareMethodse}
        \phantomsubcaption
        \label{fig:compareMethodsf}
        \phantomsubcaption
        \label{fig:compareMethodsg}
        \phantomsubcaption
        \label{fig:compareMethodsh}
   \end{subfigure}
\caption{Comparison of the homogenization framework proposed in this work with methods in the literature. Sub-plots show the variation of effective elastic constants over the fold angles $\theta_0$ between $0^\circ$ and $80^\circ$, including (a)~in-plane Poisson's ratio $\nu_{12}$, effective Young's moduli (b)~$E_1$, (c)~$E_2$, (d)~shear modulus $G_{12}$, (e)~out-of-plane Poisson's ratio $\nu^b_{12}$, bending moduli (f)~$M_1$, (g)~$M_2$, and (h)~twisting modulus $T_{12}$. $E_1$, $E_2$, $G_{12}$, $M_1$, $M_2$, and $T_{12}$ are normalized against the base material Young's modulus $E_{\text{base}}$. Insets show zoomed-in regions of the plots. A subset of data points is shown for visual clarity. \legendlinestar{asymptotic} Asymptotic plate FE; \legendlinetriangle{energy} Energy-based plate FE; \legendmarker{detailedFE} Detailed plate FE; \legendlinediamond{BH} Homogenized/detailed bar-and-hinge; \StmCourbe{rigid,dashed}{} Rigid panel.}
\label{fig:compareMethods}
\end{figure}

In \Cref{fig:compareMethodsa}, the Poisson's ratio $\nu_{12}$ closely aligns for $\theta > 10^\circ$, with value increasing with $\theta$. 
For $0^\circ<\theta<10^\circ$, the trend of both numerical models (bar-and-hinge and plate FE) differ from that of the rigid-panel model, with the former decreasing and the latter increasing. The monotonic increase in the rigid-panel model originates from the assumption that the origami metamaterial behaves as a 1-DoF mechanism. Its kinematic relation fully prescribes the mechanical response in \Cref{rigidPanelPoisson}. On the other hand, the two numerical models account for the base material properties. When $\theta$ approaches $0^\circ$, the origami metamaterial unfolds to a flat sheet of the base material, i.e., Mylar in this work. Correspondingly, the Poisson's ratio from numerical models approaches that of Mylar. As $\theta$ varies, the base material properties and origami kinematics play a competing role in the metamaterial response. With the increase of $\theta$, the kinematic relationship gradually takes dominance. The Poisson's ratio first decreases and then increases to align with the response of a pure mechanism, forming a local minimum at $\theta=10^\circ$ under the given parameters. Additionally, both the plate FE model and the bar-and-hinge model confirm that $\nu^b_{12}$ in \Cref{fig:compareMethodse} is close to the negative of $\nu_{12}$ in \Cref{fig:compareMethodsa}.

\Cref{fig:compareMethodsb,fig:compareMethodsc} show that the in-plane moduli of all methods share the same trend. The Young's moduli $E_1,E_2$ of the plate FE model are lower than others. For example, at a given fold angle $\theta_0=30^\circ$, the moduli of the bar-and-hinge model are close to 1.5 times those of the plate FE. The discrepancy grows as $\theta_0$ becomes smaller and the Miura origami unfolds towards a flat sheet. The bar-and-hinge and the rigid-panel models align well with each other for most of the cases, when $10^\circ<\theta_0<80^\circ$. The shear modulus $G_{12}$ of plate FE and bar-and hinge models match in their values. The rigid-panel model is not included since it does not allow shearing.

The out-of-plane response involves bending and twisting. In \Cref{fig:compareMethodsf}, the bending modulus $M_1$ of the plate FE model first increases then decreases as the Miura sheet unfolds, while the bar-and-hinge model has a monotonically decreasing trend. In \Cref{fig:compareMethodsg}, $M_2$ of both models agree in their trends. Only when the sheet folds beyond $\theta_0>60^\circ$ their difference in values start to grow. $T_{12}$ of both methods also qualitatively agree in \Cref{fig:compareMethodsh}. The result of the bar-and-hinge model is almost 3 times that of the plate FE, and this difference is consistent for most $\theta_0$.

\subsection{Role of crease stiffness}
This section investigates the role played by the constituent material in the effective elastic properties of an origami metamaterial. Some governing factors include the panel material properties and the crease formation process. Our paper adopts a simple crease model~\citep{Filipov2015} which neglects crease width and thickness and assumes a linear torsional stiffness evenly distributed along the crease line. The stiffness ranges from \num[]{0.1} to \qty{1.3}{N.mm} for each \qty{20}{mm} long crease. To put numbers in context, the crease stiffness $K_{\text{cr}}$ of a typical origami metamaterial is smaller than or similar to the panel bending stiffness $K_{\text{p}}$. The panel bending stiffness is expressed as 
\begin{equation}K_p =\Bigl(0.55 - 0.42 \frac{\sum \alpha}{\pi}\Bigr) \frac{E_\text{base} t^3}{12(1 - \nu^2)}\Bigl(\frac{D_s}{t}\Bigr)^{1/3},
\end{equation}
which is a function of the length of panel diagonal $D_s$, the panel thickness $t$, the base material Young's modulus $E_{\text{base}}$ and Poisson's ratio $\nu$, and the supplementary angle of a short diagonal corner $\alpha$~\citep{Filipov2017}. For the current panel geometry, $\sum \alpha=2\pi/3$. Models with varying crease stiffness are constructed, all with the parameters listed in \Cref{MiuraProperties}. 

\begin{figure}[htbp]
\centering
\includegraphics[width=0.9\linewidth]{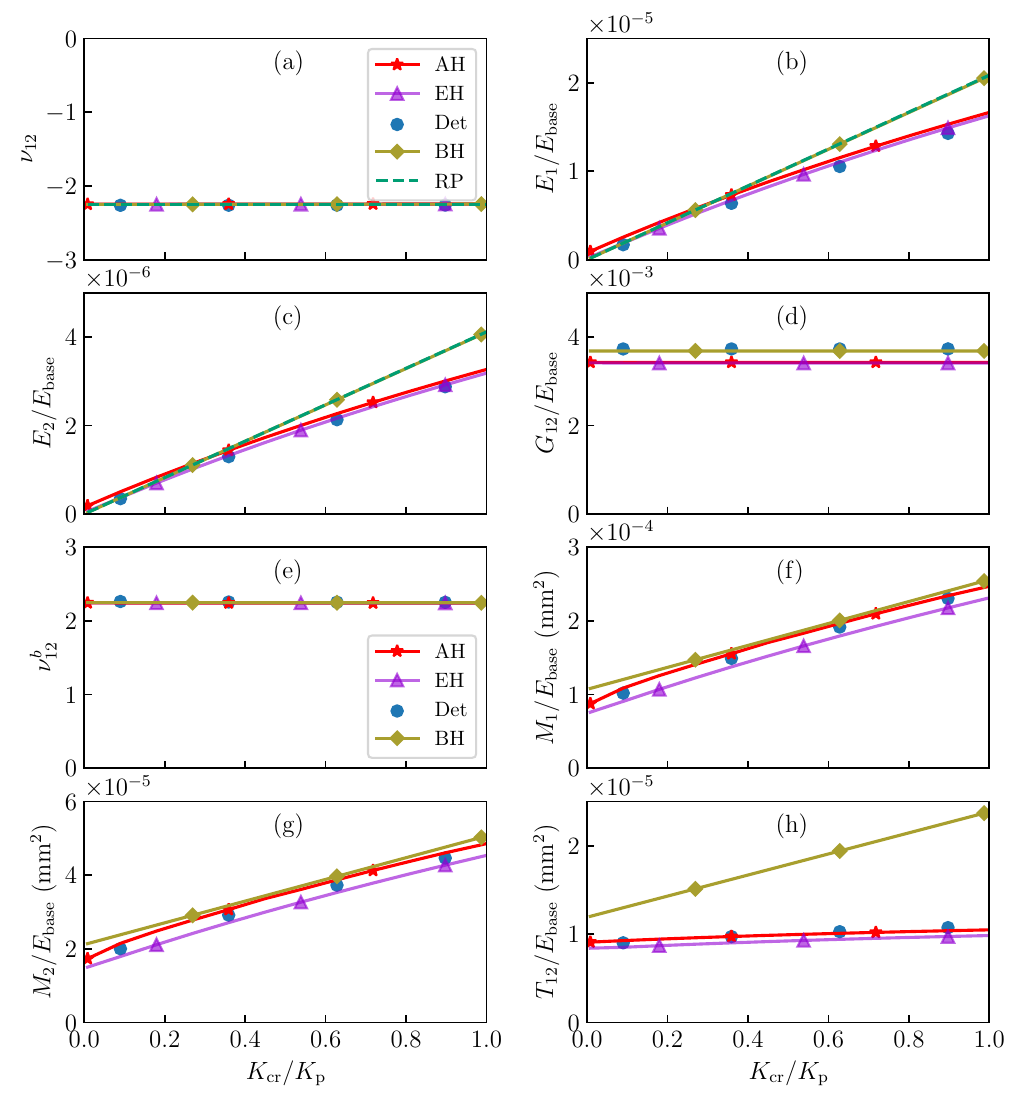}
    \begin{subfigure}{0.1\textwidth}
        \phantomsubcaption
        \label{fig:Kcra}
        \phantomsubcaption
        \label{fig:Kcrb}
        \phantomsubcaption
        \label{fig:Kcrc}
        \phantomsubcaption
        \label{fig:Kcrd}
        \phantomsubcaption
        \label{fig:Kcre}
        \phantomsubcaption
        \label{fig:Kcrf}
        \phantomsubcaption
        \label{fig:Kcrg}
        \phantomsubcaption
        \label{fig:Kcrh}
   \end{subfigure}
\caption{Variation of effective elastic constants with respect to crease stiffness, including (a)~in-plane Poisson's ratio $\nu_{12}$, effective Young's moduli (b)~$E_1$, (c)~$E_2$, (d)~shear modulus $G_{12}$, (e)~out-of-plane Poisson's ratio $\nu^b_{12}$, bending moduli (f)~$M_1$, (g)~$M_2$, and (h)~twisting modulus $T_{12}$. $E_1$, $E_2$, $G_{12}$, $M_1$, $M_2$, and $T_{12}$ are normalized against the base material Young's modulus $E_{\text{base}}$. Crease stiffness $K_{\text{cr}}$ is normalized against panel bending stiffness $K_p$. A subset of data points is shown for visual clarity. \legendlinestar{asymptotic} Asymptotic plate FE; \legendlinetriangle{energy} Energy-based plate FE; \legendmarker{detailedFE} Detailed plate FE; \legendlinediamond{BH} Homogenized/detailed bar-and-hinge; \StmCourbe{rigid,dashed}{} Rigid panel.}
\label{fig:Kcr}
\end{figure}

\Cref{fig:Kcr} presents the effective elastic constants for varying values of crease stiffness. Except for the in-plane and bending Poisson's ratios, all constants are normalized by $E_{\text{base}}$. The crease stiffness is normalized by the panel bending stiffness as $K_{\text{cr}}/K_{\text{p}}$. A small ratio between crease and panel bending stiffnesses indicates that the origami sheet favors folding along crease lines, while a ratio close to 1 suggests that both crease folding and panel deformation have significant influence on the overall response of the origami sheet. Results of all relevant models, including the plate FE models with asymptotic and energy-based homogenization, the detailed plate FE model, the bar-and-hinge model, and the rigid-panel model, are all overlayed on the same plots.

Considering each effective elastic constants, \Cref{fig:Kcra,fig:Kcre} shows that the crease stiffness has negligible influence on both in-plane and bending Poisson's ratios $\nu_{12}$, $\nu^b_{12}$. This suggests that when the crease stiffness is smaller than or equal to the panel bending stiffness, the folding kinematics of Miura origami is consistent. In \Cref{fig:Kcrb,fig:Kcrc}, effective Young's moduli $E_1$ and $E_2$ increase monotonically with the crease stiffness, demonstrating that the folding behavior of each crease has an influence on the global response of the metamaterial. \Cref{fig:Kcrd} shows that the crease stiffness does not affect the effective shear stiffness $G_{12}$ significantly, which confirms the observation that panel shearing is the main contributor to global shearing~\citep{vasudevan2024homogenization}. The bending moduli $M_1$ and $M_2$ increase with the crease stiffness as in \Cref{fig:Kcrf,fig:Kcrg}, meaning the stiffer the creases, the more difficult the origami sheet bends. The twisting modulus in \Cref{fig:Kcrh} has only a minor increase as the crease stiffness grows, if we consider the result of the detailed plate FE model. The creases do not influence the twisting modulus as much as the bending moduli, suggesting that panels contribute considerably to the global twisting behavior.

Although the global trends of the plate FE, the bar-and-hinge, and the rigid-panel models agree in general, there are some noticeable differences between their predictions. To begin with, the results of $\nu_{12}$, $\nu^b_{12}$, $G_{12}$, $M_1$, and $M_2$ show excellent agreement. \Cref{fig:Kcrd} shows that $G_{12}$ results of the bar-and-hinge model is closer to the baseline than the homogenized plate FE models. Discrepancy appears in $E_1$ and $E_2$ in \Cref{fig:Kcrb,fig:Kcrc}. The analytical results of the rigid-panel model overlap those of the bar-and-hinge model. They indicate that both $E_1$ and $E_2$ are linear functions of the crease stiffness. On the other hand, both the detailed and homogenized plate FE models predict that as the crease stiffness approaches the panel bending stiffness, results in the literature tend to overestimate $E_1$ and $E_2$, especially when the ratio $K_{\text{cr}}/K_{\text{p}}$ is above 0.4. Acknowledging that the rigid-panel model does not allow twisting, we compare the bar-and-hinge model with the plate FE models in \Cref{fig:Kcrh} and observe more deviation between them. The bar-and-hinge model indicates a linear relationship between the twisting modulus $T_{12}$ and the crease stiffness, with a slope larger than that of the plate FE models. This implies that to improve the accuracy of the effective elastic constants $E_1$, $E_2$, and $T_{12}$, we need to consider the relationship between crease and panel properties. Similarly, during mechanical simulations, it is also important to verify how significantly creases and panels influence the global behavior, while considering the applied loads and boundary conditions as another key factor.

\section{Discussion}
\label{discussion}
In this work, we present a homogenization framework for rigid and non-rigid origami metamaterials with panels modeled as plates. The framework applies to two independent methods, the asymptotic and the energy-based homogenization. The backbone of the framework unifying various homogenization methods is to apply appropriate boundary conditions to the unit cell, compute the resulting strain energy, and solve for the effective elastic constants using the constitutive relation. The majority of origami simulations in the literature include prescribed soft deformation modes according to the fold pattern. Regardless of the \emph{in-situ} loads and boundary conditions, the mechanical response of the origami metamaterial always follows its soft modes. Our framework lifts this constraint by allowing the origami panels to deform as thin plates. We demonstrate the result accuracy over varying initial fold angles and crease stiffness.

\begin{figure}[!ht]
\centering
\includegraphics[]{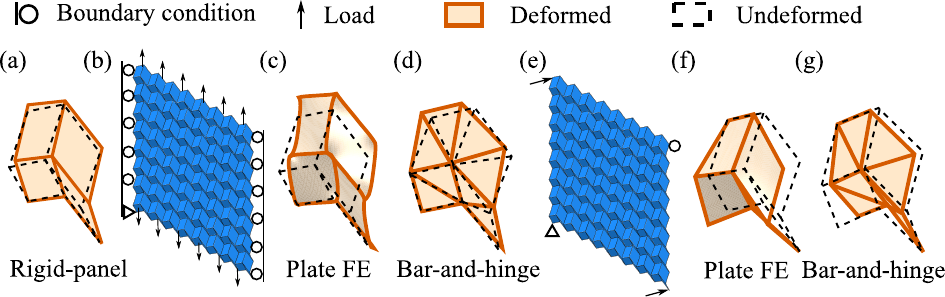}
\begin{subfigure}{0.1\textwidth}
        \phantomsubcaption
        \label{fig:UCdefa}
        \phantomsubcaption
        \label{fig:UCdefb}
        \phantomsubcaption
        \label{fig:UCdefc}
        \phantomsubcaption
        \label{fig:UCdefd}
        \phantomsubcaption
        \label{fig:UCdefe}
        \phantomsubcaption
        \label{fig:UCdeff}
        \phantomsubcaption
        \label{fig:UCdefg}
\end{subfigure}
\caption{Unit cell deformation modes in various simulation methods. (a)~Miura unit cell with rigid panels has one DoF. (b)~Stretched origami sheet under lateral constraints. Deformed unit cells at the center of the stretched sheet modeled with (c)~plate FE and (d)~bar-and-hinge plotted with identical scale factor. (e)~Twisted origami sheet. Deformed unit cells at the center of the twisted sheet modeled with (f)~plate FE and (g)~bar-and-hinge plotted with identical scale factor.}
\label{fig:UCdef}
\end{figure}

To highlight the capabilities of our framework, we compare results with other existing origami models. \Cref{fig:compareMethods,fig:Kcr} show that a Miura sheet modeled with plate FE is more flexible than those with bar-and-hinge and rigid panels. The discrepancy between models is attributed to their applicable ranges of deformation modes. In general, the plate FE models (both detailed and homogenized) tend to be more compliant than the other models. This is true for both in-plane and out-of-plane responses. This result is attributed to a richer capture of deformation modes by the plate FE model, which allows more deformation modes in origami panels and creases than other models. As an example, \Cref{fig:UCdefa} shows that the rigid-panel model can only fold along its degree of freedom. The bar-and-hinge model, in addition to folding, can also stretch the panels or bend the panels along their diagonals. The plate FE model simulates all of the above and other non-rigid deformation modes such as bending and flexural behavior of creases and panels, along with localized twists at vertices. In contrast to the scaled deformation fields corresponding to unit strain or unit curvature in \Cref{fig:AHdiff,EHdefModes}, we present a realistic response of the detailed model, zoomed in on a unit cell in \Cref{fig:UCdefc,fig:UCdeff}. All figures show response in the linear elastic regime.

\subsection{Stretching under constraints}
To visualize the deformation modes of a Miura sheet, we apply a set of loads and boundary conditions in \Cref{fig:UCdefb} to the detailed plate FE and detailed bar-and-hinge models. There are forces applied at the top and bottom vertices, stretching the specimen in-plane. The left and right edges have roller boundary conditions. The bottom left corner is pinned to prevent rigid body motion. The applied boundary conditions prevent the origami sheet from unfolding along its degree of freedom, creating additional strain in the material. Zooming in on one of the unit cells at the center of the specimen, the plate FE model in \Cref{fig:UCdefc} shows a combination of bending and twisting in the panels that resembles warpage. Correspondingly, the initially straight creases deform into C-shapes that follow the curvature of the panels. At the vertex where corners of the panels meet, there is localized bending at the corners, which do not necessarily bend towards the same direction. Compared with the deformed bar-and-hinge unit cell in \Cref{fig:UCdefd} under the same loads and boundary conditions, the plate FE model exhibits a more significant change in shape, thus higher flexibility overall.

\subsection{Twisting}
\Cref{fig:UCdefe} examines the twisting deformation mode. The bottom left corner is pinned, and the top right corner is fixed in the out-of-plane direction. Loads are applied on the other two corners. The unit cell at the center deforms as in \Cref{fig:UCdeff,fig:UCdefg}. Twisting in each panel is captured with plate FE. In the bar-and-hinge model, the diagonal bar in each panel prevents twisting in certain directions. For example, the top-left panel in \Cref{fig:UCdefg} struggles to twist along the applied load due to the placement of its diagonal bar. Switching to another variation of the bar-and-hinge model with additional diagonal bars~\citep{Filipov2017} can potentially give more accurate results.

In summary, we compare the plate FE and the bar-and-hinge models under two sets of loads and boundary conditions to explain the discrepancies in the effective properties. Our result show consistent accuracy across a spectrum of initial fold angles and crease stiffnesses. We attribute the advantage of the proposed framework to a wide range of non-rigid deformation modes of the unit cell.

\section{Conclusion}
\label{conclusion}
This work presents a homogenization framework to evaluate the linear elastic properties of an origami metamaterial through an effective computational model. The framework can model both non-rigid and rigid-foldable origami. It presents and compares two homogenization methods, asymptotic and energy-based homogenization. For demonstration, we implement both on a non-rigid Miura origami sheet and find its equivalent continuum model as a Kirchhoff-Love plate. A step-by-step procedure is provided to implement both homogenization methods. The results are benchmarked with a fully resolved numerical model of a finite Miura origami tessellation. The role of the initial fold angle is investigated through a parametric study. Both the base material properties and the folding kinematics have an influence on the effective properties. The variation of effective properties is also compared with those obtained from the rigid-panel and bar-and-hinge methods. Averaging the errors across a wide range of fold angles, the effective shear modulus has errors within \qty{8}{\%} for all models. The effective twisting modulus shows considerable difference, with the error of the bar-and-hinge model 12 times that of the homogenization framework presented here. An origami sheet modeled with plates is more flexible than the literature, a result attributed to the variety of non-rigid deformation modes that plate elements can capture. Finally, a parametric study on crease stiffness demonstrates that both the literature and the proposed framework are sufficiently accurate for the effective in-plane and out-of-plane Poisson's ratios, the shear modulus, and the bending moduli, whereas the proposed framework is recommended for the effective Young's moduli and the twisting modulus.

Although our paper examines the basic Miura origami folded from a sheet of isotropic material, our framework applies to a wide range of origami patterns and base materials. Additionally, it can be easily implemented with the aid of any FE software package. The framework can be used for various applications, including origami-based soft robots, foldable antennas, and aircraft structures with periodicity. It is possible to extend the framework to cellular origami metamaterials such as the stacked Miura origami. In the context of Continuum Mechanics, it is straightforward to find the 3D analogy to our in-plane effective stiffness coefficients $A^H_{\alpha\beta\gamma\delta}$ following \Cref{AHsubsection} or \Cref{EHsubsection}. Future work may involve extending the framework to nonlinear homogenization to incorporate the analysis of the curved-crease origami, such as the Miura lens-box pattern ~\citep{Mirzajanzadeh2025}.

\section{Methods}
This manuscript is accompanied with a set of ABAQUS and Python scripts recovering the central calculations in this work~\citep{github_wendy}.

\section{Declaration of competing interest}
The authors declare that they have no known competing financial interests or personal relationships that could have appeared to influence the work reported in this paper.

\section{Acknowledgments}
XL acknowledges support from Fonds de Recherche du Québec - Nature et technologies (Grant No. 348668). DP acknowledges the Canada Research Chair Program (Grant No. 258679), and the Natural Sciences and Engineering Research Council of Canada (Grant No. 208241). ML acknowledges the Natural Sciences and Engineering Research Council of Canada (Grant No. 401559).

XL gratefully acknowledges Morad Mirzajanzadeh and Filippo Agnelli for helpful discussions on the methodology and results.
\appendix

\section{Mesh convergence test}
To verify the proposed homogenization framework, we compute the effective elastic constants of a single-layered Miura origami sheet. Mesh convergence is checked for both asymptotic and energy-based homogenization. The results are first compared with a fully detailed FE model consisting of 49 unit cells. Next, we consider the limiting case of very stiff origami panels and compare it with an analytical model from the literature \citep{wei2013geometric}.

\begin{figure}[!ht]
\centering
\includegraphics[width=0.9\linewidth]{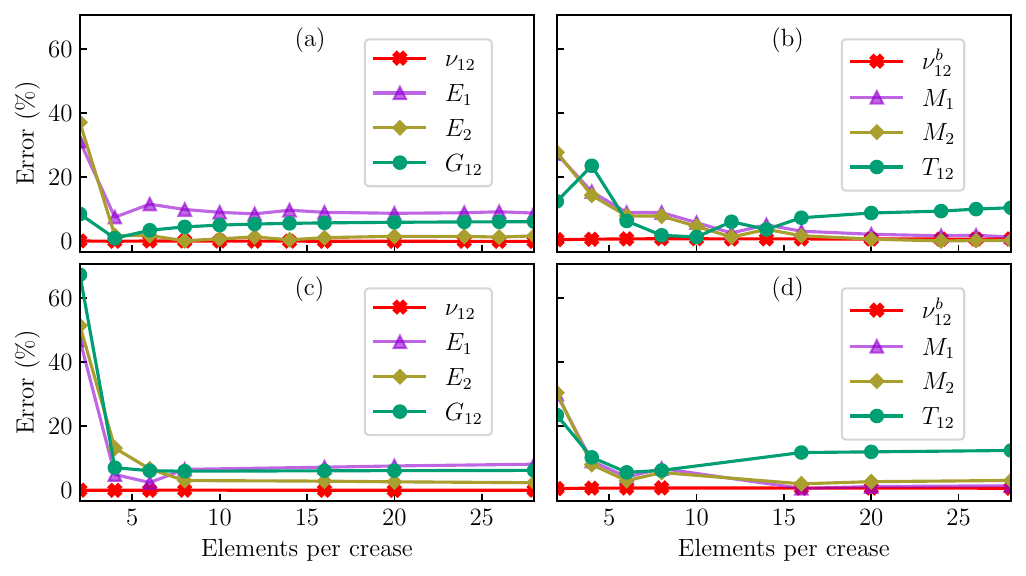}
\caption{Mesh convergence test and validation with detailed FE model. (a)~In-plane and (b)~out-of-plane effective properties from energy-based homogenization. (c)~In-plane and (d)~out-of-plane effective properties from asymptotic homogenization.}
\label{convergenceAndTheory}
\end{figure}

The geometric and material parameters of the origami metamaterial is given in \Cref{MiuraProperties}. Mesh convergence of asymptotic and energy-based homogenization methods are compared in \Cref{convergenceAndTheory} at a crease stiffness of \qty{0.05}{N}. We increase the number of elements along each side of the rhombic panel and plot the variation of $x_1$-direction Young's modulus $E_1$. The mesh convergence plot shows that the energy-based homogenization method converges at 16 elements on each side of the panel, while asymptotic homogenization requires 28 elements. In subsequent analysis, we ensure that both homogenization methods are converged with a mesh density of 28 elements per side.

\section{Error analysis}
\label{errors}
\begin{figure}[!ht]\centering
\includegraphics[scale=0.9]{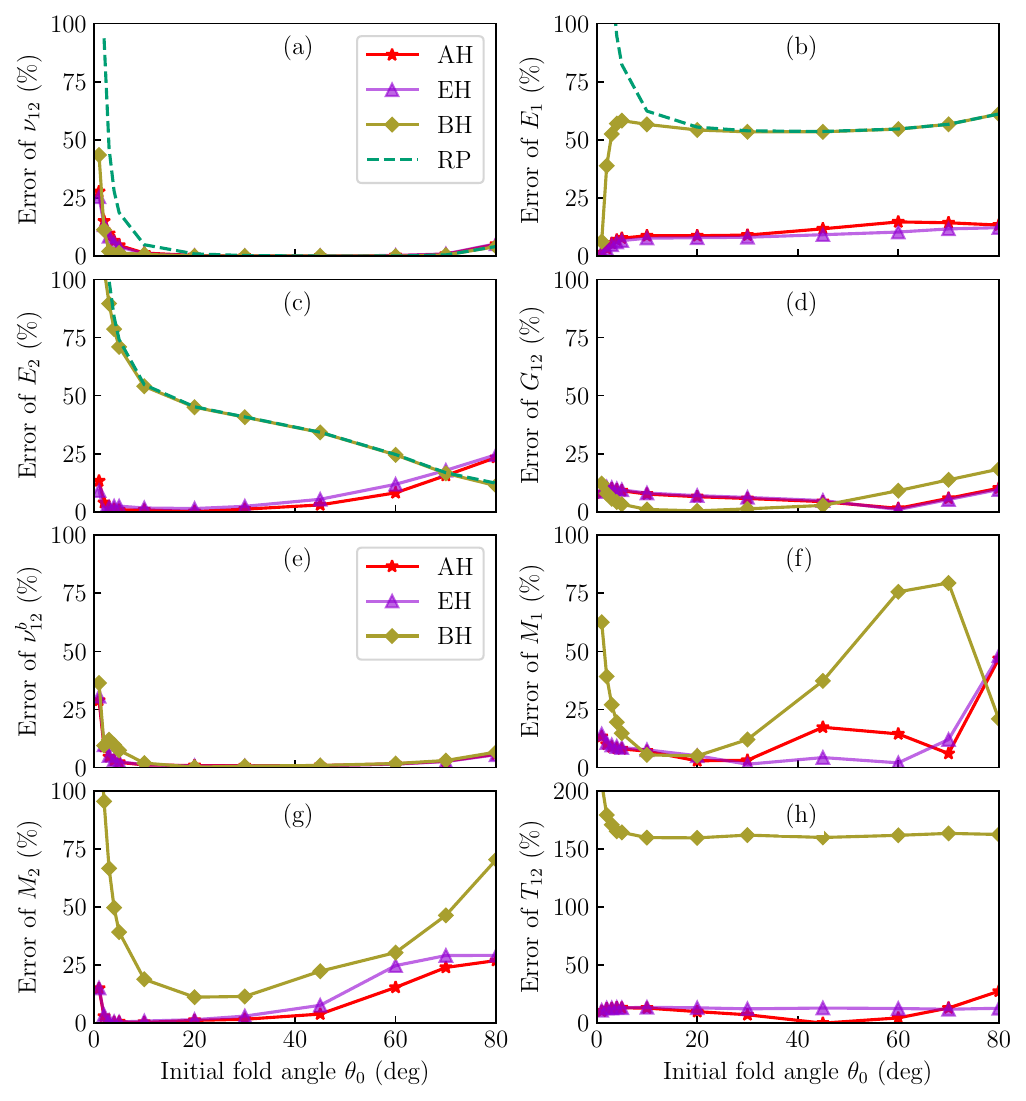}
\caption{Errors of homogenization and theoretical models compared to the detailed plate FE model under various initial fold angles $\theta_0$, including (a)~in-plane Poisson's ratio $\nu_{12}$, effective Young's moduli (b)~$E_1$, (c)~$E_2$, (d)~shear modulus $G_{12}$, (e)~out-of-plane Poisson's ratio $\nu^b_{12}$, bending moduli (f)~$M_1$, (g)~$M_2$, and (h)~twisting modulus $T_{12}$.} \legendlinestar{asymptotic} Asymptotic plate FE; \legendlinetriangle{energy} Energy-based plate FE; \legendlinediamond{BH} Homogenized/detailed bar-and-hinge; \StmCourbe{rigid,dashed}{} Rigid panel.
\label{fig:errorTheta}
\end{figure}

\begin{figure}[!ht]
    \centering
    \includegraphics[width=0.9\linewidth]{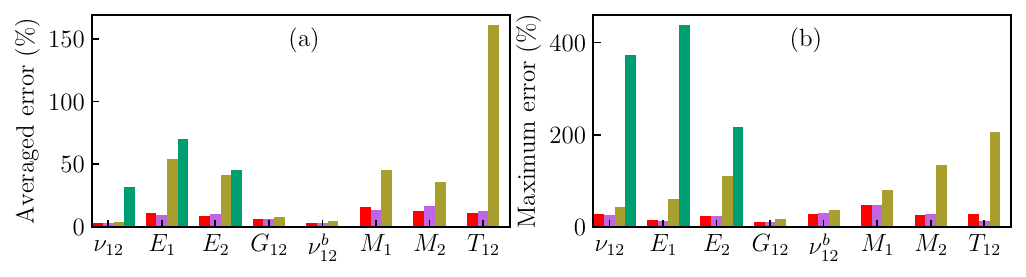}
    \caption{Comparison of errors between the proposed homogenization framework and the literature. (a)~Averaged error and (b)~maximum error of in-plane and out-of-plane elastic constants. Both plots describe errors within a range of fold angles between $0^\circ$ and $80^\circ$, while other parameters are identical to those in \Cref{MiuraProperties}. \StmRectangle{asymptotic} Asymptotic plate FE; \StmRectangle{energy} Energy-based plate FE; \StmRectangle{BH} Homogenized/detailed bar-and-hinge; \StmRectangle{rigid} Rigid panel.} 
    \label{fig:errorBar}
\end{figure}

We consider the detailed plate FE model as the baseline. We compare its effective elastic constants with those of two homogenized plate FE models, a bar-and-hinge model~\citep{vasudevan2024homogenization}, and a rigid-panel model~\citep{wei2013geometric}. In addition to the varying parameter of fold angle, which is in the range of $(0^\circ,80^\circ)$, all models share identical geometry and crease stiffness, as summarized in the first six columns of \Cref{MiuraProperties}. The rigid-panel model differs from the rest in terms of panel flexibility. Due to the nature of the rigid panels, the in-plane shear modulus and all out-of-plane properties are unbounded. Only the in-plane Poisson's ratio and Young's moduli are compared with those of other models. The relative error between each model and the baseline is quantified as 
\begin{equation}
    \text{Error}(\theta_0)=\frac{|P\text{(model)}-P\text{(detailed plate FE)}|}{|P\text{(detailed plate FE)|}}\times \qty{100}{\%},
\end{equation}
where $P$ represents any of the elastic constants $\nu_{12}$, $E_1$, $E_2$, $\nu^b_{12}$, $M_1$, $M_2$, and $T_{12}$. Their errors under varying initial fold angles $\theta_0$ are presented in \Cref{fig:errorTheta}. 
The error over the fold angles is averaged as
\begin{equation}
    \text{Averaged error}=\sqrt{\frac{1}{\theta_0^{\max}-\theta_0^{\min}}\int^{\theta_0^{\max}}_{\theta_0^{\min}}\text{Error}^2(\theta_0)\dd\theta_0},
\end{equation}
where $\theta_0^{\min}=0^\circ$ and $\theta_0^{\max}=80^\circ$. The maximum error in the range of fold angles between $0^\circ$ and $80^\circ$ is defined as
\begin{equation}
    \textup{Maximum error}=\max_{\theta_0 \in (0^\circ,80^\circ)} \textup{Error}(\theta_0).
\end{equation}
Errors of all elastic constants are compared in \Cref{fig:errorBar}.
On average, Poisson's ratios $\nu_{12}$, $\nu^b_{12}$ and shear modulus $G_{12}$ are consistently accurate across all models. For the rest of the properties, the asymptotic and energy-based plate FE methods have relatively consistent ranges of error. They show an advantage in terms of accuracy when calculating the twisting modulus $T_{12}$ in particular. The maximum errors of asymptotic homogenization and energy-based homogenization are both within \qty{53}{\%}.

\section{Effective compliance matrices of Miura origami}
\label{complMtxResult}
To provide an intuitive description of the mechanical response of a Miura sheet, we report the effective compliance matrix, which is the inverse of the stiffness coefficient matrix in \Cref{plateConst}. Small numerical values in the matrix have been simplified to 0 because they are 2 to 4 orders of magnitude smaller than other entries, reflecting the planes of symmetry and hence anisotropy of the unit cells. The compliance matrix from asymptotic homogenization reads

\begin{equation}
 \begin{bmatrix}
e_{11} \\ 
e_{22} \\ 
2e_{12} \\ 
\kappa_{11} \\ 
\kappa_{22} \\ 
2\kappa_{12}
\end{bmatrix}
=
\begin{bmatrix}
10.6 & 23.8 & 0 & -0.0589 & 0.133 & -0.0369 \\
23.8 & 54.2 & 0 & -0.133 & 0.300 & -0.0847 \\
0 & 0 & 0.0729 & 0 & 0 & 0 \\
-0.0589 & -0.133 & 0 & 0.787 & -1.76 & -0.0377 \\
0.133 & 0.300 & 0 & -1.76 & 4.02 & 0.0865 \\
-0.0369 & -0.0847 & 0 & -0.0377 & 0.0865 & 22.8
\end{bmatrix}
 \begin{bmatrix}
\sigma_{11} \\ 
\sigma_{22} \\ 
\sigma_{12} \\ 
\mu_{11} \\ 
\mu_{22} \\ 
\mu_{12}
\end{bmatrix},
\end{equation}
while for energy-based homogenization,
\begin{equation}
 \begin{bmatrix}
e_{11} \\ 
e_{22} \\ 
2e_{12} \\ 
\kappa_{11} \\ 
\kappa_{22} \\ 
2\kappa_{12}
\end{bmatrix}
=
\begin{bmatrix}
10.8 & 24.2 & 0 & -0.00493 & 0.0102 & -0.179 \\
24.2 & 55.1 & 0 & -0.0103 & 0.0211 & -0.405 \\
0 & 0 & 0.0732 & 0 & 0 & 0 \\
-0.00493 & -0.0103 & 0 & 0.817 & -1.83 & -0.0726 \\
0.0102 & 0.0211 & 0 & -1.83 & 4.17 & 0.165 \\
-0.179 & -0.405 & 0 & -0.0726 & 0.165 & 23.98
\end{bmatrix}
 \begin{bmatrix}
\sigma_{11} \\ 
\sigma_{22} \\ 
\sigma_{12} \\ 
\mu_{11} \\ 
\mu_{22} \\ 
\mu_{12}
\end{bmatrix}.
\end{equation}
The models of both matrices use geometric and material parameters in \Cref{MiuraProperties}. The units in the matrix are \unit[]{MPa^{-1}} for in-plane terms, \unit[]{mm/N} for in-plane and out-of-plane coupling terms, and \unit[]{N^{-1}} for out-of-plane terms.

\section{Flowcharts of homogenization framework}
\subsection{Asymptotic homogenization workflow}
\begin{description}
\item[Inputs] Unit cell geometry and material properties.
\item[Step 1] Prescribe deformation fields $\boldsymbol \chi^0$ corresponding to unit strain or curvature according to \Cref{unitStrainDisp,unitStrainDisp2} on every node of the unit cell. Compute the initial strain loading~$\mathbf{f}^0$.
\item[Step 2] Apply periodic boundary conditions on the unit cell boundaries according to \Cref{AHPBC}. Numerically solve the microscopic problem to obtain the characteristic displacement fields~$\boldsymbol \chi^*$.
\item[Step 3] Find the difference between the unit strain displacement field $\boldsymbol \chi^0$ and the characteristic displacement field~$\boldsymbol \chi^*$. Apply the difference as nodal displacements on the unit cell and compute its strain energy $U$.
\item[Step 4] Compute the effective stiffness coefficients $A_{\alpha \beta \gamma \delta}^H$, $B_{\alpha \beta \gamma \delta}^H$, $D_{\alpha \beta \gamma \delta}^H$ from the unit cell strain energy $U$.
\item[Homogenized problem] The full origami metamaterial is simplified into a thin plate with anisotropic material properties $A_{\alpha \beta \gamma \delta}^H$, $B_{\alpha \beta \gamma \delta}^H$, $D_{\alpha \beta \gamma \delta}^H$.
\end{description}

\subsection{Energy-based homogenization workflow}
\begin{description}
\item[Inputs] Unit cell geometry and material properties.
\item[Step 1] Apply periodic boundary condition to the unit cell boundaries according to \Cref{EHipe11,EHipe22,EHipe12,EHoopk11,EHoopk22,EHoopk12}.
\item[Step 2] Numerically find the unit cell strain energy $U$ of each deformation mode in Step~1.
\item[Step 3] Compute the effective stiffness coefficients $A_{\alpha \beta \gamma \delta}^H$, $B_{\alpha \beta \gamma \delta}^H$, $D_{\alpha \beta \gamma \delta}^H$ from the unit cell strain energy $U$.
\item[Homogenized problem] The full origami metamaterial is simplified into a thin plate with anisotropic material properties $A_{\alpha \beta \gamma \delta}^H$, $B_{\alpha \beta \gamma \delta}^H$, $D_{\alpha \beta \gamma \delta}^H$.
\end{description}

\begin{figure}[htbp]
\centering
\includegraphics[width=0.9\linewidth]{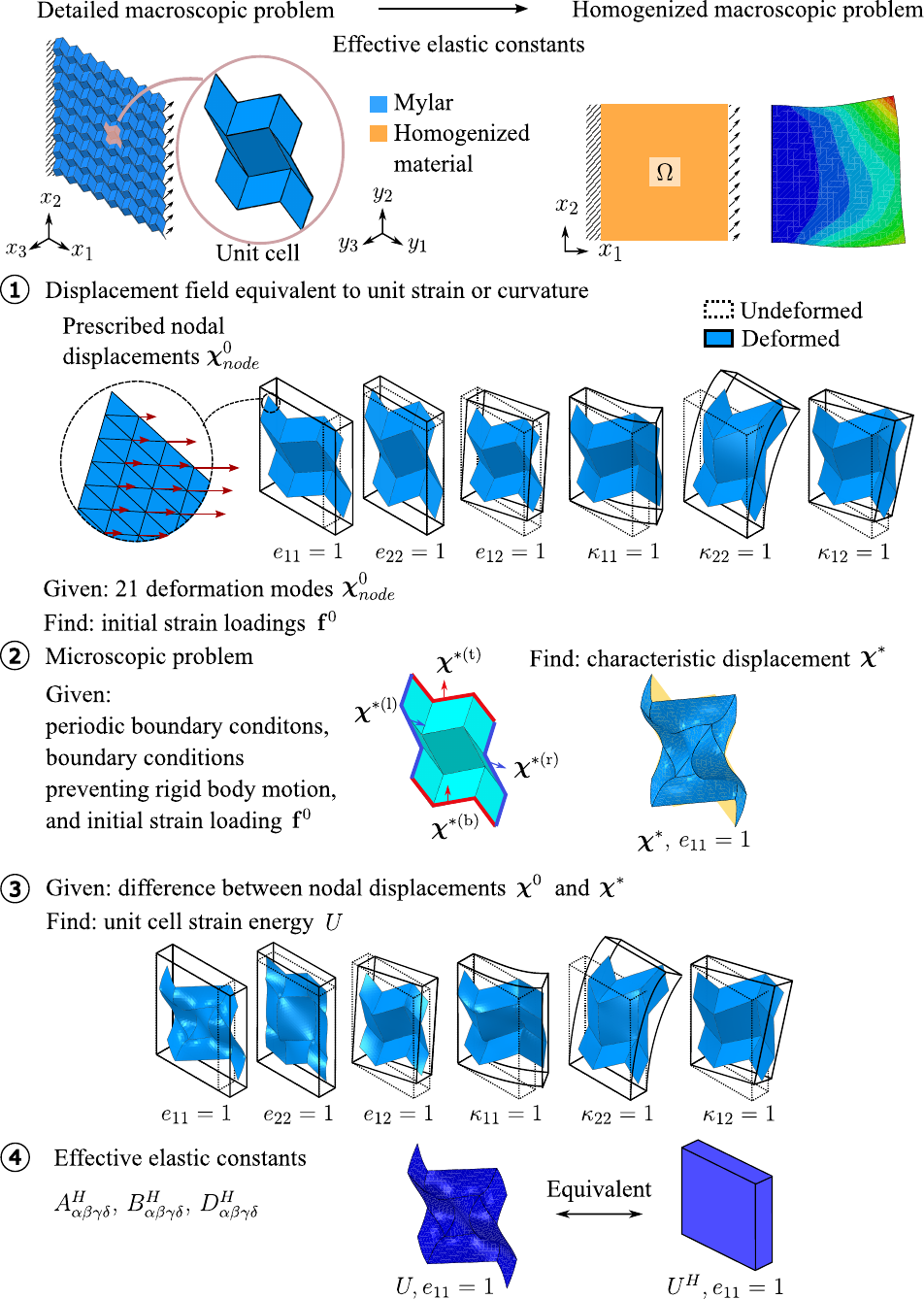}
\caption{Asymptotic homogenization flowchart.}
\label{AHFlowchart}
\end{figure}

\begin{figure}[htbp]
\centering
\includegraphics[width=0.9\linewidth]{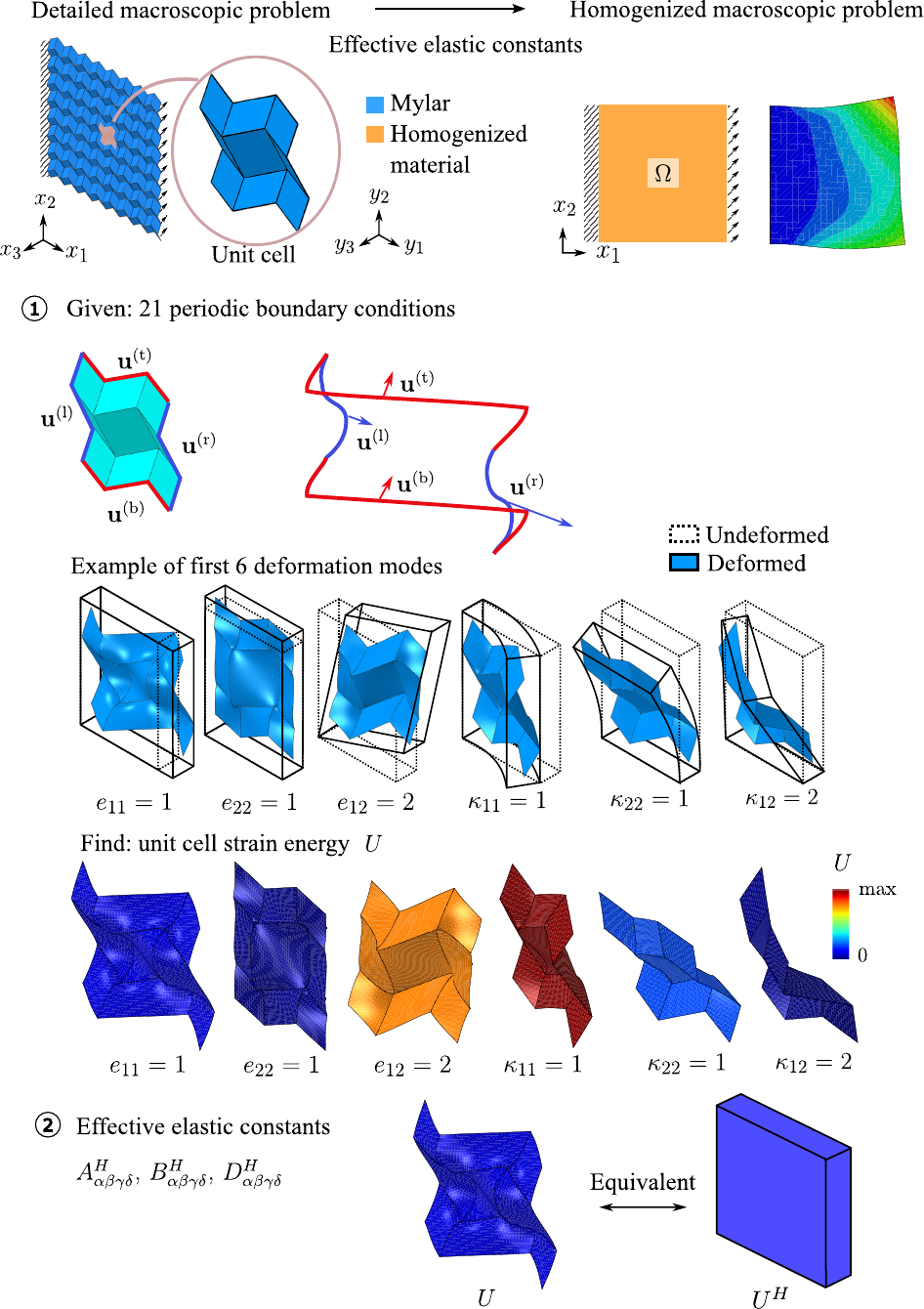}
\caption{Energy-based homogenization flowchart.}
\label{EHFlowchart}
\end{figure}

\Cref{AHFlowchart,EHFlowchart} present a summary and comparison of the numerical implementation of asymptotic homogenization and energy-based homogenization, including the required loads and boundary conditions, examples of deformation modes and examples of unit cell strain energy.

\section{Simulation methods for detailed origami metamaterials and unit cells}
\label{simMethods}

This section characterizes the linear elastic properties of an origami metamaterial, taking the Miura pattern as an example. The pattern tessellates in two directions, creating a planar metamaterial in the form of a single-layered sheet. \Cref{fig:frameworka} shows the geometric parameters that defines the origami fold pattern and the initial fold state. $a$ and $b$ denote the panel side lengths, $\gamma$ is the sector angle, $\theta$ is the fold angle, and $t$ is the panel thickness. The material properties in a unit cell is characterized separately for the creases and the panels. The creases are simplified into torsional springs with a torsional stiffness of $K_{\textup{cr}}$ per crease. The panels are made of a linear elastic material with Young's modulus $E$ and Poisson's ratio $\nu$. We select their values as those of Mylar, a material widely used to fabricate origami specimens in the literature~\citep{Filipov2017}. The geometric and material properties are summarized in \Cref{MiuraProperties}.

Some of the most widely-adopted origami simulation methods include the rigid-panel, the bar-and-hinge, and the plate FE models. \Cref{rigidPanel,barandhinge,plateFE} present each of their formulations.

\subsection{Rigid-panel model} \label{rigidPanel}
 Assuming that the kinematic deformation modes are dominant in the structural behaviors of the metamaterial, the simplest rigid-panel model requires that the origami panels acts as rigid bodies, and creases as frictionless hinges. When the origami metamaterial folds, its panels rotate around each other following the creases. The Miura origami resembles a 1 degree-of-freedom (DoF) mechanism. Since the panels are perfectly rigid, it is only possible to stow and deploy in-plane. Taking the fold angle $\theta$ as the only DoF, we describe the in-plane Poisson's ratio as~\citep{schenk2013geometry}:
 \begin{equation}
 \label{rigidPanelPoisson}
     \nu_{xy}=-\frac{e_y}{e_x}=-(\cos\theta\tan\gamma)^2
 \end{equation}
Elastic responses appear when we consider the fold stiffness. By assuming that all elastic energy is stored in the creases, we simplify the folding process to rigid panels rotating around elastic hinges. The effective elastic constants in the two in-plane directions are~\citep{wei2013geometric}:
\begin{subequations} \begin{align}
    &E_x=4K_{\textup{cr}}\frac{(1-\sin^2\gamma\sin^4(\rho/2))^2+\cos^2\gamma}{(1-\sin^2\gamma\sin^4(\rho/2))^{\frac{1}{2}}\cos\gamma\sin^2\gamma\sin\rho},\\
    &E_y=4K_{\textup{cr}}\frac{(1-\sin^2\gamma\sin^4(\rho/2))^2+\cos^2\gamma}{(1-\sin^2\gamma\sin^4(\rho/2))^{2}\sin\gamma\cos(\rho/2)},
\end{align} \end{subequations}
where $\rho$, chosen as the DoF in \citep{wei2013geometric}, is the dihedral angle between adjacent panels.
Due to panel rigidity, the Miura origami metamaterial does not deform in shear. Consequently the in-plane shear modulus $G_{xy}$ is unbounded.

\subsection{Bar-and-hinge model}\label{barandhinge}
The bar-and-hinge model approximates the origami folding process with a system of pin-jointed trusses and torsional springs~\citep{Schenk2011}. By incorporating crease folding, panel bending, stretching, and shearing, the model efficiently describes essential deformation modes of an origami metamaterial~\citep{Filipov2017}. As shown in \Cref{fig:compareMethods}, the model depicts folding and bending with rotation along the crease and diagonal bars respectively, while stretching and shearing are equivalent to a combination of bar extension and compression. Considering axial stiffness of the bars and torsional stiffness for folding and bending, we can numerically compute the elastic properties of the metamaterial. Following the procedure introduced in \citep{vasudevan2024homogenization}, the first step is to identify the force-displacement relation 
\begin{equation}
\label{equilibrium}
\mathbf{K}\mathbf{u}=\mathbf{f},
\end{equation}
where $\mathbf{u}$ is the nodal displacement vector consisting of three translational DoF of each node, $\mathbf{f}$ collects the external applied forces, and the stiffness matrix $\mathbf{K}$ is constructed from the geometry and mechanical properties of bars and torsional elastic hinges~\citep{Filipov2017}:
\begin{equation}
    \mathbf{K}=\begin{bmatrix}
        \mathbf{C}\\\mathbf{J}_B\\\mathbf{J}_F
    \end{bmatrix}^T\begin{bmatrix}
        \mathbf{D}_S & \mathbf{0} & \mathbf{0}\\
        \mathbf{0} & \mathbf{D}_B & \mathbf{0}\\
        \mathbf{0} & \mathbf{0} & \mathbf{D}_F
    \end{bmatrix}\begin{bmatrix}
        \mathbf{C}\\\mathbf{J}_B\\\mathbf{J}_F
    \end{bmatrix}.
\end{equation}
The entries of $\mathbf{K}$ include the compatibility matrix $\mathbf{C}$ of the bar framework, the Jacobian matrices $\mathbf{J}_B$, $\mathbf{J}_F$ identifying the bending and folding hinges, and stiffness parameters of the bars and hinges $\mathbf{D}_S$, $\mathbf{D}_B$, $\mathbf{D}_F$, all of which detailed in \citep{Schenk2011,Filipov2017}. The second step is to solve \Cref{equilibrium} with appropriate loads and boundary conditions applied on the fully detailed origami metamaterial under investigation~\citep{vasudevan2024homogenization}. The bar-and-hinge model is a tessellation of Miura pattern with 20 unit cells on each edge. The loads and boundary conditions correspond to uniaxial extensions in the two in-plane directions $x$ and $y$, and an in-plane shear test. Finally, obtain the effective elastic constants from the following definitions~\citep{vasudevan2024homogenization}:
\begin{equation}
\label{vxydet}
    \nu_{xy}=-\frac{e_y}{e_x},\quad
    E_x=\frac{f_x}{A_xe_x},
\quad
    E_y=\frac{f_y}{A_ye_y},
\quad
    G_{xy}=\frac{f_{xy}}{2A_xe_{xy}},
\end{equation}
where $E_x,E_y$ are the in-plane Young's modulus, $G_{xy}$ is the in-plane shear modulus, $f_x,f_y,f_{xy}$ are applied total loads on the boundary, $A_x,A_y$ are areas of cross sections perpendicular to $x,y$ axes, and $e_x,e_y,e_{xy}$ are strains in their respective axial and shear directions.

\subsection{Plate FE model}
\label{plateFE}
The plate FE method is a widely-adopted modeling strategy for thin-shelled structures including origami metamaterials. It is a preferred method in engineering design when the detailed distribution of stress and strain is required. For example, in aerospace or medical applications, engineers need such information from an FE model to minimize unpredicted failure during operation. Another advantage is that the plate FE does not involve as many assumptions on origami deformation modes as other methods. Take crease folding as an example, the rigid-panel and the bar-and-hinge models require that creases remain straight during folding, while the plate FE model allows them to curve arbitrarily. However, modeling in such details inevitably comes with significant computational costs. In many cases it becomes impractical to meet the requirements on time and resources.

Here we adopt the plate FE model introduced in \citep{Filipov2015}, with a unit cell shown in \Cref{fig:compareMethods}. Panels are represented with plate FE, creases are represented with torsional springs, and the origami metamaterial is an assembly of both types of elements. The nodal DoF of a plate element at node $i$ is
$\mathbf{u}_i=[u \; v \; w \; \theta_x \; \theta_y]^T$, including translational displacements $u,v$, deflection $w$, and rotations $\theta_x ,\theta_y$. The element stiffness matrix is 
\begin{equation}
    \mathbf{K}^e=\iiint_V\mathbf{B}^T\mathbf{D}\mathbf{B}\dd v,
\end{equation}
where $V$ is the element volume, $\mathbf{B}$ is the matrix specifying strain-displacement relation, and $\mathbf{D}$ is the matrix of elastic coefficients for an element under plane stress.
A crease is represented by a row of torsional spring elements, each connecting a pair of overlapping nodes from the two adjacent panels. The torsional stiffness of the springs sums up to the crease stiffness. The nodal DoF of a torsional spring element is $\mathbf{u}_i=[u \; v \; w \; \theta_x\; \theta_y\; \theta_z]^T$,
including translational displacements in three directions $u,v,w$, and rotations $\theta_x , \theta_y, \theta_z$. The element stiffness matrix is
\begin{equation}
    \mathbf{K}^e=\begin{bmatrix}
    \mathbf{K}_n& -\mathbf{K}_n\\
    -\mathbf{K}_n&\mathbf{K}_n
    \end{bmatrix}\quad\text{with}
\quad
\mathbf{K}_n=\begin{bmatrix}
    k_{\text{lg}} & 0 & 0 & 0 & 0 & 0 \\
    0 & k_{\text{lg}} & 0 & 0 & 0 & 0 \\
    0 & 0 & k_{\text{lg}} & 0 & 0 & 0 \\
    0 & 0 & 0 & K_{\textup{cr}}/n & 0 & 0\\
    0 & 0 & 0 & 0 & k_{\text{lg}} & 0\\
    0 & 0 & 0 & 0 & 0 & k_{\text{lg}}
        \end{bmatrix},
\end{equation}
where $k_{\text{lg}}$ are large values of translational and rotational stiffness compared to adjacent elements, and $K_{\textup{cr}}/n$ is the crease stiffness averaged across $n$ spring elements along the crease. In addition, to prevent detachment and tearing at the creases, we add kinematic constraints on the relative nodal displacements of each torsional spring. Each set of nodes $a,b$ of a spring element is subject to translational constraints $u_{a}-u_{b}=0$, $v_{a}-v_{b}=0$, $w_{a}-w_{b}=0$,
and rotational constraints $\theta_{ya}-\theta_{yb}=0$, $\theta_{za}-\theta_{zb}=0$.
The only unconstrained DoF $\theta_x$ lies on its local $x$ axis that aligns with the crease, meaning the torsional spring can only fold along the crease instead of detaching in any other directions. The constraints can be applied using Lagrange multipliers, for example. Finally we transform all element stiffness matrices and constraints to the global coordinates and assemble them to obtain the global stiffness matrix $\mathbf{K}$. Similar to the procedure in \Cref{barandhinge}, we solve \Cref{equilibrium} under three sets of loads and boundary conditions. We implement the above procedure in the commercial FE software ABAQUS, where panels are meshed with 3-node triangular plate/shell elements (S3) and connected at the creases using hinge connector elements (CONN3D2). A mesh size of $a/16$ was selected after conducting a mesh convergence test. Since we are interested in the linear elastic response, we run the simulation with a general static solver without geometric nonlinearity (Nlgeom is off). The result of displacement and reaction forces give the effective elastic constants as detailed in \Cref{validation}.

\section{Experimental setup}
\label{experiment}
\begin{figure}[!ht]
    \centering
    \includegraphics[scale=0.9]{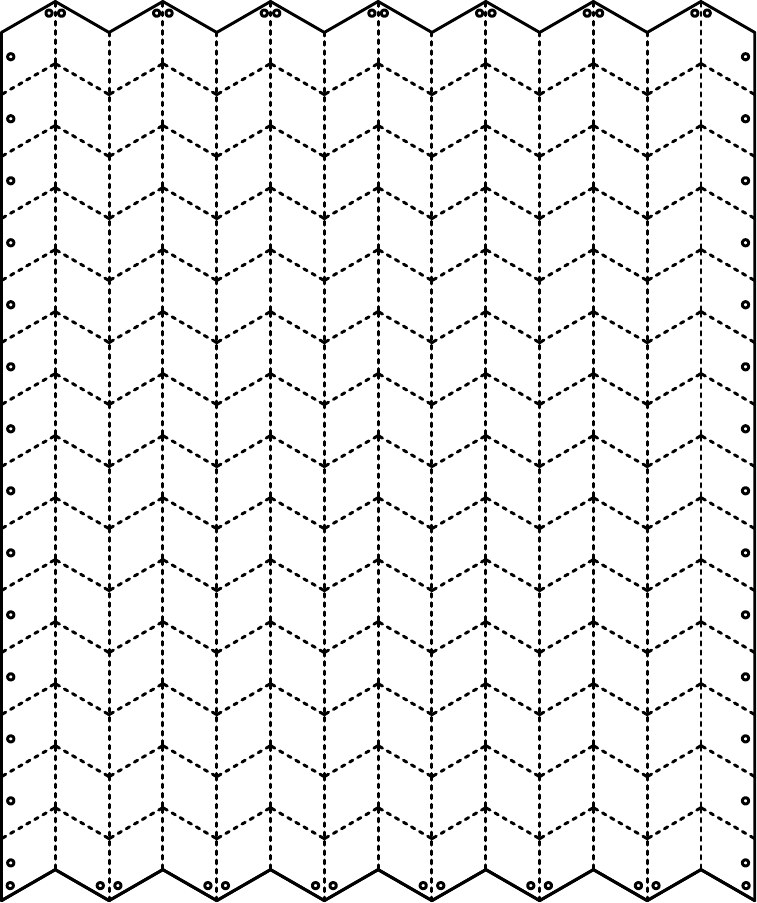}
    \caption{Cut pattern of Miura specimen.}
    \label{fig:cutPattern}
\end{figure}

The specimen used for the compression test in \Cref{fig:E11exp} is a 7 by 7 Miura
tessellation. It is fabricated from cardboard via laser cutting (CM1290 laser cutter, SignCut Inc.) and folded by hand. \Cref{fig:cutPattern} shows the cut pattern. \Cref{MiuraCardboard} summarizes its geometric and material properties. The geometric properties are directly measured from the specimen, while the material properties are cited from the literature~\citep{Mirzajanzadeh2025}.
\begin{table}[ht]
    \centering\small
    \begin{tabular}{lccc} \hline 
         Panel side length& $a$ &mm  & 20\\ 
         Panel side length& $b$& mm  & 20\\ 
         Sector angle& $\gamma$ &$^\circ$ & 60\\ 
         Initial fold angle& $\theta_0$ &$^\circ$ & 57\\
         Panel thickness& $t$& mm & 0.28\\ 
         Unit length crease stiffness& $K_{\textup{cr}}/a$ &N & 0.032\\ 
         Base material Young's modulus& $E$& MPa & 7900\\
         Base material Poisson's ratio& $\nu$ && 0.3\\ \hline 
    \end{tabular}
    \caption{Geometric and material parameters of the cardboard Miura specimen.}
    \label{MiuraCardboard}
\end{table}

The compression test is performed with a STEP Lab electrodynamic tester (STEP Engineering S.r.l., Resana, Treviso, Italy) using a \qty{100}{N} load cell (AEP Transducers, Congnento, Italy) and two aluminum plates as holders. The test is displacement-controlled with a strain rate of $\qty{5.26e-3}{\per\second}$ and a sampling frequency of \qty{50}{Hz}. We conducted 6 compression tests under the same rate.

Next, we designed a qualitative experiment to verify the homogenized results in \Cref{fig:macro}, where the left boundary of the metamaterial is fixed, and the right boundary has an applied load pointing \ang[]{45} up to the right. A 7 by 7 Miura tessellation is laser cut from a cardboard and folded from flat. The geometric parameters of the flat pattern is identical to those in \Cref{MiuraCardboard}. The left boundary of the specimen is fully fixed by gluing to a wooden panel. The right boundary is cut with an array of circular holes \qty{2}{mm} in diameter. A straight metal wire passes through the holes to act as a slider, so that the right boundary of the specimen can fold freely in-plane. A cotton string attached to slider applies an external load that evenly distribute along the slider. The load points \ang[]{45} upward between $x_1$ and $x_2$ axes. \Cref{fig:macroi} shows the experimental setup of the fixed and slider boundary conditions, with the specimen at rest. \Cref{fig:macroj} shows the direction of applied load on the slider, along with the deformed specimen.

\section{Effective properties of curve crease origami}
\label{curveCreaseEff}
\begin{figure}
    \centering
    \includegraphics[]{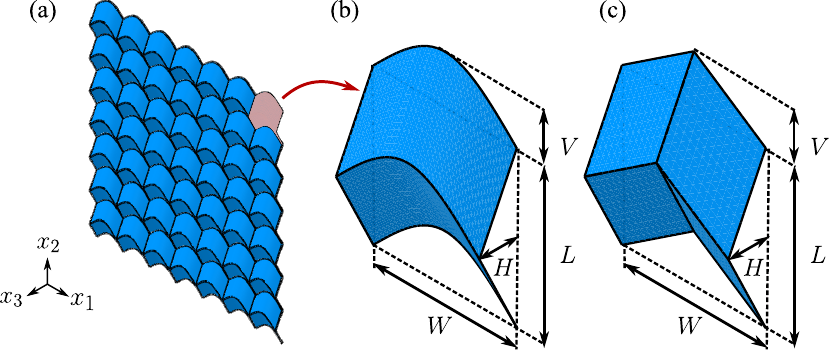}
    \begin{subfigure}{0.1\textwidth}
    \phantomsubcaption
    \label{fig:curvedMiuraa}
    \phantomsubcaption
    \label{fig:curvedMiurab}
    \end{subfigure}
    \caption{Homogenization of curve crease origami. (a)~Curve crease origami tessellation. (b)~Unit cell and its geometric parameters. (c)~A standard Miura unit cell for comparison.}
    \label{fig:curvedMiura}
\end{figure}

Curve crease origami is a typical non-rigid origami which applies to our framework. We select a pattern from the literature~\citep{liu2024design} with translational symmetry. \Cref{fig:curvedMiuraa} shows the origami tessellation. \Cref{fig:curvedMiurab} shows the details of its unit cell. Each curved panel of the unit cell has two straight edges and two curved edges. The curved edges are defined as splines passing through a front point, a midpoint, and an end point. \Cref{curvedMiuraProp} summarizes the geometry and base material properties. We report the effective compliance matrix below. Small numerical values in the matrix have been simplified to 0 because they are 2 to 4 orders of magnitude smaller than other entries, reflecting the planes of symmetry and hence anisotropy of the unit cells. The effective compliance matrix using asymptotic homogenization is
\begin{equation}
\begin{bmatrix}
e_{11} \\ 
e_{22} \\ 
2e_{12} \\ 
\kappa_{11} \\ 
\kappa_{22} \\ 
2\kappa_{12}
\end{bmatrix}
=
    \begin{bmatrix}
23.8   & 44.7   & 0  & 5.19  &  -9.80  &  -0.331 \\
44.7   & 84.8   & 0   & 9.80  & -18.5   &  -0.640 \\
0 & 0 &  0.0772  & 0 &  0 &  0 \\
5.19  & 9.80  & 0 &  1.46  & -2.74  & -0.166 \\
-9.80   & -18.5   &  0 & -2.74  &  5.21  &  0.322 \\
-0.331  & -0.640  &  0  & -0.166 &  0.322 & 12.2
\end{bmatrix}
\begin{bmatrix}
\sigma_{11} \\ 
\sigma_{22} \\ 
\sigma_{12} \\ 
\mu_{11} \\ 
\mu_{22} \\ 
\mu_{12}
\end{bmatrix}.
\end{equation}
The effective compliance matrix using energy-based homogenization is
\begin{equation}
\begin{bmatrix}
e_{11} \\ 
e_{22} \\ 
2e_{12} \\ 
\kappa_{11} \\ 
\kappa_{22} \\ 
2\kappa_{12}
\end{bmatrix}
=
    \begin{bmatrix}
24.2   & 45.6   & 0 &  5.25  & -9.92  & -0.492 \\
45.6   & 89.7   & 0  &  9.92  & -19.4  & -0.568 \\
0 & 0 &  0.0789  &  0 &  0 & 0 \\
5.25   &  9.92  &  0 &  1.54  & -2.90  & -0.126 \\
-9.92  & -19.4  &  0  & -2.90  &  5.68  &  0.153 \\
-0.492 & -0.568 & 0   & -0.126 &  0.153  & 15.1
\end{bmatrix}
\begin{bmatrix}
\sigma_{11} \\ 
\sigma_{22} \\ 
\sigma_{12} \\ 
\mu_{11} \\ 
\mu_{22} \\ 
\mu_{12}
\end{bmatrix}.
\end{equation}
The effective compliance matrix of the curve crease origami metamaterial is distinct from that of Miura in \Cref{complMtx}. This is because the curve crease unit cell has mirror symmetry, with one plane of symmetry parallel to the $x_2-x_3$ plane. The non-zero terms on the top right and the bottom left sections of the compliance matrix indicate coupling between in-plane strain (stress) and out-of-plane bending moments (curvature).

\begin{table}
    \centering\small
    \begin{tabular}{lccc} \hline 
         Unit cell length& $L$& mm & 36.06\\ 
         Unit cell width& $W$& mm & 33.28\\ 
         Unit cell height& $H$& mm & 8.66\\ 
         Spline $x_2$-length& $V$& mm & 11.09\\
         Panel thickness& $t$& mm & 0.13\\ 
         Base material Young's modulus& $E$ &MPa & 4000\\
         Base material Poisson's ratio& $\nu$& & 0.38\\ \hline 
    \end{tabular}
    \caption{Geometric and material parameters of curve crease origami metamaterial.}
    \label{curvedMiuraProp}
\end{table}

\bibliographystyle{elsarticle-harv} 
{\small\setlength{\bibsep}{0pt}\bibliography{cas-refs.bib}}

@misc{github_wendy,
  author = {Xuwen Li},
  title = {{OMhomogenization} [Software solution]},
  year = {2025},
  url = {https://archive.softwareheritage.org/swh:1:dir:01abb7dfbf60d374fb74eb02c847244d781c44e7;origin=https://github.com/Wendyli119/OMhomogenization;visit=swh:1:snp:0b26b14b0844eb2bd00a71821c8047a5c06b8d18;anchor=swh:1:rev:0c1a568120f91110162477f6d0cf509f5926f21e},
  note = {{S}oftware {H}eritage archive}
}

@incollection{Schenk2011,
  author    = {M. Schenk and S. Guest},
  title     = {Origami folding: A Structural Engineering Approach},
  year      = {2011},
  month     = {apr},
  pages     = {291--304},
  doi       = {10.1201/b10971-28},
  booktitle = {Origami 5},
  publisher = {A K Peters/{CRC} Press},
}

@Article{Filipov2015,
  author    = {Evgueni T. Filipov and Tomohiro Tachi and Glaucio H. Paulino},
  journal   = {Proceedings of the National Academy of Sciences},
  title     = {Origami tubes assembled into stiff, yet reconfigurable structures and metamaterials},
  year      = {2015},
  iissn      = {0027-8424},
  month     = {sep},
  number    = {40},
  pages     = {12321-12326},
  volume    = {112},
  doi       = {10.1073/pnas.1509465112},
  publisher = {Proceedings of the National Academy of Sciences},
}

@Article{Overvelde2017,
  author    = {Johannes T. B. Overvelde and James C. Weaver and Chuck Hoberman and Katia Bertoldi},
  journal   = {Nature},
  title     = {Rational design of reconfigurable prismatic architected materials},
  year      = {2017},
  iissn      = {0028-0836},
  month     = {jan},
  number    = {7637},
  pages     = {347-352},
  volume    = {541},
  doi       = {10.1038/nature20824},
  publisher = {Springer Science and Business Media {LLC}},
}

@Article{Hassani1998,
  author    = {B. Hassani and E. Hinton},
  journal   = {Computers {\&} Structures},
  title     = {A review of homogenization and topology optimization I—homogenization theory for media with periodic structure},
  year      = {1998},
  iissn      = {0045-7949},
  month     = {dec},
  number    = {6},
  pages     = {707-717},
  volume    = {69},
  doi       = {10.1016/s0045-7949(98)00131-x},
  publisher = {Elsevier {BV}},
}

@Article{Nassar2017,
  author    = {H. Nassar and A. Lebée and L. Monasse},
  journal   = {Proceedings of the Royal Society A: Mathematical, Physical and Engineering Sciences},
  title     = {Curvature, metric and parametrization of origami tessellations: theory and application to the eggbox pattern},
  year      = {2017},
  month     = {jan},
  number    = {2197},
  pages     = {20160705},
  volume    = {473},
  doi       = {10.1098/rspa.2016.0705},
  pmid      = {28265194},
  publisher = {The Royal Society},
  uurl       = {https://www.semanticscholar.org/paper/e72753bba0d33d880378a3031d7b05996c5929b7},
  venue     = {Proceedings of the Royal Society A},
}

@Article{Nassar2022,
  author        = {H. Nassar and A. Lebée and Emily Werner},
  journal       = {Extreme Mechanics Letters},
  title         = {Strain compatibility and gradient elasticity in morphing origami metamaterials},
  year          = {2022},
  month         = {may},
  pages         = {101722},
  volume        = {53},
  archiveprefix = {arXiv},
  doi           = {10.1016/j.eml.2022.101722},
  eprint        = {2207.08752},
  publisher     = {Elsevier {BV}},
  uurl           = {https://www.semanticscholar.org/paper/f60c65150628e636ddeee0d3f82e02d5c069abd8},
  venue         = {Extreme Mechanics Letters},
}

@Article{Filipov2017,
  author    = {E. Filipov and Ke Liu and T. Tachi and M. Schenk and G. Paulino},
  journal   = {International Journal of Solids and Structures},
  title     = {Bar and hinge models for scalable analysis of origami},
  year      = {2017},
  month     = {oct},
  pages     = {26--45},
  volume    = {124},
  doi       = {10.1016/J.IJSOLSTR.2017.05.028},
  publisher = {Elsevier {BV}},
  uurl       = {https://www.semanticscholar.org/paper/fb79cd73ab55c763a853950519451b772bf38d57},
}

@Article{Cheng2013,
  author    = {Geng-Dong Cheng and Yuan-Wu Cai and Liang Xu},
  journal   = {Acta Mechanica Sinica},
  title     = {Novel implementation of homogenization method to predict effective properties of periodic materials},
  year      = {2013},
  iissn      = {0567-7718},
  month     = {jul},
  number    = {4},
  pages     = {550-556},
  volume    = {29},
  doi       = {10.1007/s10409-013-0043-0},
  publisher = {Springer Science and Business Media {LLC}},
}

@Article{Hu2021,
  author    = {Y. C. Hu and Y. X. Zhou and K. W. Kwok and K. Y. Sze},
  journal   = {International Journal of Mechanics and Materials in Design},
  title     = {Simulating flexible origami structures by finite element method},
  year      = {2021},
  iissn      = {1569-1713},
  month     = {mar},
  number    = {4},
  pages     = {801-829},
  volume    = {17},
  doi       = {10.1007/s10999-021-09538-w},
  publisher = {Springer Science and Business Media {LLC}},
}

@Article{gattas2015behaviour,
  author    = {Gattas, JM and You, Zhong},
  journal   = {International Journal of Solids and Structures},
  title     = {The behaviour of curved-crease foldcores under low-velocity impact loads},
  year      = {2015},
  month     = {jan},
  pages     = {80--91},
  volume    = {53},
  doi       = {10.1016/j.ijsolstr.2014.10.019},
  publisher = {Elsevier},
}

@Article{hu2020folding,
  author    = {Hu, Yucai and Liang, Haiyi},
  journal   = {International Journal of Solids and Structures},
  title     = {Folding simulation of rigid origami with Lagrange multiplier method},
  year      = {2020},
  month     = {oct},
  pages     = {552--561},
  volume    = {202},
  doi       = {10.1016/j.ijsolstr.2020.06.016},
  publisher = {Elsevier},
}

@Article{chi2022bistable,
  author    = {Chi, Yinding and Li, Yanbin and Zhao, Yao and Hong, Yaoye and Tang, Yichao and Yin, Jie},
  journal   = {Advanced Materials},
  title     = {Bistable and multistable actuators for soft robots: Structures, materials, and functionalities},
  year      = {2022},
  month     = {mar},
  number    = {19},
  pages     = {2110384},
  volume    = {34},
  doi       = {10.1002/adma.202110384},
  publisher = {Wiley Online Library},
}

@Article{liu2017nonlinear,
  author    = {Liu, Ke and Paulino, Glaucio H},
  journal   = {Proceedings of the Royal Society A: Mathematical, Physical and Engineering Sciences},
  title     = {Nonlinear mechanics of non-rigid origami: an efficient computational approach},
  year      = {2017},
  month     = {oct},
  number    = {2206},
  pages     = {20170348},
  volume    = {473},
  doi       = {10.1098/rspa.2017.0348},
  publisher = {The Royal Society Publishing},
}

@Article{Cheung2014,
  author    = {Kenneth C. Cheung and Tomohiro Tachi and Sam Calisch and Koryo Miura},
  journal   = {Smart Materials and Structures},
  title     = {Origami interleaved tube cellular materials},
  year      = {2014},
  iissn      = {0964-1726},
  month     = aug,
  number    = {9},
  pages     = {094012},
  volume    = {23},
  doi       = {10.1088/0964-1726/23/9/094012},
  publisher = {IOP Publishing},
}

@article{schenk2013geometry,
  title={Geometry of Miura-folded metamaterials},
  author={Schenk, Mark and Guest, Simon D},
  journal={Proceedings of the National Academy of Sciences},
  volume={110},
  number={9},
  pages={3276--3281},
  year={2013},
  publisher={National Acad Sciences},
  doi = {10.1073/pnas.1217998110},
}

@article{vasudevan2024homogenization,
  title={Homogenization of non-rigid origami metamaterials as Kirchhoff-Love plates},
  author={Vasudevan, Siva P and Pratapa, Phanisri P},
  journal={International Journal of Solids and Structures},
  pages={112929},
  year={2024},
volume = {300},
  publisher={Elsevier},
  doi = {10.1016/j.ijsolstr.2024.112929}
}

@article{zheng2022vam,
  title={VAM-based reduced plate model for composite sandwich folded plate ({CSFP}) with V-shaped folded cores},
  author={Zheng, Shi and Yifeng, Zhong and Rong, Liu and Xiao, Peng},
  journal={Thin-Walled Structures},
  volume={170},
  pages={108601},
  year={2022},
  publisher={Elsevier},
  doi = {10.1016/j.tws.2021.108601}
}

@article{wei2013geometric,
  title={Geometric mechanics of periodic pleated origami},
  author={Wei, Zhiyan Y and Guo, Zengcai V and Dudte, Levi and Liang, Haiyi Y and Mahadevan, Lakshminarayanan},
  journal={Physical review letters},
  volume={110},
  number={21},
  pages={215501},
  year={2013},
  publisher={APS},
  doi = {10.1103/PhysRevLett.110.215501},
}

@book{reddy2006theory,
  title={Theory and analysis of elastic plates and shells},
  author={Reddy, Junuthula Narasimha},
  year={2006},
  publisher={CRC press},
  doi = {10.1201/9780849384165}
}

@article{Caillerie1984,
author = {Caillerie, D. and Nedelec, J. C.},
title = {Thin elastic and periodic plates},
journal = {Mathematical Methods in the Applied Sciences},
volume = {6},
number = {1},
pages = {159-191},
doi = {10.1002/mma.1670060112},
uurl = {https://onlinelibrary.wiley.com/doi/abs/10.1002/mma.1670060112},
eeprint = {https://onlinelibrary.wiley.com/doi/pdf/10.1002/mma.1670060112},
year = {1984}
}

@article{Wang2009,
    author = {Wang, Congyu and Feng, Liang and Jasiuk, Iwona},
    title = {Scale and Boundary Conditions Effects on the Apparent Elastic Moduli of Trabecular Bone Modeled as a Periodic Cellular Solid},
    journal = {Journal of Biomechanical Engineering},
    volume = {131},
    number = {12},
    pages = {121008},
    year = {2009},
    month = {11},
    iissn = {0148-0731},
    doi = {10.1115/1.4000192},
    uurl = {https://doi.org/10.1115/1.4000192},
    eeprint = {https://asmedigitalcollection.asme.org/biomechanical/article-pdf/131/12/121008/5769866/121008\_1.pdf},
}

@book{kalamkarov1997analysis,
  title={Analysis, design and optimization of composite structures},
  author={Kalamkarov, Alexander L and Kolpakov, Alexander G},
  volume={1},
  year={1997},
  publisher={Wiley},
note={\textsc{isbn: }9780471971894},
}

@article{cai2014novel,
  title={Novel numerical implementation of asymptotic homogenization method for periodic plate structures},
  author={Cai, Yuanwu and Xu, Liang and Cheng, Gengdong},
  journal={International Journal of Solids and Structures},
  volume={51},
  number={1},
  pages={284--292},
  year={2014},
  publisher={Elsevier},
doi={10.1016/j.ijsolstr.2013.10.003}
}

@article{eskandari2024unravelling,
  title={Unravelling Size-Dependent and Coupled Properties in Mechanical Metamaterials: A Couple-Stress Theory Perspective},
  author={Eskandari, Shahin and Shahryari, Benyamin and Akbarzadeh, Abdolhamid},
  journal={Advanced Science},
  volume={11},
  number={13},
  pages={2305113},
  year={2024},
  publisher={Wiley Online Library},
doi={10.1002/advs.202305113}
}

@book{reddy1996mechanics,
  title={Mechanics of Laminated Composite Plates: Theory and Analysis},
  author={Reddy, J. N.},
  year={1996},
  publisher={CRC-Press},
note={\textsc{isbn}: 9780849331015},
edition ={2}
}

@article{
Mirzajanzadeh2025,
author = {Morad Mirzajanzadeh  and Damiano Pasini },
title = {Reprogrammable curved-straight origami: Multimorphability and volumetric tunability},
journal = {Science Advances},
volume = {11},
number = {17},
pages = {eadu4678},
year = {2025},
doi = {10.1126/sciadv.adu4678},
uURL = {https://www.science.org/doi/abs/10.1126/sciadv.adu4678},
eeprint = {https://www.science.org/doi/pdf/10.1126/sciadv.adu4678}}

@article{melancon2021multistable,
  title={Multistable inflatable origami structures at the metre scale},
  author={Melancon, David and Gorissen, Benjamin and Garc{\'\i}a-Mora, Carlos J and Hoberman, Chuck and Bertoldi, Katia},
  journal={Nature},
  volume={592},
  number={7855},
  pages={545--550},
  year={2021},
  publisher={Nature Publishing Group UK London},
doi={10.1038/s41586-021-03407-4}
}

@article{Fang2024,
author = {Fang, Hongbin  and Wu, Haiping  and Liu, Zuolin  and Zhang, Qiwei  and Xu, Jian },
title = {Evaluating dynamic models for rigid-foldable origami: unveiling intricate bistable dynamics of stacked-Miura-origami structures as a case study},
journal = {Philosophical Transactions of the Royal Society A: Mathematical, Physical and Engineering Sciences},
volume = {382},
number = {2283},
pages = {20240014},
year = {2024},
doi = {10.1098/rsta.2024.0014},

uURL = {https://royalsocietypublishing.org/doi/abs/10.1098/rsta.2024.0014},
eeprint = {https://royalsocietypublishing.org/doi/pdf/10.1098/rsta.2024.0014}
}

@inproceedings{li2019vacuum,
  title={A vacuum-driven origami “magic-ball” soft gripper},
  author={Li, Shuguang and Stampfli, John J and Xu, Helen J and Malkin, Elian and Diaz, Evelin Villegas and Rus, Daniela and Wood, Robert J},
  booktitle={2019 International Conference on Robotics and Automation (ICRA)},
  pages={7401--7408},
  year={2019},
  organization={IEEE},
doi={10.1109/ICRA.2019.8794068}
}

@article{yang2024complex,
  title={Complex deformation in soft cylindrical structures via programmable sequential instabilities},
  author={Yang, Yi and Read, Helen and Sbai, Mohammed and Zareei, Ahmad and Forte, Antonio Elia and Melancon, David and Bertoldi, Katia},
  journal={Advanced Materials},
  volume={36},
  number={46},
  pages={2406611},
  year={2024},
  publisher={Wiley Online Library},
doi={10.1002/adma.202406611}
}

@article{xue2024rigid,
  title={Rigid-flexible coupled origami robots via multimaterial 3D printing},
  author={Xue, Wenbo and Sun, Zechu and Ye, Haitao and Liu, Qingjiang and Jian, Bingcong and Wang, Yanjie and Fang, Hongbing and Ge, Qi},
  journal={Smart Materials and Structures},
  volume={33},
  number={3},
  pages={035004},
  year={2024},
  publisher={IOP Publishing},
doi={10.1088/1361-665X/ad212c}
}

@article{li2023general,
  title={A general formulation for simulating the dynamic deployment of thick origami},
  author={Li, Jihui and Li, Qingjun and Sun, Tongtong and Zhu, Zhiwei and Deng, Zichen},
  journal={International Journal of Solids and Structures},
  volume={274},
  pages={112279},
  year={2023},
  publisher={Elsevier},
doi={10.1016/j.ijsolstr.2023.112279}
}

@article{zhang2022kirigami,
  title={Kirigami-based metastructures with programmable multistability},
  author={Zhang, Xiao and Ma, Jiayao and Li, Mengyue and You, Zhong and Wang, Xiaoyan and Luo, Yu and Ma, Kaixue and Chen, Yan},
  journal={Proceedings of the National Academy of Sciences},
  volume={119},
  number={11},
  pages={e2117649119},
  year={2022},
  publisher={National Academy of Sciences},
doi={10.1073/pnas.2117649119}
}

@article{li2019origamiMetawall,
  title={Origami metawall: Mechanically controlled absorption and deflection of light},
  author={Li, Min and Shen, Lian and Jing, Liqiao and Xu, Su and Zheng, Bin and Lin, Xiao and Yang, Yihao and Wang, Zuojia and Chen, Hongsheng},
  journal={Advanced science},
  volume={6},
  number={23},
  pages={1901434},
  year={2019},
  publisher={Wiley Online Library},
doi={10.1002/advs.201901434}
}

@article{harris2021impact,
  title={Impact response of metallic stacked origami cellular materials},
  author={Harris, JA and McShane, GJ},
  journal={International Journal of Impact Engineering},
  volume={147},
  pages={103730},
  year={2021},
  publisher={Elsevier},
doi={10.1016/j.ijimpeng.2020.103730}
}

@article{oudghiri2022effective,
  title={Effective linear wave motion in periodic origami structures},
  author={Oudghiri-Idrissi, Othman and Guzina, Bojan B},
  journal={Computer Methods in Applied Mechanics and Engineering},
  volume={399},
  pages={115386},
  year={2022},
  publisher={Elsevier},
doi={10.1016/j.cma.2022.115386}
}

@article{zhang2024propagation,
  title={Propagation of solitary waves in origami-inspired metamaterials},
  author={Zhang, Quan and Rudykh, Stephan},
  journal={Journal of the Mechanics and Physics of Solids},
  volume={187},
  pages={105626},
  year={2024},
  publisher={Elsevier},
doi={10.1016/j.jmps.2024.105626}
}

@article{yasuda2019origami,
  title={Origami-based impact mitigation via rarefaction solitary wave creation},
  author={Yasuda, Hiromi and Miyazawa, Yasuhiro and Charalampidis, Efstathios G and Chong, Christopher and Kevrekidis, Panayotis G and Yang, Jinkyu},
  journal={Science advances},
  volume={5},
  number={5},
  pages={eaau2835},
  year={2019},
  publisher={American Association for the Advancement of Science},
doi={10.1126/sciadv.aau2835}
}

@article{castle2016additive,
  title={Additive lattice kirigami},
  author={Castle, Toen and Sussman, Daniel M and Tanis, Michael and Kamien, Randall D},
  journal={Science advances},
  volume={2},
  number={9},
  pages={e1601258},
  year={2016},
  publisher={American Association for the Advancement of Science},
doi={10.1126/sciadv.1601258}}

@article{gao2023pneumatic,
  title={Pneumatic cells toward absolute {G}aussian morphing},
  author={Gao, Tian and Bico, Jos{\'e} and Roman, Beno{\^\i}t},
  journal={Science},
  volume={381},
  number={6660},
  pages={862--867},
  year={2023},
  publisher={American Association for the Advancement of Science},
doi={10.1126/science.adi2997}
}

@article{lahiri2023folding,
  title={Folding-angle framework for structural modeling of rigid triangulated Miura-ori lattices},
  author={Lahiri, Anandaroop and Pratapa, Phanisri P},
  journal={Journal of Mechanisms and Robotics},
  volume={15},
  number={5},
  pages={051004},
  year={2023},
  publisher={American Society of Mechanical Engineers},
doi={10.1115/1.4055742}
}

@article{feng2022simplified,
  title={A simplified mechanical model of the crease in the flexible origami structures},
  author={Feng, Yongjie and Wang, Mu and Qiu, Xinming},
  journal={International Journal of Solids and Structures},
  volume={241},
  pages={111530},
  year={2022},
  publisher={Elsevier},
doi={10.1016/j.ijsolstr.2022.111530}
}

@article{sturm2015multiscale,
  title={Multiscale modeling methods for analysis of failure modes in foldcore sandwich panels},
  author={Sturm, Ralf and Schatrow, Paul and Klett, Yves},
  journal={Applied Composite Materials},
  volume={22},
  pages={857--868},
  year={2015},
  publisher={Springer},
doi={10.1007/s10443-015-9440-9}
}

@article{zhang2021study,
  title={A study on ballistic performance of origami sandwich panels},
  author={Zhang, Jianjun and Lu, Guoxing and Zhang, Ying and You, Zhong},
  journal={International Journal of Impact Engineering},
  volume={156},
  pages={103925},
  year={2021},
  publisher={Elsevier},
doi={10.1016/j.ijimpeng.2021.103925}
}

@incollection{Heimbs2013,
author="Heimbs, Sebastian",
eeditor="Abrate, Serge
and Castani{\'e}, Bruno
and Rajapakse, Yapa D. S.",
title="Foldcore Sandwich Structures and Their Impact Behaviour: An Overview",
bookTitle="Dynamic Failure of Composite and Sandwich Structures",
year="2013",
publisher="Springer Netherlands",
pages="491--544",
doi="10.1007/978-94-007-5329-7_11",
}

@article{turco2024long,
  title={The long and winding road that leads to homogenisation of Kresling origami},
  author={Turco, Emilio and Barchiesi, Emilio and dell’Isola, Francesco},
  journal={International Journal of Non-Linear Mechanics},
  volume={163},
  pages={104756},
  year={2024},
  publisher={Elsevier},
doi={10.1016/j.ijnonlinmec.2024.104756}
}

@article{lewinski1988asymptotic,
  title={Asymptotic method of homogenization of two models of elastic shells},
  author={Lewinski, T and Telega, JJ},
  journal={Archiwum Mechaniki Stosowanej},
  volume={40},
  number={5},
  pages={705--723},
  year={1988}
}

@article{jamalimehr2022rigidly,
  title={Rigidly flat-foldable class of lockable origami-inspired metamaterials with topological stiff states},
  author={ Amin Jamalimehr and  Morad Mirzajanzadeh and Abdolhamid Akbarzadeh and Damiano Pasini },
  journal={Nature communications},
  volume={13},
  number={1},
  pages={1816},
  year={2022},
  publisher={Nature Publishing Group UK London},
doi={10.1038/s41467-022-29484-1}
}

@article{Almessabi,
author = {Almessabi, Abdulrahman  and Li, Xuwen  and Jamalimehr, Amin  and Pasini, Damiano },
title = {Reprogramming multi-stable snapping and energy dissipation in origami metamaterials through panel confinement},
journal = {Philosophical Transactions of the Royal Society A: Mathematical, Physical and Engineering Sciences},
volume = {382},
number = {2283},
pages = {20240005},
year = {2024},
doi = {10.1098/rsta.2024.0005}
}

@article{
Dang2025,
author = {Xiangxin Dang  and Shujia Chen  and Ali Elias Acha  and Lei Wu  and Damiano Pasini },
title = {Shape and topology morphing of closed surfaces integrating origami and kirigami},
journal = {Science Advances},
volume = {11},
number = {18},
pages = {eads5659},
year = {2025},
doi = {10.1126/sciadv.ads5659}}

@article{JIANG2025113433,
title = {Thin-walled tubular structures integrating origami patterns and tension-dominated bulkheads for enhanced energy absorption},
journal = {Thin-Walled Structures},
volume = {215},
pages = {113433},
year = {2025},
issn = {0263-8231},
doi = {10.1016/j.tws.2025.113433},
author = {Ben Jiang and Weinan Gao and Zhimin Xie and Damiano Pasini and Huifeng Tan}
}

@article{martinez2012elastomeric,
  title={Elastomeric origami: programmable paper-elastomer composites as pneumatic actuators},
  author={Martinez, Ramses V and Fish, Carina R and Chen, Xin and Whitesides, George M},
  journal={Advanced functional materials},
  volume={22},
  number={7},
  pages={1376--1384},
  year={2012},
  publisher={Wiley Online Library},
doi={10.1002/adfm.201102978}
}

@ARTICLE{Jin2022,
  author={Jin, Tao and Li, Long and Wang, Tianhong and Wang, Guopeng and Cai, Jianguo and Tian, Yingzhong and Zhang, Quan},
  journal={IEEE Transactions on Robotics}, 
  title={Origami-Inspired Soft Actuators for Stimulus Perception and Crawling Robot Applications}, 
  year={2022},
  volume={38},
  number={2},
  pages={748-764},
  doi={10.1109/TRO.2021.3096644}}

@article{hollister1992comparison,
  title={A comparison of homogenization and standard mechanics analyses for periodic porous composites},
  author={Hollister, Scott J and Kikuchi, Noboru},
  journal={Computational mechanics},
  volume={10},
  number={2},
  pages={73--95},
  year={1992},
  publisher={Springer},
doi={10.1007/BF00369853}
}

@article{kohn1984new,
  title={A new model for thin plates with rapidly varying thickness},
  author={Kohn, Robert V and Vogelius, Michael},
  journal={International Journal of Solids and Structures},
  volume={20},
  number={4},
  pages={333--350},
  year={1984},
  publisher={Elsevier},
doi={10.1016/0020-7683(84)90044-1}
}

@article{xu2025modeling,
  title={Modeling and computation of the effective elastic behavior of parallelogram origami metamaterials},
  author={Xu, Hu and Marazzato, Frederic and Plucinsky, Paul},
  journal={arXiv preprint arXiv:2503.08894},
  year={2025}
}

@article{xu2024derivation,
  title={Derivation of an effective plate theory for parallelogram origami from bar and hinge elasticity},
  author={Xu, Hu and Tobasco, Ian and Plucinsky, Paul},
  journal={Journal of the Mechanics and Physics of Solids},
  volume={192},
  pages={105832},
  year={2024},
  publisher={Elsevier},
doi={10.1016/j.jmps.2024.105832}
}

@article{liu2024design,
  title={Design of origami structures with curved tiles between the creases},
  author={Liu, Huan and James, Richard D},
  journal={Journal of the Mechanics and Physics of Solids},
  volume={185},
  pages={105559},
  year={2024},
  publisher={Elsevier},
doi={10.1016/j.jmps.2024.105559}
}

@article{LYU2021106791,
title = {Origami-based cellular mechanical metamaterials with tunable Poisson's ratio: Construction and analysis},
journal = {International Journal of Mechanical Sciences},
volume = {212},
pages = {106791},
year = {2021},
issn = {0020-7403},
doi = {10.1016/j.ijmecsci.2021.106791},
author = {Shengnan Lyu and Bo Qin and Huichao Deng and Xilun Ding},
}

@article{Jamalimehr2026,
author = {Jamalimehr, Amin and Akbarzadeh, Abdolhamid and Pasini, Damiano},
title = {Entangled Multistable Origami with Reprogrammable Stiffness Amplification and Damping},
journal = {Advanced Functional Material},
volume = {36},
number = {5},
pages = {e05620},
doi = {10.1002/adfm.202505620},
year = {2026}
}

\end{document}